\documentclass[12pt]{article}

\pdfoutput=1

\usepackage{amsmath,amssymb,exscale}
\usepackage{epsfig,psfrag}
\usepackage{multicol}
\usepackage{multirow}
\input epsf
\newcommand{\m}[1]{\marginpar{{\tiny *}} }
\newcommand{\Dslash}{{\not \!\!D}}
\newcommand{\pslash}{{\not \!p\ }}
\newcommand{\tr}{{\rm tr}}

\def\be{\begin{equation}}
\def\ee{\end{equation}}
\def\bea{\begin{eqnarray}}
\def\eea{\end{eqnarray}}

\begin{document}

\thispagestyle{empty}
\vspace{20pt}

\hfill
\vspace{20pt}
\begin{center}
{\Large \bf Minimal Composite Higgs Models at the LHC}
\end{center}

\vspace{15pt}
\begin{center}
{\large{Marcela Carena\footnote{carena@fnal.gov}$^{,a,b,c}$, Leandro Da Rold\footnote{daroldl@cab.cnea.gov.ar}$^{,d}$, Eduardo Pont\'on\footnote{eponton@ift.unesp.br }$^{,e}$}}

\vspace{20pt}
$^{a}$\textit{Fermi National Accelerator Laboratory, P.O. Box 500, Batavia, IL 60510}
\\[0.2cm]
$^{b}$\textit{Enrico Fermi Institute, University of Chicago, Chicago, IL 60637}
\\[0.2cm]
$^{c}$\textit{Kavli Institute for Cosmological Physics, University of Chicago, Chicago, IL 60637}
\\[0.2cm]
$^{d}$\textit{CONICET, Centro At\'omico Bariloche and Instituto Balseiro\\
Av.\ Bustillo 9500, 8400, S.\ C.\ de Bariloche, Argentina} \\[0.2cm]
$^{e}$\textit{ICTP South American Institute for Fundamental Research,
and \\ Instituto de F\'isica Te\'orica - Universidade Estadual
Paulista (UNESP), \\ Rua Dr.  Bento Teobaldo Ferraz 271, 01140-070
S\~ao Paulo, SP Brazil}
\end{center}

\begin{abstract}
We consider composite Higgs models where the Higgs is a pseudo-Nambu
Goldstone boson arising from the spontaneous breaking of an
approximate global symmetry by some underlying strong dynamics.  We
focus on the $SO(5) \to SO(4)$ symmetry breaking pattern, assuming the
``partial compositeness" paradigm.  We study the consequences on Higgs
physics of the fermionic representations produced by the strong
dynamics, that mix with the Standard Model (SM) degrees of freedom.
We consider models based on the lowest-dimensional representations of
SO(5) that allow for the custodial protection of the $Z\bar b b$
coupling, \textit{i.e.}~the ${\bf 5}$, ${\bf 10}$ and ${\bf 14}$.  We
find a generic suppression of the gluon fusion process, while the
Higgs branching fractions can be enhanced or suppressed compared to
the SM. Interestingly, a precise measurement of the Higgs boson
couplings can distinguish between different realizations in the
fermionic sector, thus providing crucial information about the nature
of the UV dynamics.
\end{abstract}

\newpage
\setcounter{footnote}{0}

\section{Introduction}

The discovery of a SM-like Higgs boson with a mass of about $126~{\rm
GeV}$~\cite{Aad:2012tfa,Chatrchyan:2012ufa} represents a fundamental
step towards a better understanding of the origin of Electroweak
Symmetry Breaking (EWSB).  Measuring its couplings with higher
precision will be one of the priorities in the 14~TeV run of the LHC,
and is one of the main motivations for building a future lepton
collider.  The phenomenological description of EWSB within the SM
framework provides a benchmark against which any deviations in the
Higgs boson couplings should be compared, as such deviations could
contain the key to a more fundamental understanding of this
phenomenon.

A currently open question is whether this particle is elementary
(\textit{i.e.}~pointlike), down to distance scales much shorter than
the EW scale, or if, on the contrary, it is a composite bound state of
more fundamental degrees of freedom, whose physics should be revealed
at energies not far above the weak scale.  In either case the
discovery of this scalar particle is truly remarkable.  If it turns
out to be elementary it would be the first and only known example of
this kind in nature.  Its existence at energies low compared to
\textit{e.g.}~the Planck scale could indicate that the universe as we
know it results from a rather perplexing fine-tuning, or perhaps more
plausibly that there is a symmetry at work as exemplified by
supersymmetric scenarios.  If it turns out that the Higgs boson is a
composite state arising from some underlying strong dynamics, we would
be in a situation that also presents new characteristics compared to
other known composite scalars.  For instance, unlike the pions of QCD,
the dynamics of the Higgs boson must lead to EWSB by generating a
non-vanishing vacuum expectation value (vev) for the composite scalar.

The fact that the LHC has not observed any major deviation from the SM
in its 7-8 TeV run indicates that any new physics should be roughly
above 1~TeV (although one can think of specific examples that are less
constrained, and also examples that are significantly more
constrained).  In the context of Higgs compositeness, this means that
there must exist a scalar resonance much lighter than the other strong
resonances.  It is then natural to interpret the Higgs as a
pseudo-Nambu Goldstone boson (pNGB) arising from the spontaneous
breaking of an approximate global symmetry of the new strong
sector~\cite{GK}.  This idea has received considerable attention
lately~\cite{Contino:2003ve}.  A question of special importance
centers on the type of deviations in the Higgs properties that would
be expected in such scenarios.  This has been studied to some extent
within specific realizations of a Higgs as a pNGB, and also in the
context of an effective low-energy parametrization such as the
SILH~\cite{Giudice:2007fh} and similar
approaches~\cite{Gillioz:2012se,Corbett:2012dm,Carmi:2012in,Jenkins:2013zja}.

We will focus here on the minimal case~\footnote{The terminology
``Minimal Composite Higgs Model (MCHM)" was actually introduced in a
slightly different context in~\cite{Dobrescu:1999gv}.  Our study is
limited to more recent models based on the pNGB idea which have also
been named MCHM~\cite{Agashe:2004rs}.  Since we consider a variety of
fermionic realizations, here the ``minimality" refers specifically to
the (common) bosonic sector.} based on the $SO(5) \to SO(4)$ symmetry
breaking pattern~\cite{Agashe:2004rs}, which leads to exactly four
Nambu-Goldstone bosons and contains a custodial symmetry that ensures
that the corrections to certain electroweak observables are
sufficiently suppressed.  Although the embedding of the SM gauge
sector is fixed by the above assumption, there is still a considerable
arbitrariness in how the SM fermionic sector is embedded into the
framework.  This depends, in particular, on which $SO(5)$
representations for the fermionic resonances are generated by the
strong dynamics and would therefore be sensitive to further details of
the specific UV realization of the idea.  Our aim is to study in
detail the implications for the properties of the Higgs boson.  In
particular, we will show that if one were to measure a robust
deviation from the SM in the rates $h \to \gamma\gamma$, $h \to ZZ$
and $h \to Z\gamma$ and to a lesser extent in $h \to \tau\tau$, one
could gather indirect information regarding the quantum numbers of the
fermionic resonances.  One also expects a generic reduction of the
Higgs production cross section (in particular through gluon fusion),
as well as a suppression of all Yukawa couplings w.r.t.~the SM.

There have been a number of studies on the phenomenology of a pNGB
Higgs as well as partial compositeness.  Since the pioneer work of
Ref.~\cite{Falkowski:2007hz} studying Higgs production by gluon
fusion, many works have considered the deviations of the Higgs
couplings in this setup, exploring the dependence on the degree of
compositeness of the fermions, the scale of compositeness and their
relation with the spectrum of resonances, among other important
variables~\cite{Low:2009di,Montull:2013mla,Gillioz:2013pba}.  However
most of them have considered generic regions of the parameter space,
that could be unphysical, in the sense that either there is no EWSB,
or the decay constant of the Higgs and its vev are not separated
enough to guarantee compatibility with EW precision measurements, or
the spectrum of the lightest level of states does not reproduce the SM
one, to cite a few examples.  To ensure that these conditions are
satisfied and therefore make a realistic study of the Higgs
phenomenology, in general requires a full study of the Higgs potential
that can only be performed in a well defined model, with the risk of
loosing some generality.  One of the purposes of this work is to make
a step in that direction.  We consider a family of well defined
models, with the same pattern of symmetry breaking for the pNGB Higgs
but allowing different representations for the fields of the theory.
This still represents considerable freedom and for this reason we make
some restrictive assumptions that ensure calculability of the Higgs
potential within the framework of a two site model.  We will also
assume that at high energies the symmetry behind the pNGB is linearly
realized for the massive resonances, and for that reason we will
include massive resonances in complete SO(5) representations.  It is
possible to relax some of these assumptions, for example by
considering models with more sites, or even to allow for logarithmic
divergences of the potential.\footnote{L.D. thanks Gilad Perez for
discussions on this topic.} Nevertheless, we hope that our setup can
still capture generic features of minimal pNGB
models.\footnote{Recently, another class of pNGB models based on
four-fermion interactions has been discussed in~\cite{Cheng:2013qwa}.
Although they rely on a different breaking pattern, in principle they
could be extended to ${\rm SO}(5) / {\rm SO}(4)$, following the
analysis of~\cite{Barnard:2013zea}.} We will show that it can give
information on the size of the corrections that one can expect on the
Higgs phenomenology as well as on the wealth and direction of
corrections that follow by allowing for different representations of
the fields.

This paper is organized as follows.  In Sec.~\ref{pNGB} we review the
basic aspects of the effective two-site description of the composite
Higgs scenario.  In Sec.~\ref{sec:models} we present the details of
the specific models we study in this work, which differ in the
realization of the fermionic sector.  In Sec.~\ref{sec:corrections} we
describe the low-energy consequences of the pNGB nature of the Higgs
and the presence of the composite resonances, while in
Sec.~\ref{sec_div_VH} we discuss the properties of the Higgs
potential.  Sec.~\ref{sec:pheno} contains our numerical results, while
Sec.~\ref{sec:tuning} contains some remarks on the tuning of the
phenomenologically viable models.  We summarize and conclude in
Sec.~\ref{sec:conclusions}.  We also include four appendices:
App.~\ref{app:generators} summarizes several useful group theoretical
results, App.~\ref{app:masses} contains the mass matrices of the gauge
sector of the models, App.~\ref{sec:correlators} contains all the
correlators for the low-energy limit of the various models, and
finally App.~\ref{app:loops} summarizes how we compute the 1-loop
processes $h \to \gamma\gamma$, $h \to ZZ$ and $h \to Z\gamma$.

\section{A minimal pNGB Higgs}
\label{pNGB}

We are interested in the minimal model that can deliver the Higgs as a
pNGB resonance arising from the spontaneous breaking of a global
symmetry in a strongly coupled sector (SCFT).  We will assume that the
SCFT has an exact global symmetry that is spontaneously broken to a
subgroup by effects of the strong dynamics, with the Higgs being the
associated Nambu-Goldstone boson (NGB).  The interactions of the
fields in the SCFT with the SM fields explicitly break the global
symmetry, leading to a Higgs potential at loop level.  In this case
the degeneracy of the vacuum is uplifted and the Higgs becomes a
pNGB, leading to a natural separation between the scale
of the resonances and the Higgs mass.  Usually the gauge contributions
to the 1-loop Coleman-Weinberg potential are aligned with the EW gauge
group.  However the fermion contributions, that are expected to be
large because of the large top mass, can induce a missalignement of
the vacuum triggering EW symmetry breaking dynamically.

Ref.~\cite{Agashe:2004rs} has shown that the minimal group containing
the SM EW gauge symmetry and an unbroken custodial symmetry that can
lead to a pNGB Higgs is SO(5).  This group is spontaneously broken to
${\rm SO(4)} \simeq {\rm SU(2)}_L \times {\rm SU(2)}_R$, with the
Higgs being the NGB in the coset SO(5)/SO(4) that transforms as a
${\bf 4}$ of SO(4).  Besides the Higgs, the SCFT is assumed to lead to
vector resonances in the adjoint representation of the global group
(these are created by the Noether currents of this symmetry).  In
addition, one assumes the existence of fermion resonances, some of
which can mix with the SM degrees of freedom.  We will consider that
all the massive composite resonances are in complete irreducible
representations of SO(5), realizing the symmetry in a linear way.  All
the composite states are taken to interact with typical couplings
$g_{\rho}\gg g_{SM}$.  The SM gauge and fermion fields can be
considered as external sources probing the SCFT, {\it i.e.}:
elementary fields.  The SM particles do not interact with the Higgs at
leading order, but these interactions are mediated by the resonances
of the SCFT that mix with the elementary fields.

The gauge fields of the SM weakly gauge a subgroup of the SCFT global
symmetry.  The conserved currents of the SCFT associated to this
subgroup couple linearly with the SM gauge fields, explicitly breaking
the global symmetry.  The masses of the EW vector bosons arises from
mixing between the vector resonances created by the SCFT currents and
the SM gauge fields, as well as from the Higgs interactions.

We are also interested in partial compositeness of the SM fermions,
that can be realized if the elementary fermions couple linearly with
operators of the SCFT: ${\cal L}\supset \lambda \bar\psi {\cal
O}_\psi$.  The low energy scaling of the coupling $\lambda$ is
controlled by the dimension of the corresponding SCFT operator $D =
\text{dim}[{\cal O}_\psi]$~\cite{Contino:2004vy,Agashe:2004rs}.  For
$D>5/2$ the coupling is irrelevant leading to small mixing between the
elementary fermions and the fermionic resonances created by the SCFT
operator.  For $D<5/2$ the coupling is relevant leading to large
mixing between the elementary fermion and the resonances, and thus to
a large Yukawa coupling.  The former case leads to light states that
are mainly elementary, whereas the latter one can lead to large
fermion masses, as for the top quark, which is associated with a large
degree of compositeness.

The proper normalization of hypercharge for fermions requires the
introduction of an extra ${\rm U(1)}_X$ symmetry in the composite
sector, with the identification $Y=T^{3}_{R}+X$, where $T^{3}_{R}$ is
the diagonal generator of SU(2)$_R$.  The SU(2)$_R$ charge of the
composite operators ${\cal O}_\psi$ is not fixed, allowing for
different representations ${\bf r}_{\cal O}$ under SO(5).  However,
the stringent constraints on the corrections to the $Zb\bar b$
couplings arising from LEP and SLC require a non-trivial protection of
the $Zb_L\bar{b}_L$ coupling.  Ref.~\cite{Agashe:2006at} has shown
that there is a subgroup of the custodial symmetry ${\rm O}(3)$ that
can ensure that the corrections to this coupling are indeed
sufficiently suppressed.  This symmetry requires that the
representation ${\bf r}_{{\cal O}_q}$, where ${\cal O}_q$ is coupled
to the doublet of the third generation $q_L$, decompose under SO(4)
as: ${\bf r}_{{\cal O}_q}\simeq{\bf 4}\oplus \dots$.  The smallest
representations satisfying this condition are: ${\bf r}={\bf 5},{\bf
10},{\bf 14}$.  On the other hand, invariance of the SCFT under ${\rm
SO(5)} \times {\rm U(1)}_X$ restricts the representations of the
operators ${\cal O}_u$ and ${\cal O}_d$, coupled with $t_R$ and $b_R$
respectively.  In this work we will consider several representations
${\bf r}_{\cal O}$ subject to the above restrictions, and we will
study their impact in the Higgs phenomenology at the LHC.

The scenario described in the previous paragraphs can be realized by
considering a theory in a slice of a warped five dimensional
space-time, with the metric being AdS$_5$ near the UV. The elementary
fields and resonances can be identified with degrees of freedom on the
UV boundary and Kaluza-Klein states, respectively.  However it is
possible to capture most of the essential ingredients by considering a
theory with the first level of resonances only, as in the
elementary/composite description of Ref.~\cite{Contino:2006nn}.  At
low energies one considers an effective description with elementary
fields, one level of resonances and linear mixing between them.  This
description has more freedom than the full 5D theory, allowing for new
terms~\cite{DeCurtis:2011yx} as well as a lack of correlation between
some parameters, such as the masses of the different resonances.  It
also has a cut-off of order a few TeV. However it is able to
parametrize a family of realistic theories with a pNGB Higgs and it is
still predictive enough to explore, at the LHC, the consequences of
the symmetries protecting the Higgs potential.  In the next subsections
we will summarize a realization of this effective theory.

\subsection{Effective description: 2-site model}
\label{sec:2site}

\begin{figure}[t] 
\centering
\includegraphics[width=0.6\textwidth]{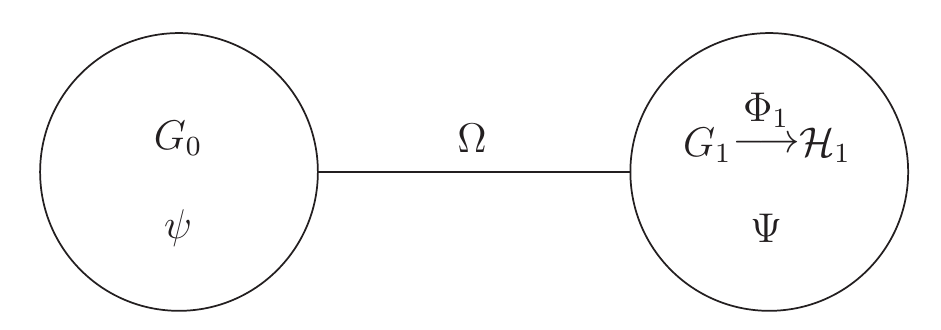}
\caption{Moose diagram of the two site theory describing the model.
$G_0$ is the SM gauge symmetry and $G_1 = {\rm SO(5)} \times {\rm
U(1)}_X$.  The (spontaneous) breaking of $G_1$ down to ${\cal H}_1 =
{\rm SO(4)} \times {\rm U(1)}_X$, is parametrized by a field $\Phi_1$,
that transforms under $G_1$ as a $5$ of SO(5) with $Q_X = 0$, and
whose vev is $\langle \Phi_1 \rangle = \{ 0,0,0,0,1 \}^T$.  The link
field transforms like $\Omega \to g_0 \Omega g_1^\dagger$, with
$g_{0,1} \in G_{0,1}$.}
\label{fig-moose}
\end{figure}

We consider the effective description of the Higgs as a pNGB arising
from a strongly coupled sector, as introduced in
Ref.~\cite{DeCurtis:2011yx} (see also~\cite{Panico:2011pw}).  The
simplest model has two sites: one called site-0 that describes
\textit{elementary} fields, and another called site-1 describing the
first level of resonances arising from the strongly coupled sector
(the \textit{composite} sector).  Site-0 contains a set of gauge and
fermion fields with the same symmetry group and fermionic
representations as the SM. We will call $G_0$ the gauge symmetry of
this site: $G_0 = {\rm SU(2)}_L\times {\rm U(1)}_Y$.\footnote{There is
also a color SU(3)$_C$ on each site, but we omit mentioning these
factors in the following.} Note that there are no elementary
\textit{scalar} fields.  On site-1 we consider a \textit{gauge}
symmetry $G_1 = {\rm SO(5)} \times {\rm U(1)}_X$, which allows to
describe effectively the lowest lying spin-1 resonances of the strong
dynamics.  Site-1 also contains several multiplets of fermion fields
in various representations of $G_1$, which will be described in detail
later.  The two sites are connected by a $\sigma$-model field
$\Omega$,\footnote{Strictly speaking, there are two link fields,
$\Omega$ and $\Omega_X$, for the ${\rm SO(5)}$ and ${\rm U(1)}_X$
factors.  These will be described in detail below.} transforming as
$\Omega \to g_0 \Omega g_1^\dagger$, with $g_{0,1} \in G_{0,1}$.  In
Fig.~\ref{fig-moose} we show the Moose diagram corresponding to this
theory.  We use lower case letters for fields on site-0 and upper case
letters for fields on site-1.

It turns out to be very convenient to extend $G_0$ to a spurious $G'_0
= {\rm SO(5)} \times {\rm U(1)}_x$.  This is achieved by introducing
non-dynamical gauge and fermion fields on site-0 that, together with
the dynamical fields that fill representation of $G_0 \subset G'_0$,
complete full representations of $G'_0$.  When one considers
\textit{all} the fields on site-0 as non-dynamical, they act as
sources for an exact global $G'_0$ symmetry, which is to be thought as
a global symmetry of the strongly coupled sector.  We assume that the
strong dynamics giving rise to the composite resonances spontaneously
breaks the ${\rm SO(5)}$ global factor down to SO(4), thus delivering
a set of NGB's in the coset SO(5)/SO(4).  These will be identified as
the composite Higgs, and are described by a field $\Phi_1$ as shown in
Fig.~\ref{fig-moose}.  The presence of the dynamical fields on site-0
explicitly breaks $G'_0$ (e.g.~by their kinetic terms, which are not
present for the spurious fields on site-0), and therefore generates a
potential for the Higgs, which becomes a pNGB. This potential is often
calculable and is one of the attractive theoretical features of these
scenarios.  The observation of a Higgs boson at the LHC and
the measurement of its mass and couplings then imposes non-trivial
constraints on the parameters of the model.

As we will see in detail below, the presence of $\Omega$ allows to
realize partial compositeness of the fermions through bilinear terms
involving a fermion $\psi$ at site-0 and a fermion $\Psi$ at site-1.
It also leads to non-zero masses for the \textit{axial} combination of
the gauge fields in sites-0 and 1, and contains the would-be NGB's
that are eaten in this process.

%

\subsection{Bosonic Sector}
\label{BosonicSector}

Let us consider first the Lagrangian describing the fields that
parametrize the ${\rm SO(5)} \to {\rm SO(4)}$ breaking.\footnote{Since
the NGB fields are neutral under $U(1)_X$, we omit this factor for
simplicity in this discussion, but it should be understood.} We denote
the unbroken generators of SO(5) [{\it i.e.}~of ${\cal H}_1 \equiv
SO(4) \simeq {\rm SU(2)}_L \times {\rm SU(2)}_R$] by $T^a$, while the
broken ones are denoted by $T^{\hat{a}}$.  For reference, we give
their explicit expressions in a convenient basis in
Appendix~\ref{app:generators}.  The NGB's are parametrized by
\begin{equation}
U(\Pi)=e^{i\Pi/f_1}\ , \qquad \Pi=\Pi^{\hat a}T^{\hat a} \ ,
\end{equation}
where $f_1$ is the corresponding decay constant.  The $G_1 = {\rm
SO(5)}$ symmetry is non-linearly realized, that is, under a $g_1 \in
G_1$ we have $U \to g_1 \, U \, h_1(g_1;\Pi)^\dagger$, where
$h_1(g_1;\Pi) \in {\cal H}_1$ is an element of the unbroken group,
that depends on the SO(5) transformation $g_1$ and the NGB fields
$\Pi$.  The leading order Lagrangian of these NGB's is
\begin{equation}
{\mathcal L}_{\rm NGB}=\frac{f_1^2}{2} \, {\cal D}^{\hat a}_\mu{\cal D}^{\mu\hat a}~,
\label{LNGB}
\end{equation}
with ${\cal D}^{\hat a}_\mu$ implicitly defined by $U^\dagger D_\mu
U=i{\cal E}^{a}_\mu T^a+{\cal D}^{\hat a}_\mu T^{\hat a}$.  The
covariant derivative contains the composite spin-1 resonances,
$A_\mu$, and leads to the interactions between these and the NGB's.
We defer the description of the interactions between the NGB's and the
fermions $\Psi$ on site-1 to the next section.

One can obtain a simpler and more explicit description of the above
sector by defining $\Phi_1=U\phi_1$, with $\phi_1^B = \delta^{B \, 5}$
($B = 1, \ldots 5$).  Under a $g_1 \in G_1$ one simply has $\Phi_1 \to
g_1 \Phi_1$, and it can be checked that the above Lagrangian can be
written as
\begin{equation}
{\mathcal L}_{\rm NGB} = \frac{f_1^2}{2}|D_\mu\Phi_1|^2~.
\label{LNGBSimple}
\end{equation}
In this form, the breaking of SO(5) down to SO(4) is simply
parametrized by $\langle \Phi_1 \rangle = \{ 0,0,0,0,1 \}^T$.

As for the Lagrangian describing the $\sigma$-model connecting the two
sites, at leading order one has:
\begin{equation}
{\mathcal L}_{\Omega} = \frac{f_\Omega^2}{4}\tr|D_\mu\Omega|^2 + \frac{f_{\Omega_X}^2}{4} |D_\mu\Omega_X|^2,
\label{LOmega}
\end{equation}
with
\bea
\Omega = e^{\sqrt{2} \, i \Pi_{\Omega} / f_\Omega}~,
\hspace{2cm}
\Omega_X = e^{\sqrt{2} \, i \Pi_{\Omega_X} / f_{\Omega_X}}~,
\eea
where $\Pi_{\Omega} = \Pi_{\Omega}^{b} \, T^b_{0-1}$ and $T^b_{0-1}$
denote the generators of ${\rm SO(5)}_0 \times {\rm SO(5)}_1 / {\rm
SO(5)}_{0+1}$, with ${\rm SO(5)}_{0+1}$ denoting the diagonal (vector)
subgroup of ${\rm SO(5)}_0 \times {\rm SO(5)}_1$.  We have also
included an additional link field $\Omega_X$ (with its decay constant
$f_{\Omega_X}$ and charges $Q_x = Q_X = 1$) for the ${\rm U(1)}_x
\times {\rm U(1)}_X$ factors.  The covariant derivatives above are
given by
\bea
D_\mu\Omega = \partial_\mu \Omega - i \tilde{a}_\mu\Omega + i\Omega \tilde{A}_\mu \ ,
\hspace{1cm}
D_\mu\Omega_X = \partial_\mu \Omega_X - i \tilde{x}_\mu\Omega_X + i\Omega_X \tilde{X}_\mu \ ,
\eea
where $\{\tilde{a}_\mu, \tilde{x}_\mu\}$ and $\{\tilde{A}_\mu,
\tilde{X}_\mu \}$ are the gauge fields of site-0 and site-1,
respectively (the tildes denote non-canonical normalization).  

Besides the terms above the bosonic Lagrangian includes the kinetic
terms for the gauge fields of $G_0$ and $G_1$:
\be
{\cal L}_{\rm gauge} = - \frac{1}{4 g^2_0} \tilde{w}^{j}_{L \, \mu \nu} \tilde{w}^{j \, \mu \nu}_L - \frac{1}{4 g^{\prime 2}_0} \tilde{b}_{\mu \nu} \tilde{b}^{\mu \nu} - \frac{1}{4 g^2_\rho} \tilde{A}^{B}_{\mu \nu} \tilde{A}^{B \, \mu \nu} - \frac{1}{4 g^{2}_X} \tilde{X}_{\mu \nu} \tilde{X}^{\mu \nu}~,
\label{Lgauge}
\ee
where $j = 1,2,3$, $B = 1, \ldots, 10$, and $\tilde{w}^{j}_{L \, \mu
\nu}$, $\tilde{b}_{\mu \nu}$ and $\{\tilde{A}^{B}_{\mu \nu},
\tilde{X}_{\mu\nu} \}$ are the field strengths of ${\rm SU(2)}_L$,
${\rm U(1)}_Y$ and ${\rm SO(5)} \times {\rm U(1)}_X$, respectively.
The embedding of ${\rm U(1)}_Y \subset {\rm SU(2)}_R \times {\rm
U(1)}_x$ on site-0 is obtained by the identifications $\tilde{w}^3_{R
\, \mu} = \tilde{x}_\mu = \tilde{b}_\mu$
so that $b_\mu$ couples to $Y = T^3_R + Q_X$ with coupling $g'_0 = g_0
g_x / \sqrt{g^2_0 + g^2_x}$.~\footnote{Here the fields are normalized
according to ${\cal L}_{\rm gauge} \supset -1 / (4g_0^2) \,
\tilde{w}^{3}_{R \, \mu \nu} \tilde{w}^{3 \, \mu \nu}_R - 1 / (4g_x^2)
\, \tilde{x}_{\mu \nu} \tilde{x}^{\mu \nu}$, while $\tilde{b}_\mu$ is
normalized as in Eq.~(\ref{Lgauge}).} The relation between the
couplings $g_0$ and $g^\prime_0$ and their SM counterparts will be
specified below, and similarly for the relation between the elementary
gauge fields $\tilde{w}^\mu_L$ and $\tilde{b}^\mu$, and the SM gauge
fields $W^\mu_L$ and $B^\mu$.  We assume that the couplings
characterizing the interactions of the composite spin-1 fields,
$g_\rho$ and $g_X$, are large but still perturbative.

The physical field content of the theory becomes evident in unitary
gauge, where the would-be NGB's eaten by the composite $A_\mu$'s are
set to zero.  This is achieved by a gauge transformation $g_1=\Omega$
(and using $\Omega_X$ for the $U(1)_X$ factor).  The physical NGB's
are then fully parametrized by
\begin{equation}
\Phi \equiv \Omega \, \Phi_1 = \frac{1}{h}\sin \frac{h}{f_h} \, (h_1,h_2,h_3,h_4,h \cot\frac{h}{f_h})^T \ ,
\end{equation}
with
\begin{equation}
\frac{1}{f^2_h}=\frac{1}{f_\Omega^2}+\frac{1}{f_1^2} \ , \qquad h^2=\sum_a h^{\hat a}h^{\hat a}~.
\label{eq:fh}
\end{equation}

The vacuum is characterized by the variable $\epsilon=\sin(v/f_h)$,
with $v=\langle h\rangle$ and $\langle\Phi\rangle^T = (0,0,0,
\epsilon,\sqrt{1-\epsilon^2})$.

The link field Lagrangian in unitary gauge reads
\be
{\cal L}_{\Omega} = \frac{1}{4} \, f^2_\Omega \left( \tilde{a}^B_\mu - \tilde{A}^B_\mu \right)^2 + \frac{1}{4} \, f^2_{\Omega_X} \left( \tilde{x}_\mu - \tilde{X}_\mu \right)^2~,
\ee
where we allowed for all possible external source fields on site-0.
Turning on only those that are dynamical as in Eq.~(\ref{Lgauge}), we
have
\bea
{\cal L}_{\Omega} &=& \frac{1}{2} \, m^2_\rho \sum^3_{i = 1} \left( t_\theta w^{i}_{L \, \mu} - A^i_{L \, \mu} \right)^2 + \frac{1}{2} \, m^2_\rho \left( t_\theta w^{3}_{R \, \mu} - A^3_{R \, \mu} \right)^2 + \frac{1}{2} \, m^2_X \left[ (g_x / g_X) \,   x_\mu - X_\mu \right]^2
\nonumber \\
& & \mbox{} + \frac{1}{2} \, m^2_\rho \sum^2_{k = 1} A^k_{R \, \mu} A^{k \, \mu}_{R} + \frac{1}{2} \, m^2_\rho \sum^4_{a = 1} A^{\hat{a}}_{\mu} A^{\hat{a} \, \mu}~,
\eea
where we denoted by $A^i_{L \, \mu}$ and $A^i_{R \, \mu}$ the
composite spin-1 fields associated with the ${\rm SU(2)}_L$ and ${\rm
SU(2)}_R$ factors in ${\rm SO(5)}$, respectively, defined
\bea
t_\theta ~=~ \frac{g_0}{g_\rho}~, 
\eea
and
\bea
m^2_\rho ~=~ \frac{1}{2} \, g^2_\rho f^2_\Omega~, 
\hspace{1.5cm} 
m^2_X ~=~ \frac{1}{2} \, g^2_X f^2_{\Omega_X}~,
\label{mrhomX}
\eea
and rescaled the fields according to $\tilde{w}_{L, R} = g_0 w_{L,
R}$, $\tilde{A}_{L, R} = g_\rho A_{L, R}$, $\tilde{x} = g_x x$ and
$\tilde{X} = g_X X$ for canonical normalization.  Recall that
$\tilde{w}^{3}_{R \, \mu}$ and $\tilde{x}_\mu$ are written in terms of
$\tilde{b}_\mu$ as given after Eq.~(\ref{Lgauge}).  By going to the
mass eigenbasis, we can then identify (in the limit that $\langle h
\rangle = 0$), the following massless fields:
\be
W^{i}_{L \, \mu} = c_\theta w^{i}_{L \, \mu} + s_\theta A^{i}_{L \, \mu}~,
\hspace{1cm} 
\textrm{for}~~i = 1,2,3~,
\label{Wboson}
\ee
and
\be
B_{\mu} = \frac{1}{\sqrt{1 + t^2_{\theta'_\rho} + t^2_{\theta'_X}}} \left[ b_{\mu} + t_{\theta'_\rho} A^{3}_{R \, \mu} + t_{\theta'_X} X_{\mu} \right]~,
\label{Bboson}
\ee
where $t_{\theta'_\rho} = g'_0/g_\rho$ and $t_{\theta'_X} =g'_0 / g_X$.
These are then identified with the SM gauge fields, and acquire masses
when $\langle h \rangle = v$.  Indeed, one finds that
\begin{equation}
\label{eq-vSM}
m_Z \approx \frac{1}{2} \sqrt{g^2 + g^{\prime
2}} \, \epsilon \, f_h~,
\hspace{1cm}
\textrm{hence}
\hspace{1cm}
v_{SM} = 246~{\rm GeV} \simeq \epsilon \, f_h~.
\end{equation}
One can also identify the SM gauge couplings:
\be
g = c_\theta g_0 = \left( \frac{1}{g^{2}_0} + \frac{1}{g^2_\rho} \right)^{-1/2}~,
\hspace{1cm}
g' = \frac{g'_0}{\sqrt{1 + t^2_{\theta'_\rho} + t^2_{\theta'_X}}}
= \left( \frac{1}{g^{\prime 2}_0} + \frac{1}{g^2_\rho} + \frac{1}{g^2_X} \right)^{-1/2}~.
\label{SMggp}
\ee
We note here, for later use, that in the case that $g_X = g'_0 g_\rho
/ \sqrt{g^2_0 - g^{\prime 2}_0}$ one has that $t_\theta = g_0 / g_\rho
= g_x / g_X$, \textit{i.e.} the ratios of elementary to composite
couplings in the two sites coincide for the SO(5) and ${\rm U(1)}_X$
factors.  In this case the usual Weinberg angle coincides with the
naive elementary Weinberg angle: $t_W = g' / g = g'_0 / g_0$.

The combinations orthogonal to Eqs.~(\ref{Wboson}) and (\ref{Bboson})
are massive even in the absence of the Higgs vev.  For the ${\rm
SU(2)}_L \times {\rm SO(5)}$ factor one finds states
$\tilde{\rho}^i_{L \, \mu} = c_\theta A^{i}_{L \, \mu} - s_\theta
w^{i}_{L \, \mu}$ (i = 1,2,3) with mass $m_{\tilde{\rho}}^2 = (1 +
t^2_\theta) m^2_\rho$;  the other fields in ${\rm SO(5)}$, that do not
mix with elementary fields, correspond to two (charged) fields in
${\rm SU(2)}_R \subset {\rm SO(5)}$ with mass $m_\rho$, and four
fields associated with the broken ${\rm SO(5)}/{\rm SO(4)}$
generators, with squared masses
\be
m^2_a = \frac{1}{2} \, g^2_\rho (f^2_\Omega + f_1^2)~,
\label{ma}
\ee
the latter term arising from ${\cal L}_{\rm NGB}$ in
Eq.~(\ref{LNGBSimple}).  There are also two massive neutral resonances
arising from the ``hypercharge" gauge sector.  Assuming that $m_X =
m_\rho$, the expressions for the latter simplify considerably and one
finds that the state $\propto t_{\theta'_X} A^{3}_{R \, \mu} -
t_{\theta'_\rho} X_\mu$ has mass $m_\rho$ while the state $\propto
t_{\theta'_\rho} A^{3}_{R \, \mu} + t_{\theta'_X} X_\mu -
(t^2_{\theta'_\rho} + t^2_{\theta'_X}) \, b_{\mu}$ has mass squared
$[1 + t^2_{\theta'_\rho} + t^2_{\theta'_X} ] \, m^2_\rho$.  All of the
above states receive small corrections when $\langle h \rangle$ is
turned on.  For completeness, we give the full mass matrices in
App.~\ref{app:masses}.

\subsection{Fermionic Sector}

On site-0 we consider a set of massless chiral fields $\psi$ with the
same quantum numbers as the fermions of the SM. As explained earlier,
often these will be extended to full $G'_0$ multiplets by the
introduction of additional fermionic sources.  On site-1 we include a
set of massive Dirac fermions $\Psi^{(r)}$ arising from the strong
dynamics, transforming in different representations $r$ of $G_1$.  The
fermions on site-0 and site-1 can be connected by the $\sigma$-model
fields $\Omega$ and $\Omega_X$.  Similarly, fermions in different
representations on site-1 can be connected by the NGB fields in $U$.
The generic form of the fermion Lagrangian at quadratic order in the
fermion fields that we consider in this work is
\begin{equation}
{\cal L}_f=i\bar\psi \Dslash_0 \psi+\bar\Psi^{(r)}(i\Dslash_1-m_r)\Psi^{(r)}+m^{(rs)}\bar\Psi^{(r)}_LUP^{(rs)}U^\dagger\Psi^{\prime (s)}_R+\Delta^{(r)}\bar\psi^{(r)} \Omega \, [\Omega_X]^{q_r} \Psi^{(r)}+{\rm h.c.}~
\label{LfermionGeneric}
\end{equation}
Here $D_{\mu \, 0}$ and $D_{\mu \, 1}$ are the covariant derivatives
on sites-0 and 1 ({\it i.e.}~carrying the corresponding elementary or
composite gauge fields) and $P^{(rs)}$ is a projector in the space of
representations of ${\cal H}_1$.  Note that besides the ``diagonal"
fermion masses, $m_r$, the NGB's can allow additional ``non-diagonal"
mass terms coupling different fermion representations.  From the point
of view of the fermion field content, these bear some similarity with
the Yukawa terms of the SM. By some simple algebraic manipulations,
this term can be written in terms of the field $\Phi$ plus mixing
terms between composite fermions in the same representation of SO(5).
In the next section we will show them explicitly for each fermion
embedding.  The last term in Eq.~(\ref{LfermionGeneric}) leads to
mixing between the elementary and composite fields, and realizes the
idea of partial compositeness in the fermion sector.  This term is
only written for pairs of elementary and composite fermions with the
same quantum numbers under $G'_0$ and $G_1$ [here $q_r$ denotes the
common charge of $\psi^{(r)}$ under $U(1)_x$ and $\Psi^{(r)}$ under
$U(1)_X$].  Note that this last term violates the $G'_0 \times G_1$
symmetry explicitly only after the non-dynamical source fields in
$\psi^{(r)}$ are set to zero.  The precise form of the above
Lagrangian depends on the representations of the fermionic resonances
which would be determined by the strongly coupled UV completion.  In
the absence of such an explicit theory, we will study several
possibilities based on the lowest dimensional representations of
SO(5).  We will provide the detailed forms of the Lagrangians in
Sec.~\ref{sec:models}.

A comment regarding the structure of the third term that contains the
interactions between fermions and the NGB's parametrized by $U$ is in
order.  As will be discussed in Sec.~\ref{sec_div_VH} and
explicitly shown in Sec.~\ref{sec:models}, we will not consider
the most general mass terms.  Rather, in order to obtain a finite
Higgs potential $V_H$ we have imposed some constraints.  By $\Psi_L$
we mean the Left-handed component of the fields $\Psi$ on site-1 that
mix with the fields $\psi_L$ on site-0, whereas $\Psi'_R$ is the
Right-handed component of the fields $\Psi$ on site-1 that mix with
$\psi_R$ on site-0.  Therefore $m^{(rs)}$ will only connect
$\Psi^{(r)}_L$ and $\Psi^{\prime (s)}_R$, but there are neither terms
of type $\bar\Psi^{(r)}_RUP^{(rs)} U^\dagger\Psi^{\prime (s)}_L$ nor
of type $\bar\Psi^{(r)}_L U P^{(rs)}U^\dagger\Psi^{(s)}_R$.  

Also, to avoid large corrections to $Zb_L\bar b_L$ we will embed $Q$,
the composite multiplet mixing with $q_L$, in a multiplet such that:
$T_L=T_R$ and $T^{3}_L = T^{3}_R = -1/2$ for $Q_d$, with $T_{L,R}$ the
${\rm SU(2)}_{L,R}$ generators and $Q_d$ the component mixing with
$b_L$.  This means that $Q$ contains a $({\bf 2},{\bf 2})$ of ${\rm
SU(2)}_L \times {\rm SU(2)}_R$.  The smallest irreducible
representations of SO(5) satisfying this condition are the fundamental
${\bf 5}$, the adjoint (antisymmetric) ${\bf 10}$ and the (symmetric)
${\bf 14}$.  The U(1)$_X$ charge is fixed by demanding that the
correct hypercharge be reproduced, where $Y=T^{3}_R+X$, leading to
$X=2/3$.  For the composite multiplet $U$ ($D$) mixing with $u_R$
($d_R$) we will consider several possibilities, but we will choose
those that allow to write a Yukawa term $\bar Q \Phi^n U$ ($\bar Q
\Phi^n D$) that is a singlet of $G_1$ and contain a ${\bf 1}_{2/3}$
(${\bf 1}_{-1/3}$) of ${\rm SU(2)}_L \times {\rm U(1)}_Y$.  We will
consider the following models: MCHM$_5$ (all the fermions in ${\bf
5}$), MCHM$_{10}$ (all the fermions in ${\bf 10}$), and models
involving more than one representation: MCHM$_{10-5-10}$,
MCHM$_{5-5-10}$, MCHM$_{5-10-10}$, MCHM$_{14-14-10}$ and
MCHM$_{14-1-10}$, with notation MCHM$_{Q-U-D}$ (see also
Refs.~\cite{Carena:2006bn,Carena:2007ua,Csaki:2008zd,Pomarol:2012qf,Pappadopulo:2013vca}).

Since the BR of the Higgs decaying to $\tau^+\tau^-$ is not
negligible, we will also consider the leptonic sector.  For each
generation we include two multiplets of composite fermions: $L$ and
$E$, mixing with the elementary leptons $\ell_L$ and $e_R$
respectively.  These composite leptons are singlets of ${\rm SU(3)}_C$
and, for each model, we choose their SO(5) embedding copying that of
$Q$ and $D$, again with $X$ chosen to obtain $Y=T^{3}_{R}+X$.

\subsection{The Low-Energy Effective Theory}
\label{sec:EFT}

In order to make contact with measurements at current energies, it is
useful to integrate out the heavy resonances in the previous model.
We will present in this section the result of integrating out the
spin-1 resonances, which is common to the various models we consider
and illustrates the general procedure.  In Sec.~\ref{sec:models} we
present the result of integrating out the heavy fermionic sector in
the different models of interest.

In order to simplify the computations it is useful to start with all
elementary fields as non-dynamical and filling complete $G'_0 = {\rm
SO(5)} \times {\rm U(1)}_x$ representations, as discussed in
Subsection~\ref{sec:2site} above.  Since in this limit the full theory
has an exact global ${\rm SO(5)} \times {\rm U(1)}_X$ symmetry,
corresponding to the diagonal group of $G'_0 \times G_1$ (due to the
vev of the link fields), the effective theory for these external
sources must take a fully ${\rm SO(5)} \times {\rm U(1)}_X$ form.
Listing all the invariant terms that are quadratic in the external
gauge fields, we must obtain (in momentum space):
\be
{\cal L}_{\rm eff}^{\rm sources} = \frac{1}{2} \, \Pi^{(0)}_A \, {\rm tr}(\tilde{a}_\mu \tilde{a}^\mu) + \frac{1}{2} \, \Pi^{(2)}_A \, \Phi^T \tilde{a}_\mu \tilde{a}^\mu \Phi + \frac{1}{2} \, \Pi^{(0)}_X \, \tilde{x}_\mu \tilde{x}^\mu~,
\label{LeffGeneral}
\ee
for some functions $\Pi^{(0)}_A(p^2)$, $\Pi^{(2)}_A(p^2)$ and
$\Pi^{(0)}_X(p^2)$.  In the limit that $\langle h \rangle = 0$, {\it
i.e.}~$\Phi = \{ 0,0,0,0,1 \}^T$, this becomes
\be
\left. {\cal L}_{\rm eff}^{\rm sources} \right|_{h = 0} = \frac{1}{2} \, \Pi^{(0)}_A \, \tilde{a}^j_\mu \tilde{a}^{j \, \mu} + \frac{1}{2} \left( \Pi^{(0)}_A + \frac{1}{2} \, \Pi^{(2)}_A \right) \tilde{a}^{\hat{b}}_\mu \tilde{a}^{\hat{b} \, \mu} + \frac{1}{2} \, \Pi^{(0)}_X \, \tilde{x}_\mu \tilde{x}^\mu~,
\label{Leffheq0}
\ee
where $j = 1, \ldots 6$ and $b = 1,2,3,4$ label the two SO(4)
representations in the adjoint of SO(5): ${\bf 10} = {\bf 6} + {\bf
4}$.  We can then integrate out the heavy spin-1 resonances from
${\cal L} = {\cal L}_{\rm gauge} + {\cal L}_{\Omega} + {\cal L}_{\rm
NGB}$ [Eqs.~(\ref{LNGBSimple})--(\ref{Lgauge})] in the limit $\langle h
\rangle = 0$ and in unitary gauge, and identify $\Pi^{(0)}_A$,
$\Pi^{(2)}_A$ and $\Pi^{(0)}_X$.  The equations of motion for the
heavy fields simply read
\be
\tilde{A}^j_\mu = - \frac{m^2_\rho}{p^2 - m^2_\rho} \, \tilde{a}^j_\mu~,
\hspace{1cm}
\tilde{A}^{\hat{b}}_\mu = - \frac{m^2_\rho}{p^2 - m^2_a} \, \tilde{a}^{\hat{b}}_\mu~,
\hspace{1cm}
\tilde{X}_\mu = - \frac{m^2_X}{p^2 - m^2_X} \, \tilde{x}_\mu~,
\ee
where $m_\rho$ and $m_X$ were defined in Eq.~(\ref{mrhomX}), and
$m_a$ was defined in Eq.~(\ref{ma}).  Replacing back
in the original Lagrangian, we find
\be
\Pi^{(0)}_A = \hat{\Pi}_6~,
\hspace{1cm}
\Pi^{(2)}_A = 2( \hat{\Pi}_{4} - \hat{\Pi}_6)~,
\hspace{1cm}
\Pi^{(0)}_X = \hat{\Pi}_X~,
\ee
where
\be
\hat{\Pi}_6 = \frac{p^2 m^2_\rho}{g^2_\rho (p^2 - m^2_\rho)}~,
\hspace{1cm}
\hat{\Pi}_{4} = \frac{m^2_\rho (p^2 + m^2_\rho - m^2_a)}{g^2_\rho (p^2 - m^2_a)}~,
\hspace{1cm}
\hat{\Pi}_X = \frac{p^2 m^2_X}{g^2_X (p^2 - m^2_X)}~.
\ee
Going back to Eq.~(\ref{LeffGeneral}) evaluated for an arbitrary Higgs
configuration, and keeping only the sources corresponding to the SM
gauge fields, as described after Eq.~(\ref{Lgauge}), one finds in an
obvious notation:
\be
{\cal L}_{\rm eff} = \frac{1}{2} \, \sum^3_{i=1} \Pi_{\tilde{w}^i_L} \tilde{w}^i_{L \, \mu} \tilde{w}^{i \, \mu}_L
+ \Pi_{\tilde{w}^3_L \, \tilde{b}} \, \tilde{w}^3_{L \, \mu} b^{\mu}
+ \frac{1}{2} \, \Pi_{\tilde{b}} \, \tilde{b}_{\mu} \tilde{b}^{\mu}~,
\ee
where
\bea
\Pi_{\tilde{w}^i_L} = \Pi^{(0)}_A + \frac{1}{4} \, \Pi^{(2)}_A \sin^2(h/f_h)~,
\hspace{1cm}
\Pi_{\tilde{w}^3_L \, \tilde{b}} = - \frac{1}{4} \, \Pi^{(2)}_A \sin^2(h/f_h)~,
\nonumber \\ [0.4 em]
\Pi_{\tilde{b}} = \Pi^{(0)}_X + \Pi^{(0)}_A  + \frac{1}{4} \, \Pi^{(2)}_A \sin^2(h/f_h)~.
\hspace{2cm}
\eea
These correlators, which are valid to all orders in momentum as well
as on the Higgs vev will be useful when evaluating the Higgs potential
in Sec.~\ref{sec_div_VH}. 

\section{Models based on the ${\bf 1}$, ${\bf 5}$, ${\bf 10}$ and ${\bf 14}$ Reps.~of SO(5)}
\label{sec:models}

In this section, we present a summary of the models we consider in
this work, which differ in the SO(5) representations of the fermionic
resonances arising from the strongly interacting sector.  We start
with a few general comments, and then describe each model in turn.
The reader may want to read only the first part of this section and
skip to Sec.~\ref{sec:corrections}, coming back to
Subsections~\ref{sec:MCHM5}-\ref{sec:MCHM14110} only if further
details are desired.

In unitary gauge the fermion Lagrangian can be written as:
\begin{eqnarray}
\label{Lfermions}
{\cal L}_f&=&\sum_{\psi=q_L,u_R,d_R} Z_\psi \bar\psi i\Dslash \psi + \bar q_L \Delta_{q} Q_R + \bar u_R \Delta_u U_L + \bar d_R \Delta_d D_L + {\rm h.c.} \\ 
&+& \sum_{\Psi=Q,U,D} \bar\Psi (i\Dslash-m_\Psi) \Psi 
+  m_{y_u} \bar Q_L U_R + m_{y_d} \bar Q_L D_R + {\cal L}_y(Q_L,U_R,D_R, \Phi) + {\rm h.c.}
\nonumber
\end{eqnarray}
Depending on the fermion embedding, the terms $m_{y_u} \bar Q_L U_R +
m_{y_d} \bar Q_L D_R$ can contain a gauge singlet or not.  They are
present only in the former case.  The explicit form of the Yukawa
terms also depends on the fermion embedding, and will be specified for
each model below.\footnote{These Yukawa interactions are not yet the
SM Yukawa interactions, but will give rise to them.  Therefore, we
will refer to them as ``proto-Yukawa" interactions.} For the MCHM$_5$
it is necessary to include two different composite fermions $Q^u$ and
$Q^d$ that mix with the elementary doublet $q_L$.  In this case, we
replace $\bar q_L \Delta_{q} Q_R \to \bar q_L\Delta_{q^u} Q^u_R + \bar
q_L\Delta_{q^d} Q^d_R$ and $m_{y_u} \bar Q_L U_R + m_{y_d} \bar Q_L
D_R \to m_{y_u} \bar Q^u_L U_R + m_{y_d} \bar Q^d_L D_R$ above.
However, for the other models a single $Q$ is sufficient, as written
in Eq.~(\ref{Lfermions}).

Integrating out the composite resonances we obtain an effective theory
involving the elementary degrees of freedom only, in complete analogy
to the procedure presented in Sec.~\ref{sec:EFT} for the spin-1
case.  The fermions are in complete irreducible representations $r_5$
of SO(5).  However, due to the spontaneous breaking ${\rm SO(5)} \to
{\rm SO(4)}$ in the composite sector, each fermion is in general split
into several irreducible representations $r_4$ of SO(4):
$\psi^{r_5}=\sum_{r_4}\alpha_{r_5,r_4}\psi^{r_4}$, with
$\alpha_{r_5,r_4}$ the coefficients associated to the decomposition.
Thus, before EWSB, and taking $\langle\Phi\rangle=\Phi_0$ ({\it
i.e.}~$h=0$), one can write the effective Lagrangian as:
\begin{eqnarray}
\label{Leff-fermions-SO4}
\left. {\cal L}_{\rm eff} \right|_{h = 0} = \sum_{\psi=q_L,u_R,d_R}\sum_{r_4}\bar \psi^{(r_4)} \pslash (Z_\psi+\hat\Pi_\psi^{(r_4)}) \psi^{(r_4)} 
+ \sum_{\psi=u,d}\sum_{r_4}\bar q_L^{(r_4)} \hat M_{\psi^{(r_4)}} \psi_R^{(r_4)} + {\rm h.c.}
\end{eqnarray}
The explicit form of the correlators $\hat\Pi_\psi^{r_4}$ and $\hat
M_\psi^{r_4}$ are given in the appendix for the different models.

It is then simple to compare to the correlators of an effective
Lagrangian, ${\cal L}_{\rm eff}$, written in fully SO(5) invariant
form with the help of an arbitrary $\Phi$ (one should list all
possible ${\rm SO(5)} \times {\rm U(1)}_X$ invariant operators that
are quadratic in the external fermionic sources, which depends on the
specific model in question).  If one then retains the SM degrees of
freedom only, the effective Lagrangian for the elementary fermions
takes the form
\begin{eqnarray}
\label{Leff-fermions}
{\cal L}_{\rm eff}&=& \bar u_L \pslash (Z_q+\Pi_{u_L}) u_L + \bar d_L \pslash (Z_q+\Pi_{d_L}) d_L + \bar u_R \pslash (Z_u+\Pi_{u_R}) u_R + \bar d_R \pslash (Z_d+\Pi_{d_R}) d_R \nonumber \\ [0.4em]
& & \mbox{} + \bar u_L M_u u_R + \bar d_L M_d d_R + {\rm h.c.}
\end{eqnarray}
The correlators $\Pi_\psi$ and $M_\psi$ can be expressed in terms of
the correlators of the SO(4) symmetric theory $\hat\Pi_\psi^{r_4}$ and
$\hat M_\psi^{r_4}$, and have an explicit (and generally simple)
dependence on $s_h = \sin h / f_h$ and $c_h = \cos h / f_h$.  We show
below the full expressions for each specific model.

The spectrum of fermions that mix with the SM ones (as well as the
masses of the SM degrees of freedom) is given by the zeroes of the
quadratic operator
\begin{equation}\label{eq-spectrum}
{\rm Zero}\left\{p^2[Z_q+\Pi_{\psi_L}(p^2)][Z_\psi+\Pi_{\psi_R}(p^2)]-|M_\psi(p^2)|^2\right\} \ , \qquad \psi=u,d \ .
\end{equation}

The SM states, being lighter than the compositeness scale, can be
obtained by expanding Eq.~(\ref{eq-spectrum}) to ${\cal O}(p^2)$,
leading to
\begin{equation}\label{eq-spectrum-0}
m_\psi^{(0)}\simeq |M_\psi(0)|\left\{[Z_q+\Pi_{\psi_L}(0)][Z_\psi+\Pi_{\psi_R}(0)] - \left. 2 |M_\psi(0)| \frac{d |M_\psi(p^2)|}{d p^2} \right|_{p^2 = 0} \right\}^{-1/2} \ , \qquad \psi=u,d \ ,
\end{equation}
We have used the superindex
$(0)$ for the lightest states, since in the absence of mixings they
are massless.

Similarly, the Yukawa coupling of these states to (a single) Higgs
boson can be obtained by differentiating with respect to $v$:
\bea
\label{eq-y}
y_\psi^{(0)} &\simeq& \frac{dm_\psi^{(0)}}{dv} \ , \qquad \psi=u,d \ .
\eea
This coupling depends on the model, but since the vev dependence of
the correlators is simple (it is encoded in $s_h$ and $c_h$ in the
formulas given in the following subsections), we can derive simple
expression in terms of the correlators, that will be given for each
model below.

A very important combination for the phenomenology is the function
$y^{(0)}_\psi/m^{(0)}_\psi$.  To leading order in $\epsilon$ it can be
approximated by:
\bea
\label{eq-yom}
\frac{y_\psi^{(0)}}{m_\psi^{(0)}} &\simeq& \frac{F_\psi(\epsilon)}{\epsilon \, f_h} \left[1+{\cal O}(\epsilon^2) \right] ~, \qquad \psi=u,d \ .
\eea
where the $F_\psi(\epsilon)$ depends only on $\epsilon$ (as well as on
the fermion representation) and will be given in
Sec.~\ref{sec:corrections}.~\footnote{The are exceptions to this
statement, with additional dependence on the Yukawa couplings on the
r.h.s.~of Eq.~(\ref{eq-yom}).  We consider one such detailed example
in this work and mention a few others.  However, in certain limits the
above discussion often applies.} The ${\cal O}(\epsilon^2)$ correction
(which also depends on other microscopic parameters) determines the
deviation compared with the simple and compact leading approximation.
The above relation is intimately connected to certain sum rules that
have been already observed in the
literature~\cite{Falkowski:2007hz,Low:2010mr,Azatov:2011qy}.  We will
comment further on this in Sec.~\ref{sec:pheno}.

As will be shown below, different models lead to different sizes for
the ${\cal O}(\epsilon^2)$ term.  Multiplying Eq.~(\ref{eq-yom}) by
$v_{SM}$, and using Eq.~(\ref{eq-vSM}), we can obtain the ratio
between the Yukawa couplings in the MCHM and in the SM:
\bea
\label{eq-yoy}
\frac{y_\psi^{(0)}}{y_\psi^{SM}} &\simeq& F_\psi(\epsilon) \left[1+{\cal O}(\epsilon^2) \right] \ , \qquad \psi=u,d \ ,
\eea
showing that deviations from $F_\psi(\epsilon)$ are suppressed by
${\cal O}(\epsilon^2)$.  This correction depends also on the fermionic
mixings in the following way: ${\cal O}(\epsilon^2 s_{\psi_L}^2,
\epsilon^2 s_{\psi_R}^2)$, requiring in general the mixing of both
chiralities to be small to ensure extra suppression factors.  However,
for some models the structure inherited from the fermion embedding is
such that the correction involves just one chirality to leading order:
${\cal O}(\epsilon^2 s_{\psi_L}^2)$ or ${\cal O}(\epsilon^2
s_{\psi_R}^2)$.  In those cases an extra suppression can be achieved
with small mixing for one chirality only.  Note also that the above
corrections do not take the form claimed in \cite{Falkowski:2007hz},
i.e.~${\cal O}(\epsilon^2 m_\psi^2) \sim {\cal O}(\epsilon^2 \,
s_{\psi_L}^2 s_{\psi_R}^2)$, where $m_\psi$ denotes the mass of the SM
field (this has also been observed in Ref.~\cite{Montull:2013mla}).
Thus, the bottom quark, in particular, can give corrections that are
larger than expected, as will be illustrated in Sec.~\ref{sec:pheno}.

\subsection{MCHM$_5$}
\label{sec:MCHM5}

In this model we consider 4 composite fermions for each generation:
$Q^u,U\sim {\bf 5}_{2/3}$ and $Q^d,D\sim {\bf 5}_{-1/3}$, where the
subindex denotes the $U(1)_X$ charge.  In unitary gauge the Yukawa
terms of the fermion Lagrangian~(\ref{Lfermions}) read:
\bea
{\cal L}_y &=& y_u (\bar Q^u_L \Phi)(\Phi^\dagger U_R) + y_d (\bar Q^d_L \Phi)(\Phi^\dagger D_R) \ .
\label{LY5}
\eea
In this case $q_L$ mixes with two composite fermions: $Q^u$ and $Q^d$.
The bottom mass can result from small $\Delta_{q^d}$ and/or small
$\Delta_{d}$.

The correlators of the effective Lagrangian~(\ref{Leff-fermions}) are: 
\begin{align}
\Pi_{u_L}&= \Pi_{q^u}^0 + \Pi_{q^d}^0 + \Pi_{q^u}^1 \frac{s_h^2}{2} \ ,\qquad &
\Pi_{d_L}&= \Pi_{q^u}^0 + \Pi_{q^d}^0 + \Pi_{q^d}^1 \frac{s_h^2}{2} \ ,\nonumber\\
\Pi_{u_R}&=\Pi_u^0+\Pi_u^1 c_h^2 \ ,\qquad &
\Pi_{d_R}&=\Pi_d^0+\Pi_d^1 c_h^2 \ ,\nonumber\\
M_u&= m_u^1 \, \frac{s_h c_h}{\sqrt{2}} \ ,\qquad &
M_d&= m_d^1 \, \frac{s_h c_h}{\sqrt{2}} \ .
\end{align}
where the $\Pi^i_\psi$ are defined by
\bea
{\cal L}^{\rm sources}_{\rm eff} &=& 
\bar{q}^u_L \pslash \Pi^{0}_{q^u} q^u_L + \bar{q}^d_L \pslash \Pi^{0}_{q^d} q^d_L + \bar{u}_R \pslash \Pi^{0}_{u} u_R + \bar{d}_R \pslash \Pi^{0}_{d} d_R 
+ (\bar{q}^u_L \Phi) \pslash \Pi^{1}_{q^u} (\Phi^\dagger q^u_L) 
\nonumber \\ [0.4em]
& & \mbox{} 
+ (\bar{q}^d_L \Phi) \pslash \Pi^{1}_{q^d} (\Phi^\dagger q^d_L) + 
(\bar{u}_R \Phi) \pslash \Pi^{1}_{u} (\Phi^\dagger u_R) + (\bar{d}_R \Phi) \pslash \Pi^{1}_{d} (\Phi^\dagger d_R) 
\label{LeffSO5}
\\ [0.4em]
& & \mbox{} 
+ m_u^0 \, \bar q^u_L u_R + m_d^0 \, \bar q^d_L d_R
+ m_u^1 (\bar q^u_L \Phi)(\Phi^\dagger u_R) + m_d^1 (\bar q^d_L \Phi)(\Phi^\dagger d_R) + {\rm h.c.}
\nonumber 
\eea
The superindex ``sources" serves as a reminder that here the $q^u_L$,
$q^d_L$, $u_R$ and $d_R$ fill complete SO(5) multiplets and that all
components are to be treated as external sources.  One must still add
``bare" kinetic terms for the dynamical fields on site-0,
\textit{i.e.}~those with SM quantum numbers, as in
Eq.~(\ref{Leff-fermions}).  Since a ${\bf 5}$ of SO(5) decomposes
under SO(4) as ${\bf 5}\sim {\bf 1}+{\bf 4}$, one finds
\begin{align}
\Pi_{q^u}^0 &= \hat\Pi_{q^{u(4)}}\ ,\qquad &
\Pi_{q^d}^0 &= \hat\Pi_{q^{d(4)}}\ ,\qquad &
\Pi_d^0 &= \hat\Pi_{d^{(4)}}\ ,\nonumber\\
\Pi_{q^u}^1 &= \hat\Pi_{q^{u(1)}}-\hat\Pi_{q^{u(4)}}
\ ,\qquad &
\Pi_{q^d}^1 &= \hat\Pi_{q^{d(1)}}-\hat\Pi_{q^{d(4)}}
\ ,\qquad &
\Pi_d^1 &= \hat\Pi_{d^{(1)}}-\hat\Pi_{d^{(4)}}\ ,\nonumber\\
\Pi_u^0 &= \hat\Pi_{u^{(4)}} \ ,\qquad &
m_u^0 &= \hat M_{u^{(4)}} \ , \qquad &
m_d^0 &= \hat M_{d^{(4)}} \ ,\\
\Pi_u^1 &= \hat\Pi_{u^{(1)}}-\hat\Pi_{u^{(4)}}\ , \qquad &
m_u^1 &= \hat M_{u^{(1)}}-\hat M_{u^{(4)}} \ ,\qquad &
m_d^1 &= \hat M_{d^{(1)}}-\hat M_{d^{(4)}}\ .
\nonumber
\end{align}
where the hatted correlators are given in Appendix~\ref{app:MCHM5}.

Using these correlators we can compute the prediction for
$y_\psi^{(0)}/m_\psi^{(0)}$: 
\bea
\label{eq-yomt5}
\frac{y_t^{(0)}}{m_t^{(0)}}-\frac{F_t}{s_h f_h} &\simeq& \frac{s_h}{f_h}\frac{ 2 |m_u^1(0)| |m_u^{1}(0)|^{\prime} - [Z_u+\Pi_u^0(0)+\Pi_u^1(0)]\Pi_{q^u}^1+2[Z_q+\Pi_q^0(0)]\Pi_u^1}{2[Z_u+\Pi_u^0(0)+\Pi_u^1(0)][Z_q+\Pi_q^0(0)]} ~, \\
\label{eq-yomb5}
\frac{y_b^{(0)}}{m_b^{(0)}}-\frac{F_b}{s_h f_h} &\simeq& \frac{s_h}{f_h}\frac{ 2 |m_d^1(0)| |m_d^{1}(0)|^{\prime} - [Z_d+\Pi_d^0(0)+\Pi_d^1(0)]\Pi_{q^d}^1+2[Z_q+\Pi_q^0(0)]\Pi_d^1}{2[Z_d+\Pi_d^0(0)+\Pi_d^1(0)][Z_q+\Pi_q^0(0)]} \ ,
\eea
where $|m_\psi^i(0)|^{\prime} \equiv \left.  | d \, m_\psi^i(p^2) | /
dp^2 \right|_{p^2 = 0}$.  Taking into account that
$\Pi_\psi^j\sim\Delta_\psi^2$ and $M_\psi\sim\Delta_q\Delta_\psi$,
Eq.~(\ref{eq-yomb5}) shows that the ${\cal O}(s^2_h)$ correction to
$y_b$ in this model is small.  By expressing $\Delta_\psi$ in terms of
the elementary-composite mixing angles, one sees that the correction
is suppressed by $s_{q^d}^2$ or $s_d^2$.  By choosing both of them
small, we expect $y_b^{(0)} / m_b^{(0)}$ to be well approximated by
$F_b / s_h f_h$ in this model.  On the other hand,
Eq.~(\ref{eq-yomt5}) shows that the corrections to $y_t$ do not have
any extra suppression factor in general, since the top mass requires
both, $s_{q^u}$ and $s_u \sim {\cal O}(1)$.  This property has
important consequences for the phenomenology: one can expect
corrections to loop-induced processes that depend on $y_t$ [gluon
fusion, $h \to \gamma\gamma$ to be discussed in Sec.~\ref{sec:pheno}]
of ${\cal O}(s^2_h)$.  The size of these corrections is similar for
all the models.  Since all of them require $s_{q}$ and $s_u\lesssim
1$, there can be differences of ${\cal O}(1)$ between them arising
from the different embeddings and regions of the parameter space
selected.

\subsection{MCHM$_{10}$}

From now on, we consider 3 composite fermions for each generation.  In
this model: $Q,U,D\sim {\bf 10}_{2/3}$.  In unitary gauge the Yukawa
terms of the fermion Lagrangian~(\ref{Lfermions}) read: 
\bea
{\cal L}_y &=& y_u \Phi^\dagger \bar Q_L U_R \Phi + y_d \Phi^\dagger \bar Q_L D_R \Phi \ .
\label{LY10}
\eea
In this case $q_L$ mixes with a single composite fermion $Q$ and,
therefore, the bottom mass requires small $\Delta_{d}$.  In this model
the interactions $\Phi^\dagger \bar U_L D_R \Phi$ and $\Phi^\dagger
\bar U_R D_L \Phi$ are also compatible with the symmetries.  However
they lead to a logarithmically divergent Higgs potential, and we do
not include them.  Note also that we do not include terms of the form
$\epsilon_{ABCDE} \, \Phi^A \bar{Q}_L^{BC} U_R^{DE}$, etc., which
would break a LR symmetry, and have been studied in \cite{Azatov:2013ura}.

The correlators of the effective Lagrangian~(\ref{Leff-fermions}) are: 
\begin{align}
\Pi_{u_L}&=\Pi_q^0+\Pi_{q}^1 \left(\frac{c_h^2}{2}+\frac{s_h^2}{4}\right) \ ,\qquad &
\Pi_{d_L}&=\Pi_q^0+\Pi_{q}^1 \frac{c_h^2}{2} \ ,\nonumber\\
\Pi_{u_R}&=\Pi_u^0+\Pi_u^1 \frac{s_h^2}{4} \ ,\qquad &
\Pi_{d_R}&=\Pi_d^0+\Pi_d^1 \frac{s_h^2}{4} \ ,\\
M_u&= - \, m_u^1 \, \frac{s_h c_h}{4} \ ,\qquad &
M_d&= -m_d^1 \, \frac{s_h c_h}{2\sqrt{2}} \ .
\nonumber
\end{align}
where the $\Pi^i_\psi$ are now defined by [see also comments
following Eq.~(\ref{LeffSO5})]
\bea
{\cal L}^{\rm sources}_{\rm eff} &=& 
{\rm Tr} \left[ \bar{q}_L \pslash \Pi^{0}_{q} q_L + \bar{u}_R \pslash \Pi^{0}_{u} u_R + \bar{d}_R \pslash \Pi^{0}_{d} d_R \right]
\nonumber \\ [0.4em]
& & \mbox{} 
+ \Phi^\dagger \bar{q}_L \pslash \Pi^{1}_{q} \, q_L \Phi 
+ \Phi^\dagger \bar{u}_R \pslash \Pi^{1}_{u} \, u_R \Phi + \Phi^\dagger \bar{d}_R \pslash \Pi^{1}_{d} \, d_R \Phi
\label{LeffSO10}
\\ [0.4em]
& & \mbox{} +
{\rm Tr} \left[m_u^0 \, \bar q_L u_R + m_d^0 \, \bar q_L d_R \right]
+ m_u^1 \, \Phi^\dagger \bar q_L u_R \Phi + m_d^1 \, \Phi^\dagger \bar q_L d_R\Phi + {\rm h.c.}
\nonumber 
\eea
Since a ${\bf 10}$ of SO(5)
decomposes under SO(4) as ${\bf 10}\sim {\bf 4}+{\bf 6}$, we find
\begin{align}
\Pi_q^0&= \hat\Pi_{q^{(6)}}\ ,\qquad &
\Pi_u^0&= \hat\Pi_{u^{(6)}}\ ,\qquad &
\Pi_d^0&= \hat\Pi_{d^{(6)}}\ ,\nonumber\\
\Pi_{q}^1&= 2(\hat\Pi_{q^{(4)}}-\hat\Pi_{q^{(6)}})\ ,\qquad &
\Pi_u^1&= 2(\hat\Pi_{u^{(4)}}-\hat\Pi_{u^{(6)}})\ ,\qquad &
\Pi_d^1&= 2(\hat\Pi_{d^{(4)}}-\hat\Pi_{d^{(6)}})\ ,\nonumber\\
m_u^0&= \hat M_{u^{(6)}} \ ,\qquad &
m_d^0&= \hat M_{d^{(6)}}\ , \nonumber\\
m_u^1&= 2(\hat M_{u^{(4)}}-\hat M_{u^{(6)}}) \ ,\qquad &
m_d^1&= 2(\hat M_{d^{(4)}}-\hat M_{d^{(6)}})\ .
\end{align}
where the hatted correlators are given in Appendix~\ref{app:MCHM10}.

The prediction for $y_\psi/m_\psi$ is:
\bea
\label{eq-yomt10}
\frac{y_t^{(0)}}{m_t^{(0)}}-\frac{F_t}{s_h f_h} &\simeq& \frac{s_h}{f_h}\frac{|m_u^1(0)| |m_u^{1}(0)|^{\prime}+[2Z_u+2\Pi_u^0(0)-\Pi_u^1(0)]\Pi_q^1-2[Z_q+\Pi_q^0(0)]\Pi_u^1}{4[Z_u+\Pi_u^0(0)][2Z_q+2\Pi_q^0(0)+\Pi_q^1(0)]} ~,\\
\label{eq-yomb10}
\frac{y_b^{(0)}}{m_b^{(0)}}-\frac{F_b}{s_h f_h} &\simeq& \frac{s_h}{f_h}\frac{2 |m_d^1(0)| |m_d^{1}(0)|^{\prime}+4[Z_d+\Pi_d^0(0)]\Pi_q^1-[2Z_q+2\Pi_q^0(0)+\Pi_q^1(0)]\Pi_d^1}{4[Z_d+\Pi_d^0(0)][2Z_q+2\Pi_q^0(0)+\Pi_q^1(0)]} \ .
\eea
Eq.~(\ref{eq-yomb10}) shows that the ${\cal O}(s_h^2)$ corrections to
$y_b$ in this model can be sizable.  This is because there is a term
suppressed by $s_q^2$ only, but $s_q\sim 1$ to reproduce the top mass.
Thus, we find a suppression by $s_h^2$ only.

\subsection{MCHM$_{10-5-10}$}

In this model: $Q,D\sim{\bf 10}_{2/3}$ and $U\sim{\bf 5}_{2/3}$.  In
unitary gauge the Yukawa terms of the fermion
Lagrangian~(\ref{Lfermions}) read:
\bea
{\cal L}_y &=& y_u \Phi^\dagger\bar Q_L U_R  + y_d \Phi^\dagger\bar Q_L D_R \Phi \ .
\label{LY10510}
\eea

The correlators of the effective Lagrangian~(\ref{Leff-fermions}) are:
\begin{align}
\Pi_{u_L}&=\Pi_q^0+\Pi_{q}^1 \left(\frac{c_h^2}{2}+\frac{s_h^2}{4}\right) ,\qquad &
\Pi_{d_L}&=\Pi_q^0+\Pi_{q}^1 \frac{c_h^2}{2} \ ,\nonumber\\
\Pi_{u_R}&=\Pi_u^0+\Pi_u^1 c_h^2 \ ,\qquad &
\Pi_{d_R}&=\Pi_d^0+\Pi_d^1 \frac{s_h^2}{4} \ ,\\
M_u&= -m_u^1 \, \frac{s_h}{2} \ ,\qquad &
M_d&= -m_d^1 \, \frac{s_h c_h}{2\sqrt{2}} \ .
\nonumber
\end{align}
where the $\Pi^i_\psi$ are defined in analogy to Eqs.~(\ref{LeffSO5})
and (\ref{LeffSO10}), with the $\Phi$-dependent terms following the
structure displayed in Eq.~(\ref{LY10510}) for the Yukawa terms in
this model [see also comments following Eq.~(\ref{LeffSO5})].
Expanding the Higgs potential in powers of $s_h$ and $\Delta_\psi$,
the contribution of $M_u$ to the quartic coupling is of order ${\cal
O}(\Delta^8_\psi)$ and the only contributions of order ${\cal
O}(\Delta^4_\psi)$ are from $\Pi_L$ and $\Pi_R$.  Therefore, in this
model we expect a small self-coupling and a very light Higgs.  This
fact is reflected in the tuning of the model which, after requiring
the proper Higgs mass, is one order of magnitude larger than in the
other models.  A sizable quartic coupling demands very large mixings
for the top quark, inducing departures from the analytical
approximations for the Yukawa couplings.  This also affects the bottom
since the $b_L$ mixing is equal to the $t_L$ mixing in this model.

Using the previous decompositions of ${\bf 5}$ and ${\bf 10}$ of SO(5)
under SO(4) one finds:
\begin{align}
\Pi_q^0&= \hat\Pi_{q^{(6)}}\ ,\qquad &
\Pi_u^0&= \hat\Pi_{u^{(4)}}\ ,\qquad &
\Pi_d^0&= \hat\Pi_{d^{(6)}}\ ,\nonumber \\
\Pi_{q}^1&= 2(\hat\Pi_{q^{(4)}}-\hat\Pi_{q^{(6)}})\ ,\qquad &
\Pi_u^1&= \hat\Pi_{u^{(1)}}-\hat\Pi_{u^{(4)}}\ ,\qquad &
\Pi_d^1&= 2(\hat\Pi_{d^{(4)}}-\hat\Pi_{d^{(6)}})\ ,\nonumber \\
m_u^0&= 0 \ ,\qquad &
m_d^0&= \hat M_{d^{(6)}}\ , \nonumber \\
m_u^1&= \sqrt{2}\hat M_{u^{(4)}} \ ,\qquad &
m_d^1&= 2(\hat M_{d^{(4)}}-\hat M_{d^{(6)}})\ .
\end{align}
where the hatted correlators are given in
Appendix~\ref{app:MCHM10510}.

The prediction for $y_\psi/m_\psi$ is:
\bea
\label{eq-yomt10-5-10}
\frac{y_t^{(0)}}{m_t^{(0)}}-\frac{F_t}{s_h f_h} &\simeq& \frac{s_h}{f_h}\frac{2 |m_u^1(0)| |m_u^{1}(0)|^{\prime}+[Z_u+\Pi_u^0(0)+3\Pi_u^1(0)]\Pi_q^1+4[Z_q+\Pi_q^0(0)]\Pi_u^1}{2[Z_u+\Pi_u^0(0)+\Pi_u^1(0)][2Z_q+2\Pi_q^0(0)+\Pi_q^1(0)]} ~,\\
\label{eq-yomb10-5-10}
\frac{y_b^{(0)}}{m_b^{(0)}}-\frac{F_b}{s_h f_h} &\simeq& \frac{s_h}{f_h}\frac{2 |m_d^1(0)| |m_d^{1}(0)|^{\prime}+4[Z_d+\Pi_d^0(0)]\Pi_q^1-[2Z_q+2\Pi_q^0(0)+\Pi_q^1(0)]\Pi_d^1}{4[Z_d+\Pi_d^0(0)][2Z_q+2\Pi_q^0(0)+\Pi_q^1(0)]} \ .
\eea
$y_b/m_b$ in this model is exactly as in the MCHM$_{10}$ when
expressed in terms of the correlators, although the correlators
themselves are different in both models.  This can be understood
because the bottom mass arises from the coupling between $q$ and $d$,
that share the same embedding in both models.

\subsection{MCHM$_{5-5-10}$}

In this model: $Q,U\sim{\bf 5}_{2/3}$ and $D\sim{\bf 10}_{2/3}$.  In
unitary gauge the Yukawa terms of the fermion
Lagrangian~(\ref{Lfermions}) read:
\bea
{\cal L}_y &=& y_u (\bar Q_L \Phi) (\Phi^\dagger U_R) + y_d \bar Q_L D_R \Phi \ .
\label{LY5510}
\eea

The correlators of the effective Lagrangian~(\ref{Leff-fermions}) are:
\begin{align}
\Pi_{u_L}&=\Pi_q^0+\Pi_{q}^1 \frac{s_h^2}{2} ,\qquad &
\Pi_{d_L}&=\Pi_q^0 \ ,\nonumber\\
\Pi_{u_R}&=\Pi_u^0+\Pi_u^1 c_h^2 \ ,\qquad &
\Pi_{d_R}&=\Pi_d^0+\Pi_d^1 \frac{s_h^2}{4} \ ,\\
M_u&= m_u^1 \, \frac{s_h c_h}{\sqrt{2}} \ ,\qquad &
M_d&= m_d^1 \, \frac{s_h}{2} \ .
\nonumber
\end{align}
where the $\Pi^i_\psi$ are defined in analogy to Eqs.~(\ref{LeffSO5})
and (\ref{LeffSO10}), with the $\Phi$-dependent terms following the
structure displayed in Eq.~(\ref{LY5510}) for the Yukawa terms in this
model [see also comments following Eq.~(\ref{LeffSO5})].  Using the
previous decompositions of ${\bf 5}$ and ${\bf 10}$ of SO(5) under
SO(4):
\begin{align}
\Pi_q^0&= \hat\Pi_{q^{(4)}}\ ,\qquad &
\Pi_u^0&= \hat\Pi_{u^{(4)}}\ ,\qquad &
\Pi_d^0&= \hat\Pi_{d^{(6)}}\ ,\nonumber \\
\Pi_{q}^1&= \hat\Pi_{q^{(1)}}-\hat\Pi_{q^{(4)}}\ ,\qquad &
\Pi_u^1&= \hat\Pi_{u^{(1)}}-\hat\Pi_{u^{(4)}}\ ,\qquad &
\Pi_d^1&= 2(\hat\Pi_{d^{(4)}}-\hat\Pi_{d^{(6)}})\ ,\nonumber \\
m_u^0&= \hat M_{u^{(4)}} \ ,\qquad &
m_d^0&=0\ , \nonumber \\
m_u^1&= \hat M_{u^{(1)}}-\hat M_{u^{(4)}} \ ,\qquad &
m_d^1&= \sqrt{2}\hat M_{d^{(4)}}\ .
\end{align}
where the hatted correlators are given in Appendix~\ref{app:MCHM5510}.

The prediction for $y_\psi/m_\psi$ is:
\bea
\label{eq-yomt5-5-10}
\frac{y_t^{(0)}}{m_t^{(0)}}-\frac{F_t}{s_h f_h} &\simeq& \frac{s_h}{f_h}\frac{2 |m_u^1(0)| |m_u^{1}(0)|^{\prime}+2[Z_q+\Pi_q^0(0)]\Pi_u^1-[Z_u+\Pi_u^0(0)+\Pi_u^1(0)]\Pi_q^1}{2[Z_u+\Pi_u^0(0)+\Pi_u^1(0)][Z_q+\Pi_q^0(0)]} \ ;\\
\label{eq-yomb5-5-10}
\frac{y_b^{(0)}}{m_b^{(0)}}-\frac{F_b}{s_h f_h} &\simeq& \frac{s_h}{f_h}\frac{2 |m_d^1(0)| |m_d^{1}(0)|^{\prime}-[Z_q+\Pi_q^0(0)]\Pi_d^1}{4[Z_d+\Pi_d^0(0)][Z_q+\Pi_q^0(0)]} \ .
\eea
For the top quark we obtain a result similar to the MCHM$_5$.
Eq.~(\ref{eq-yomb5-5-10}) shows that the ${\cal O}(s_h^2)$ corrections
to $y_b^{(0)}/m_b^{(0)}$ in this model is also suppressed by
$s_d^2\ll1$.

\subsection{MCHM$_{5-10-10}$}

In this model: $Q\sim{\bf 5}_{2/3}$ and $U,D\sim{\bf 10}_{2/3}$.  In
unitary gauge the Yukawa terms of the fermion
Lagrangian~(\ref{Lfermions}) read:
\bea
{\cal L}_y &=& y_u \bar Q_L U_R \Phi  + y_d \bar Q_L D_R \Phi \ .
\label{LY51010}
\eea
In this model the interactions $\Phi^\dagger \bar U_L D_R \Phi$ and
$\Phi^\dagger \bar U_R D_L \Phi$ are also compatible with the
symmetries. However they lead to a logarithmically divergent Higgs
potential, therefore we will not include them.

The correlators of the effective Lagrangian~(\ref{Leff-fermions}) are:
\begin{align}
\Pi_{u_L}&=\Pi_q^0+\Pi_{q}^1 \frac{s_h^2}{2} ,\qquad &
\Pi_{d_L}&=\Pi_q^0 \ ,\nonumber\\
\Pi_{u_R}&=\Pi_u^0+\Pi_u^1 \frac{s_h^2}{4} \ ,\qquad &
\Pi_{d_R}&=\Pi_d^0+\Pi_d^1 \frac{s_h^2}{4} \ ,\\
M_u&= -m_u^1 \, \frac{s_h}{2\sqrt{2}} \ ,\qquad &
M_d&= m_d^1 \, \frac{s_h}{2} \ .
\nonumber
\end{align}
where the $\Pi^i_\psi$ are defined in analogy to Eqs.~(\ref{LeffSO5})
and (\ref{LeffSO10}), with the $\Phi$-dependent terms following the
structure displayed in Eq.~(\ref{LY51010}) for the Yukawa terms in
this model [see also comments following Eq.~(\ref{LeffSO5})].  Since
the Higgs dependence on $M_u$ is the same as in the MCHM$_{10-5-10}$,
the behavior of the Higgs potential and the top Yukawa are similar.

Using the previous decompositions of ${\bf 5}$ and ${\bf 10}$ of SO(5)
under SO(4):
\begin{align}
\Pi_q^0&= \hat\Pi_{q^{(4)}}\ ,\qquad &
\Pi_u^0&= \hat\Pi_{u^{(6)}}\ ,\qquad &
\Pi_d^0&= \hat\Pi_{d^{(6)}}\ ,\nonumber \\
\Pi_{q}^1&= \hat\Pi_{q^{(1)}}-\hat\Pi_{q^{(4)}}\ ,\qquad &
\Pi_u^1&= 2(\hat\Pi_{u^{(4)}}-\hat\Pi_{u^{(6)}})\ ,\qquad &
\Pi_d^1&= 2(\hat\Pi_{d^{(4)}}-\hat\Pi_{d^{(6)}})\ ,\nonumber \\
m_u^0&= 0 \ ,\qquad &
m_d^0&= 0\ , \nonumber \\
m_u^1&= \sqrt{2}\hat M_{u^{(4)}} \ ,\qquad &
m_d^1&= \sqrt{2}\hat M_{d^{(4)}}\ .
\end{align}
where the hatted correlators are given in Appendix~\ref{app:MCHM51010}.

The prediction for $y_\psi/m_\psi$ is:
\bea
\label{eq-yomt5-10-10}
\frac{y_t^{(0)}}{m_t^{(0)}}-\frac{F_t}{s_h f_h} &\simeq& \frac{s_h}{f_h}\frac{|m_u^1(0)| |m_u^{1}(0)|^{\prime} - [Z_q+\Pi_q^0(0)]\Pi_u^1-2[Z_u+\Pi_u^0(0)]\Pi_u^1}{4[Z_u+\Pi_u^0(0)][Z_q+\Pi_q^0(0)]} ~,\\
\label{eq-yomb5-10-10}
\frac{y_b^{(0)}}{m_b^{(0)}}-\frac{F_b}{s_h f_h} &\simeq& \frac{s_h}{f_h}\frac{2 |m_d^1(0)| |m_d^{1}(0)|^{\prime}-[Z_q+\Pi_q^0(0)]\Pi_d^1}{4[Z_d+\Pi_d^0(0)][Z_q+\Pi_q^0(0)]} \ .
\eea
$y_b/m_b$ in this model is exactly as in the MCHM$_{5-5-10}$ when
expressed in terms of the correlators, although the correlators
themselves are different in both models.  This can be understood,
again, because the bottom mass arises from the coupling between $q$
and $d$, which share the same embedding in both models.
Eq.~(\ref{eq-yomb5-10-10}) shows that the ${\cal O}(s_h^2)$
corrections to $y_b^{(0)}/m_b^{(0)}$ in this model is also suppressed
by $s_d^2\ll1$.

\subsection{MCHM$_{14-14-10}$}
\label{sec14-14-10}

In this model: $Q,U\sim{\bf 14}_{2/3}$ and $D\sim{\bf 10}_{2/3}$.  In
unitary gauge the Yukawa term of the fermion
Lagrangian~(\ref{Lfermions}) includes:
\bea
\label{14y1}
{\cal L}_y &\supset& y_u \Phi^\dagger \bar Q_L U_R \Phi + y_d \Phi^\dagger \bar Q_L D_R \Phi \ .
\label{LY141410}
\eea
The following term is also allowed by the symmetries
\begin{equation}
\label{14y2}
{\cal L}_y\supset \tilde y_u (\Phi^\dagger \bar Q_L\Phi)\ (\Phi^\dagger U_R \Phi) \ ,
\end{equation}
having potentially important consequences for the phenomenology, as
will be discussed in the next section.

The correlators of the effective Lagrangian~(\ref{Leff-fermions}) are:
\begin{align}
\Pi_{u_L}&=\Pi_q^0+\Pi_{q}^1 \left(\frac{c_h^2}{2}+\frac{s_h^2}{4}\right) + \Pi_{q}^2 s_h^2 c_h^2\ ,\qquad &
\Pi_{d_L}&=\Pi_q^0+\Pi_{q}^1 \frac{c_h^2}{2} \ ,\nonumber\\
\Pi_{u_R}&=\Pi_u^0+\Pi_u^1 \left(\frac{4}{5}c_h^2+\frac{s_h^2}{20}\right) + \Pi_u^2 \frac{(4c_h^2-s_h^2)^2}{20} \ ,\qquad &
\Pi_{d_R}&=\Pi_d^0+\Pi_d^1 \frac{s_h^2}{4} \ ,\\
M_u&= i\ m_u^1 \, \frac{3}{4\sqrt{5}}s_h c_h + i\ m_u^2 \, \frac{1}{2\sqrt{5}}s_h c_h (4c_h^2-s_h^2) \ ,\qquad &
M_d&= i m_d^1 \, \frac{s_h c_h}{2\sqrt{2}} \ .
\nonumber
\end{align}
where the $\Pi^i_\psi$ are defined in analogy to Eqs.~(\ref{LeffSO5})
and (\ref{LeffSO10}), with the $\Phi$-dependent terms following the
structure displayed in Eqs.~(\ref{14y1}) and (\ref{14y2}) for the
Yukawa terms in this model [see also comments following
Eq.~(\ref{LeffSO5})].  Since a ${\bf 14}$ of SO(5) decomposes under
SO(4) as ${\bf 14}\sim {\bf 1}+{\bf 4}+{\bf 9}$, we find
\begin{align}
\Pi_q^0&= \hat\Pi_{q^{(9)}}\ ,\qquad &
\Pi_u^0&= \hat\Pi_{u^{(9)}}\ ,\qquad &
\Pi_d^0&= \hat\Pi_{d^{(6)}}\ ,\nonumber \\
\Pi_{q}^1&= 2(\hat\Pi_{q^{(4)}}-\hat\Pi_{q^{(9)}})\ ,\qquad &
\Pi_u^1&= 2(\hat\Pi_{u^{(4)}}-\hat\Pi_{u^{(9)}})\ ,\qquad &
\Pi_d^1&= 2(\hat\Pi_{d^{(4)}}-\hat\Pi_{d^{(6)}})\ ,\nonumber \\
\Pi_q^2&= \frac{1}{4}(5\hat\Pi_{q^{(1)}}-8\hat\Pi_{q^{(4)}}+3\hat\Pi_{q^{(9)}})\ ,\qquad &
\Pi_u^2&= \frac{1}{4}(5\hat\Pi_{u^{(1)}}-8\hat\Pi_{u^{(4)}}+3\hat\Pi_{u^{(9)}})\ , \\
m_u^0&= \hat M_{u^{(1)}} \ ,\qquad &
m_d^0&= 0\ , \nonumber \\
m_u^1&= 2(\hat M_{u^{(4)}}-\hat M_{u^{(9)}}) \ ,\qquad &
m_d^1&= 2i\hat M_{d^{(4)}}\ ,\nonumber \\
m_u^2&= \frac{1}{4}(5\hat M_{u^{(1)}}-8\hat M_{u^{(4)}}+3\hat M_{u^{(9)}}) \ .
\nonumber
\end{align}
where the hatted correlators are given in
Appendix~\ref{app:MCHM141410}.

The prediction for $y_\psi/m_\psi$ is:
\bea
\label{eq-yomt14}
\frac{y_t^{(0)}}{m_t^{(0)}}-\frac{F_t}{s_h f_h} &\simeq& \frac{s_h}{f_h}\left\{
-2\frac{-3 |m_u^1(0)| [\Pi_q^1(0)-4\Pi_q^2(0)]+16 |m_u^2(0)| [5Z_q+5\Pi_q^0(0)+2\Pi_q^1(0)+2\Pi_q^2(0)]}{[3 |m_u^1(0)| + 8 |m_u^2(0)| ][2Z_q+2\Pi_q^0(0)+\Pi_q^1(0)]}\right.\nonumber\\
& & \hspace{-3cm}
\left. +
\frac{-[3 |m_u^1(0)| + 8|m_u^2(0)| ][3 |m_u^{1}(0)|^{\prime} + 8 |m_u^{2}(0)|^{\prime}] + 5[2Z_q+2\Pi_q^0(0)+\Pi_q^1(0)][3\Pi_u^1(0)+8\Pi_u^2(0)]}{[5Z_u+5\Pi_u^0(0)+4\Pi_u^1(0)+4\Pi_u^2(0)][2Z_q+2\Pi_q^0(0)+\Pi_q^1(0)]}
\right\} ~,\\
\label{eq-yomb14}
\frac{y_b^{(0)}}{m_b^{(0)}}-\frac{F_b}{s_h f_h} &\simeq& \frac{s_h}{f_h}\frac{2 |m_d^1(0)| |m_d^{1}(0)|^{\prime}+4[Z_d+\Pi_d^0(0)]\Pi_q^1-[2Z_q+2\Pi_q^0(0)+\Pi_q^1(0)]\Pi_d^1}{4[Z_d+\Pi_d^0(0)][2Z_q+2\Pi_q^0(0)+\Pi_q^1(0)]} \ .
\eea
$y_b/m_b$ in this model is exactly as in the MCHM$_{10}$ when
expressed in terms of the correlators, although the correlators
themselves are different in both models.

\subsection{MCHM$_{14-1-10}$}
\label{sec:MCHM14110}

In this model: $Q\sim{\bf 14}_{2/3}$, $U\sim{\bf 1}_{2/3}$ and
$D\sim{\bf 10}_{2/3}$: In unitary gauge the Yukawa term of the fermion
Lagrangian~(\ref{Lfermions}) reads:
\bea
{\cal L}_y &=& y_u (\Phi^\dagger\bar Q_L \Phi) U_R  + y_d \Phi^\dagger\bar Q_L D_R \Phi \ .
\label{LY14110}
\eea

The correlators of the effective Lagrangian~(\ref{Leff-fermions}) are:
\begin{align}
\Pi_{u_L}&=\Pi_q^0+\Pi_{q}^1 \left(\frac{c_h^2}{2}+\frac{s_h^2}{4}\right)+\Pi_q^2 c_h^2 s_h^2 ,\qquad &
\Pi_{d_L}&=\Pi_q^0+\Pi_{q}^1 \frac{c_h^2}{2} \ ,\nonumber\\
\Pi_{u_R}&=\Pi_u^0 \ ,\qquad &
\Pi_{d_R}&=\Pi_d^0+\Pi_d^1 \frac{s_h^2}{4} \ ,\\
M_u&= -m_u^1 \, \frac{s_h}{2} \ ,\qquad &
M_d&= -m_d^1 \, \frac{s_h c_h}{2\sqrt{2}} \ .
\nonumber
\end{align}
where the $\Pi^i_\psi$ are defined in analogy to Eqs.~(\ref{LeffSO5})
and (\ref{LeffSO10}), with the $\Phi$-dependent terms following the
structure displayed in Eq.~(\ref{LY14110}) for the Yukawa terms in
this model [see also comments following Eq.~(\ref{LeffSO5})].  Using
the previous decompositions of ${\bf 14}$ and ${\bf 10}$ of SO(5)
under SO(4):
\begin{align}
\Pi_q^0&= \hat\Pi_{q^{(9)}}\ ,\qquad &
\Pi_u^0&= \hat\Pi_{u^{(1)}}\ ,\qquad &
\Pi_d^0&= \hat\Pi_{d^{(6)}}\ ,\nonumber \\
\Pi_{q}^1&= 2(\hat\Pi_{q^{(4)}}-\hat\Pi_{q^{(9)}})\ ,\qquad &
& \qquad & \Pi_d^1&= 2(\hat\Pi_{d^{(4)}}-\hat\Pi_{d^{(6)}})\ ,\nonumber \\
\Pi_{q}^2&= \frac{1}{4}(5\hat\Pi_{q^{(1)}}-8\hat\Pi_{q^{(4)}}+3\hat\Pi_{q^{(9)}})\ , \\
m_u^0&= 0 \ ,\qquad &
m_d^0&= 0 \ , \nonumber \\
m_u^1&= \frac{\sqrt{5}}{2}\hat M_{u^{(1)}} \ ,\qquad &
m_d^1&= 2i\hat M_{d^{(4)}}\ .
\nonumber
\end{align}
where the hatted correlators are given in Appendix~\ref{app:MCHM14110}.

The prediction for $y_\psi/m_\psi$ is:
\bea
\label{eq-yomt4}
\frac{y_t^{(0)}}{m_t^{(0)}}-\frac{F_t}{s_h f_h} &\simeq& \frac{s_h}{f_h} \frac{-8 |m_u^1(0)| |m_u^{1}(0)|^{\prime}+[Z_u+\Pi_u^0(0)][\Pi_q^1(0)-4\Pi_q^2(0)]}{2[Z_u+\Pi_u^0(0)][2Z_q+2\Pi_q^0(0)+\Pi_q^1(0)]} \ ;\\
\label{eq-yomb4}
\frac{y_b^{(0)}}{m_b^{(0)}}-\frac{F_b}{s_h f_h} &\simeq& \frac{s_h}{f_h}\frac{2 |m_d^1(0)| |m_d^{1}(0)|^{\prime}+4[Z_d+\Pi_d^0(0)]\Pi_q^1-[2Z_q+2\Pi_q^0(0)+\Pi_q^1(0)]\Pi_d^1}{4[Z_d+\Pi_d^0(0)][2Z_q+2\Pi_q^0(0)+\Pi_q^1(0)]} \ .
\eea
The prediction for $y_b/m_b$ in this model is exactly as in the
MCHM$_{10}$ when expressed in terms of the correlators, although the
correlators themselves are different in both models.

\subsection{Other Models Based on the Lowest-dimensional Reps.~of SO(5)}
\label{sec:othermodels}

Although we will not provide all the details, we list here the other
possible models one can consider when using the $\bf 1$, $\bf 5$, $\bf
10$ and $\bf 14$ representations of $SO(5)$ in all possible
combinations for the quark sector (assuming the same assignments for
all the families).  Besides the cases given above, one can have an
MCHM$_{5-1-10}$, MCHM$_{14-10-10}$, MCHM$_{10-14-10}$,
MCHM$_{14-5-10}$ and MCHM$_{5-14-10}$.  This would exhaust all the
models that allow to write Yukawa couplings (in particular for the top
quark, which is hard to imagine arising from other than tree-level
effects).  For instance, the MCHM$_{10-1-X}$ does not allow to write
the operator $y_u (\Phi^\dagger \bar{Q}_L \Phi) U_R + {\rm h.c.}$
since it vanishes due to the antisymmetry of the $\bf 10$.  Some of
these models (the MCHM$_{14-5-10}$ and MCHM$_{5-14-10}$), like the
MCHM$_{14-14-10}$ described in detail in Sec.~\ref{sec14-14-10}, allow
for two Yukawa structures in the up sector, which can a priori lead to
qualitative differences with the remaining models that allow only a
single Yukawa structure.  We will study in detail only the
MCHM$_{14-14-10}$ to illustrate the possible features in such cases,
and will restrict our comments for the models mentioned in this
subsection to only a few general remarks in the following sections
(but enough to get a feel for their phenomenology).

\section{Corrections to Low-Energy Observables in the MCHM}
\label{sec:corrections}

To analyze the low-energy consequences of the model one can either
diagonalize the gauge and fermion mass matrices, explicitly including
the heavy states and their mixing with the elementary fields.  The SM
fields are then identified as the lowest lying states in the presence
of a given $\langle h \rangle$.  The latter is actually determined
dynamically as discussed in Sec.~\ref{sec_div_VH}, but the procedure
works for any fixed vev.  Finding the Higgs mass, however, requires
the minimization of the potential, and incorporating this information
will be deferred to later sections.

Alternatively, one can obtain an effective theory for the fields on
site-0, as done in Sec.~\ref{sec:EFT} for the gauge fields and in
Sec.~\ref{sec:models} for the fermion sector.  The zeroes of the
correlators thus obtained determine the spectrum of the model.  The
correlators also encode in their Higgs vev dependence information
regarding the couplings of the physical fields and the Higgs boson, as
discussed in the previous section.

Although the numerical analysis to be presented in
Sec.~\ref{sec:pheno} has been obtained by the previous methods (and we
have checked that they agree), it is useful to have a simple analytic
approximation that captures the main phenomenological features of the
Higgs sector in composite Higgs models.  To do so, one starts from the
following relation that holds in the simplest situations, which
includes most of the models we study:
\bea
\sum_n \frac{y^{(n)}_\psi}{m^{(n)}_\psi} &=& \frac{1}{2} \, \frac{d}{dh} \log {\rm det} (M^\dagger_\psi M_\psi) ~=~ \frac{1}{s_h f_h} \, F_\psi(s_h)~,
\label{BasicRelation}
\eea
where $m^{(n)}_\psi$ and $y^{(n)}_\psi$ are the mass and the Yukawa
coupling of the n-th fermionic resonance to the Higgs, respectively,
and $M_\psi$ is the $h$-dependent mass matrix.  The fact that the
above trace depends only on $s_h = \sin(h/f_h)$, but not on other
parameters of the model~\footnote{However, one should remember that
$\langle h \rangle$ itself is determined by the effective potential,
which is calculable and depends on various microscopic parameters.
Therefore, the most precise statement is that the r.h.s.~of
Eq.~(\ref{BasicRelation}) depends on the microscopic parameters only
through $h/f_h$.} is not a general statement, but a consequence of the
particular models considered in this work.  In the simplest situation
there is just one Yukawa term that leads only to one non-trivial SO(4)
invariant for each sector, resulting in a determinant that factorizes
as $\det (M^\dagger_\psi M_\psi) = \hat{F}_\psi(s_h)\
h_\psi(y,\Delta,m)$.  Therefore, its logarithmic derivative depends
only on $s_h$ and $f_h$.  $F_\psi(s_h)$ is a model-dependent function
that depends on the representation of the fermions under
$G_1$~\cite{Falkowski:2007hz,Low:2010mr,Azatov:2011qy}.

In the general situation, for arbitrary representations of the
composite fermions, there is more than one non-trivial SO(4) invariant
arising from the Yukawa interactions in each sector.  The determinant
does not factorize in this case and its derivative generically depends
on other microscopic parameters as well, such as the composite Yukawa
couplings.  This is the case for the most general MCHM$_{14-14-10}$
discussed in Sec.~\ref{sec14-14-10}.  This could be important for the
phenomenology, since in the general case one could in principle obtain
enhancement or suppression of the gluon fusion process in different
regions of the parameter space, while there is no such freedom for the
minimal cases with just one invariant.

Under the assumption that Eq.~(\ref{BasicRelation}) holds, the
additional useful observation is that, to leading order in $\epsilon =
\sin(v/f_h)$, the sum is saturated by the zero-mode term, leading to
\bea
\frac{y^{(0)}_\psi}{m^{(0)}_\psi} &\approx& \frac{1}{\epsilon f_h} \left[ F_\psi(\epsilon) + {\cal O}(\epsilon^2 s_{\psi_L}^2) + {\cal O}(\epsilon^2 s_{\psi_R}^2) \right]~,
\label{BasicApproximation}
\eea
where $s_{\psi_L}$ and $s_{\psi_R}$ are the LH and RH
elementary-composite mixing angles, respectively.  This was explicitly
shown in Sec.~\ref{sec:models} for each model, and in
Sec.~\ref{sec:pheno} we will further show numerically that the above
approximation works reasonably well even in the top sector (we will
also discuss the cases where important deviations arise).

Except for the case considered in Sec.~\ref{sec14-14-10} and two
embeddings described in Sec.~\ref{sec:othermodels}, we find only two
different functions for the models considered in this work:
\begin{equation}
F_1=\frac{1-2\epsilon^2}{\sqrt{1-\epsilon^2}}~, \qquad\qquad
F_2=\sqrt{1-\epsilon^2}~.
\label{F1F2}
\end{equation}
The MCHM$_{14-14-10}$ presented in Sec.~\ref{sec14-14-10} is somewhat
different in that two different Yukawa structures are allowed [see
Eqs.~(\ref{14y1}) and (\ref{14y2})].  As a result, the trace involves
a function with a non-trivial dependence on these Yukawa couplings,
not just on $\epsilon$:
\begin{equation}
\frac{1}{\epsilon f_h} \, F_3 \equiv \tr(Y_uM_u^{-1}) = \frac{1}{\epsilon f_h} \, \frac{\left(6 \epsilon^2-3\right) y_u-2 \left(20 \epsilon^4-23 \epsilon^2+4\right) \tilde y_u}{\sqrt{1-\epsilon^2} \left(2 \left(5 \epsilon^2-4\right) \tilde y_u-3 y_u\right)} \ ,
\label{F3}
\end{equation}
which can change the size and sign of $F_3$.  Being $F_3$ a
homogeneous function of the Yukawa couplings, it depends only on the
ratio $r_y=\tilde y_u/y_u$.  For $r_y=0$ one recovers the $F_1$
function of the other models: $F_3|_{r_y=0}=F_1$.  In the opposite
limit we define a new function
\bea
\tilde{F}_3 \equiv \lim\limits_{r_y\to\infty}F_3 =
\frac{4-23\epsilon^2+20\epsilon^4}{\sqrt{1-\epsilon^2} \, (4-5
\epsilon^2)}~.
\label{F3tilde}
\eea
For $r_y\to\infty$ one can obtain in principle a large suppression,
since $\tilde{F}_3$ changes sign for $\epsilon\simeq 0.46$.  $F_3$
interpolates between $F_1$ and $\tilde{F}_3$ as $r_y$ varies, thus one
can expect a suppression larger than $F_1$ in the general case (see
right panel of Fig.~\ref{YukApprox}).  However there is a small region
of the parameter space where there could be an enhancement and a
violent change of sign of $F_3$, as a consequence of an accidental
cancellation in $\det M_u$ that leads to a singularity of $F_3$ 
(this has also been observed in Ref.~\cite{Montull:2013mla}). This
is connected to the existence of a very light resonance in this
region.  For $\epsilon\in(0,0.5)$ the singularity is present if
$r_y\in(-6/11,-3/8)$, thus for points of the parameter space near the
singularity the value of $F_3$ can be very large, changing sign across
the singularity.  Although a large correction in any direction is
possible in this model it requires tuning of the Yukawa couplings.
This large correction, being associated with a zero of $\det M_u$,
signals the presence of a very light mode in the spectrum, that can be
in conflict with bounds on top partners.  Moreover, by performing a
random scan we have checked that the points able to reproduce the
spectrum and EW constraints are usually far from the singularity.
Thus, we typically obtain a suppression as opposed to an enhancement
from this more complicated function.

Another important consequence is that the presence of two different
flavor structures leads to missalignement of Higgs coupling in LR
operators~\cite{Mrazek:2011iu}.  For anarchic models, these new
sources of flavor violation mediated by Higgs exchange are too large
compared with bounds from flavor physics, requiring extra protection.
For this reason we will perform one scan imposing $\tilde y_u=0$, and
a second one allowing $\tilde y_u\neq0$.  It turns out that the latter
ends up preferring regions with $y_u \ll \tilde y_u$, so that it is
effectively described by $\tilde{F}_3(\epsilon)$ given in
Eq.~(\ref{F3tilde}) above.

The other models mentioned in Sec.~\ref{sec:othermodels} can be
described by the same $F_i(\epsilon)$ above, except for the
MCHM$_{14-5-10}$ and MCHM$_{5-14-10}$ which lead to the following new
functions that, like the one for the MCHM$_{14-14-10}$, also depend on
the microscopic Yukawa couplings [$F_4$ and $F_5$ are defined in
analogy to Eq.~(\ref{F3})]:
\bea
F_4 &=& \frac{\sqrt{1-\epsilon ^2} \left( y_u + 2 \tilde{y}_u - 6 \tilde{y}_u \epsilon^2 \right)}{y_u + 2 \tilde{y}_u\left(1 - \epsilon^2 \right)}~, 
\hspace{1cm}
F_5 ~=~ \frac{\sqrt{1-\epsilon ^2} \left(y_u - \tilde{y}_u \left(4-15 \epsilon^2\right)\right)}{y_u - \tilde{y}_u \left(4 - 5 \epsilon ^2\right)}~.
\label{F4F5}
\eea
In the limiting cases where only one of the two Yukawa couplings is
turned on, the above become functions of $\epsilon$ only.  In such
limits, they lie between the curves for $F_1$ and $\tilde{F}_3$ in the
right panel of Fig.~\ref{YukApprox} in Sec.~\ref{sec:couplings} (they
are not shown in the figure).

\begin{table}[tb]
\begin{center}
\begin{tabular}{|c|c|c|c|c|c|c|c|c|}
\hline
\rule{0mm}{5mm}
$r$/ MCHM & 10-5-10 & 5-5-10 & 
\begin{tabular}{c} 5-10-10, \\ 5-1-10\end{tabular} & 
\begin{tabular}{c} 5,\ 10, \\ 14-1-10 \\ 14-10-10 \\ 10-14-10 \end{tabular} & 
14-14-10 & 
14-5-10 & 
5-14-10  \\ [0.3em]
\hline
\rule{0mm}{5mm}
$r_t$ & $F_2$ & $F_1$ & $F_2$ & $F_1$ & $ F_3$ & $ F_4$ & $ F_5$ \\ [0.3em]
\hline
\rule{0mm}{5mm}
$r_b$ & $F_1$ & $F_2$ & $F_2$ & $F_1$ & $ F_1$ & $ F_1$ & $ F_1$ \\ [0.3em]
\hline
\rule{0mm}{5mm}
$r_V$ & $F_2$ & $F_2$ & $F_2$ & $F_2$ & $F_2$ & $ F_2$ & $ F_2$ \\ [0.3em]
\hline
\rule{0mm}{5mm}
$r_g$ & $F_2$ & $F_1$ & $F_2$ & $F_1$ & $ F_3$ & $ F_4$ & $ F_5$ \\ [0.3em]
\hline
\end{tabular}
\end{center}
\caption{Ratio of Higgs SM and MCHM couplings, $r=c^{MCHM}/c^{SM}$,
approximated by the functions $F_i$.  $g$ stands for the loop induced
gluon coupling (we have only considered the top sector effect for
$r_g$ in this table, but in the numerical results we have included the
bottom sector as well), $\psi=t,b$ are the Yukawa couplings and
$V=W,Z$ is the coupling to the massive EW gauge bosons.  For
completeness, we include also the result for additional models that
were not described in full detail in the main text.}
\label{table-F}
\end{table}

The $F_i$ functions defined in Eqs.~(\ref{F1F2})-(\ref{F4F5}) encode
the deviations from the SM couplings, $r=c_{\rm MCHM}/c_{\rm SM}$, as
shown in Table~\ref{table-F},~\footnote{Some of these functions have
been shown previously in Refs.~\cite{Pomarol:2012qf}
and~\cite{Montull:2013mla}.} and determine the $c_i$ coefficients of
the following set of operators in the low-energy theory:
\begin{align}
&{\mathcal O}_{g} = h \, G^a_{\mu\nu}G^{a \, \mu\nu}~, \qquad 
{\mathcal O}_{\gamma} = h \, A_{\mu\nu}A^{\mu\nu}~, \qquad
{\cal O}_{Z\gamma} = h \, A_{\mu\nu}Z^{\mu\nu}~, 
\label{Off} \\
&{\cal O}_{w} = h \, W^+_\mu W^{-\mu}~, \qquad
{\cal O}_{z} = h \, Z_\mu Z^{\mu}~, \\
&{\mathcal O}_f = \bar q_L H f_R+{\rm h.c.}
\label{Of}
\end{align}
These are the leading order operators involved in Higgs production and
decay at the LHC. Since the operators ${\mathcal O}_{g}$ and
${\mathcal O}_{\gamma}$ break the shift symmetry of the pNGB Higgs and
must, therefore, involve the explicit symmetry breaking parameters
such as the SM gauge and Yukawa couplings, they are generated at loop
level.  Our computation gives the contributions to the Wilson
coefficients of these operators in the MCHM after EWSB to all order in
the Higgs vev, leading to coefficients $c_{\cal O}(v/f)$.  Expanding
these coefficients in powers of $v/f$ one can do the matching to the
Wilson coefficients of dimension-six operators which, in the basis of
Refs.~\cite{Elias-Miro:2013gya,Elias-Miro:2013mua,Pomarol:2013zra},
are
\begin{align}
&{\cal O}_H=\frac{1}{2}\left(\partial_\mu|H|^2\right)^2 ~, & {\cal O}_{y_f}=|H|^2\bar q_LHf_R ~, \nonumber \\ 
&{\mathcal O}_{GG}=|H|^2 G_{\mu\nu}G^{\mu\nu} ~, & {\mathcal O}_{BB}=|H|^2 B_{\mu\nu}B^{\mu\nu} ~, \nonumber \\
&{\cal O}_W=\frac{i}{2}\left(H^\dagger \sigma^a\overleftrightarrow D_\mu H\right)D^\nu W^a_{\mu\nu} ~,
&{\cal O}_B=\frac{i}{2}\left(H^\dagger \overleftrightarrow D_\mu H\right)\partial^\nu B_{\mu\nu} ~, \nonumber \\
&{\cal O}_{HW}=i\left(D^\mu H\right)^\dagger \sigma^a\left(D^\nu H\right) W^a_{\mu\nu} ~, 
&{\cal O}_{HB}=i\left(D^\mu H\right)^\dagger \left(D^\nu H\right) B_{\mu\nu} ~.
\label{OpsDim6}
\end{align}

By redefining the Higgs field one can show that ${\cal O}_H$
renormalizes the Higgs couplings to all the other SM fields.  ${\cal
O}_{GG}, {\cal O}_{BB}$ and ${\cal O}_-=({\cal O}_W-{\cal O}_B)-({\cal
O}_{HW}-{\cal O}_{HB})$ enter in the interactions $hgg, h\gamma\gamma$
and $hZ\gamma$, respectively, and ${\cal O}_{y_f}$ enters in $hf\bar
f$~\cite{Giudice:2007fh}.  The Wilson coefficients $c_H, c_W$ and
$c_B$ are universal for all the MCHM with SO(5)/SO(4) breaking and
have been computed in the SILH description~\cite{Giudice:2007fh}:
\begin{equation}
c_H=1 \ ; \qquad\qquad 
c_W= c_B=\frac{27\pi^2}{256}\simeq 1.0 \ .
\end{equation}
$c_y$ has been computed in~\cite{Giudice:2007fh} for the top sector in
the MCHM$_{5}$.  In general it can be obtained from the functions
$F_\psi$ that codify the deviation of the Yukawa coupling, leading to:
\begin{align}
&c_{y_t}=1 \ , \qquad {\rm for\ the\ MCHM_{5,\ 10,\ 14-14-10,\ 14-1-10,\ 5-5-10}} \ , \nonumber \\
&c_{y_t}=0 \ , \qquad {\rm for\ the\ MCHM_{10-5-10,\ 5-10-10}} \ , \nonumber \\
&c_{y_b}=1 \ , \qquad {\rm for\ the\ MCHM_{5,\ 10,\ 14-14-10,\ 14-1-10,\ 10-5-10}} \ , \nonumber \\
&c_{y_b}=0 \ , \qquad {\rm for\ the\ MCHM_{5-5-10,\ 5-10-10}} \ .
\end{align}
The coefficients $c_{g,\gamma}$ and $c_{HW,HB}$ are generated at loop
level.  Starting with ${\cal O}_g$, this operator is generated by
fermion loops.  For each fermion species there is a contribution (see
App.~\ref{app:loops})
\begin{equation}
c_{g}\propto \sum_n \frac{y_n}{m_n}A_{1/2}(\tau_n) \ , \qquad \tau_n=\frac{m_h^2}{4m_n^2} \ .
\end{equation}
For heavy fermions, $\left.A_{1/2}(\tau)\right|_{\tau \to 0} \to 4/3$.  Thus,
considering heavy resonances we obtain:
\begin{equation}
\label{eq_cg}
c_{g} \propto \frac{4}{3}\left[\tr(Y_\psi M_\psi^{-1})-\frac{y^{(0)}_\psi}{m^{(0)}_\psi}\right] + \frac{y^{(0)}_\psi}{m^{(0)}_\psi}A_{1/2}(\tau_0)~,
\end{equation}
with the index 0 referring to the would-be 0-mode, associated with the
SM mass eigenstate.  The last term is similar to the SM one, up to
corrections in the Yukawa coupling.  These corrections are important
only if the mixing is large.  Since $A_{1/2}(\tau)\to_{\tau\to \infty}
0$, this term is small for light fermions, $m_\psi \ll m_h$.  As was
shown in Sec.~\ref{sec:models}, the first term is also small if the
mixing of both, the Left and Right chiralities, is small.  For the top
quark one can take the limit $A_{1/2}(\tau_t)\to 4/3$, and
Eq.~(\ref{eq_cg}) is dominated by $4/3\ \tr(Y_t M_t^{-1})$, which is
the sum considered in Eq.~(\ref{BasicRelation}).  Thus, one can also
obtain an approximate expression for the gluon fusion process in terms
of the functions above, as shown in Table~\ref{table-F}.  For the
coupling of the Higgs to two photons, there is an additional
contribution due to the heavy spin-1 resonances.  However, a similar
sum rule applies which allows to obtain an approximate analytical
expression.  These will be studied in more detail in
Sec.~\ref{sec:pheno}, after taking into account the constraints from
the recently measured Higgs mass~\cite{measured-h-mass}, as well as
the masses of the $Z$ gauge boson and the top and bottom quarks, which
have the most important impact on the Higgs potential and the Higgs
phenomenology.

\section{Higgs potential}
\label{sec_div_VH}

Discrete models of pNGB Higgs can lead to a finite Higgs potential
under some suitable assumptions.  The degree of divergence of the
Higgs potential depends on the particular mechanism of collective
breaking, being thus model dependent.  There are at least two concepts
involved: distance between the sites where the symmetries protecting
the pNGB potential are broken, and number of symmetries broken on each
site.

The Higgs potential can be computed by the holographic method
\bea
\label{ec_VHholo}
V(h) &=& \int \! \frac{d^4p}{(2\pi)^4} \left[ \frac{6}{2}\sum^2_{i = 1} \log \Pi_{w^i_L} + \frac{3}{2}\log \left[\Pi_{w^3_L} \Pi_{b} - (\Pi_{w^3_L \, b})^2 \right]
\right.
\nonumber \\ [0.4em]
& & \hspace{1cm}
\left. \rule{8mm}{0mm}  - 2N_c\sum_\psi \log[p^2\Pi_{\psi_L} \Pi_{\psi_R} - |M_{\psi}|^2] \right] \ , 
\eea
where the correlators are obtained from Secs.~\ref{sec:EFT} and
\ref{sec:models}, taking care to add the ``bare" kinetic terms, as in
Eqs.~(\ref{Lgauge}) and (\ref{Leff-fermions}), which were not included
as part of the definition of the correlators in those sections:
\bea
\Pi_{w^i_L} = \frac{p^2}{g^2_0} + \Pi_{\tilde{w}^i_L}~,
\hspace{1cm}
\Pi_{w^3_L \, b} = \Pi_{\tilde{w}^3_L \, \tilde{b}}~,
\hspace{1cm}
\Pi_{b} = \frac{p^2}{g^{\prime 2}_0} + \Pi_{\tilde{b}}~,
\eea
and similarly for the fermionic correlators.  Equivalently, one can
use the standard expression for the Coleman-Weinberg potential in
terms of determinants involving the Higgs-dependent mass matrices of
the gauge and fermion fields.  We have checked that the same results
can be reproduced with either approach.  Note that
Eq.~(\ref{ec_VHholo}) contains the photon, although it does not
contribute to the Higgs potential, and one can regularize the
divergent constant terms by subtracting $V(0)$.

\subsection{Finiteness of the 1-loop Higgs potential}
\label{sec:finiteness}

In this subsection we illustrate in a toy example how the
inclusion/exclusion of certain operators in the Lagrangian affects the
divergence structure of the Higgs potential.  Our example is based on
the fundamental representation of SO(5), but the conclusion holds for
other representations as well.  In order to understand the structure
of divergences of the $h$-dependent terms, let us consider the 2-site
model with the following set of fields:

\noindent {\bf site 0}: An elementary fermion doublet $q_L$ and a
singlet $t_R$ of a global symmetry $G_0 = {\rm SU}(2)_L$.\footnote{For
simplicity we ignore U(1)$_Y$ in this discussion.}

\noindent {\bf site 1}: Four chiral composite fermions $Q_L, Q_R, T_L,
T_R$, each transforming in the fundamental representation of a
different global SO(5), called: $G_{Q_L}$, $G_{Q_R}$, $G_{T_L}$,
$G_{T_R}$.  In this site there is also a scalar $\Phi_1$ transforming
in the fundamental of another SO(5), called: $G_1$.  The vev of
$\Phi_1$ spontaneously breaks $G_1$ to ${\cal H}_1= {\rm SO(4)}$.

Notice that before introducing fermion masses, each chiral fermion of
the composite sector transforms independently, leading to a large
global symmetry (in fact, the symmetry is much larger, but we need
only focus on this subgroup).  The Higgs, being a NGB, is in the coset
$G_1/{\cal H}_1$.  The following operators break different symmetries:
\begin{itemize}
\item $m_{Q} \, \bar QQ$: $G_{Q_L} \times G_{Q_R} \to G_{Q_{L+R}} = {\rm SO(5)}$~,

\item $m_T \, \bar TT$: $G_{T_L} \times G_{T_R} \to G_{T_{L+R}} = {\rm SO(5)}$~,

\item $\Delta_q \, \bar q_LQ_R + {\rm h.c.}$: $G_0 \times G_{Q_R} \to G_{Q_R+0} = {\rm SU(2)}$~,

\item $\Delta_t \, \bar t_RT_L + {\rm h.c.}$: $G_0 \times G_{T_L} \to G_{T_L+0} = {\rm SU(2)}$~,

\item $y_T \, \bar Q_L\Phi_1\Phi_1^\dagger T_R + {\rm h.c.}$: $G_{Q_L} \times G_{T_R} \times G_1 \to G_{Q_L+T_R+1} = {\rm SO(5)}$~,

\item $y'_T \, \bar Q_R\Phi_1\Phi_1^\dagger T_L + {\rm h.c.}$: $G_{Q_R} \times G_{T_L} \times G_1 \to G_{Q_R+T_L+1}= {\rm SO(5)}$~.
\end{itemize}
There is some abuse of notation in the previous paragraph, since
$G_{Q_R,T_L}$ and $G_0$ have different dimensions, so that when writing
$G_{T_L+0}$ we really mean the diagonal subgroup $G_0'=$SU(2). In
addition to the above, the symmetries allow operators of the form $\bar
Q_L\Phi_1\Phi_1^\dagger Q_R + {\rm h.c.}$ or $\bar
T_L\Phi_1\Phi_1^\dagger T_R + {\rm h.c.}$, which would also lead to
divergences in the Higgs potential of the 2-site model.  With three or
more sites, these would lead to a finite 1-loop
result~\cite{Panico:2011pw,DeCurtis:2011yx}.  For illustration, we
limit the following discussion to the operators listed above.

A Higgs potential requires insertions of $y_T$ and/or $y'_T$.  Let us
consider the following cases:

\noindent {\bf (a)} $y'_T=0$: The $y_T$ term only preserves the
diagonal subgroup $G_{Q_L+T_R+1}$.  The Higgs is in the coset
$G_{Q_L+T_R+1}/{\cal H}_1$, and thus a Higgs potential requires
explicit breaking of $G_{Q_L+T_R+1}$.  This necessitates interactions
with the elementary sector, which arise from the $\Delta_q$ and/or
$\Delta_t$ terms.  However, due to their chirality structure,
insertions of $\Delta_{q,t}$ still do not break $G_{Q_L+T_R+1}$:
$G_{Q_L+T_R+1}\times$$G_0$ is broken only after additional $m_{Q,T}$
insertions.  Thus,
\begin{equation}
V_H\sim (\Delta_{q,t} m_{Q,T} y_T)^2\ .
\end{equation}

\noindent {\bf (b)} $y_T=0$: The $y'_T$ term only preserves the
diagonal subgroup $G_{Q_R+T_L+1}$ and the Higgs is in the coset
$G_{Q_R+T_L+1}/{\cal H}_1$.  In this case, insertions of $\Delta_q$
and/or $\Delta_t$ break $G_{Q_R+T_L+1}\times$$G_0$ without the need of
$m_{Q,T}$ insertions:
\begin{equation}
V_H\sim (\Delta_{q,t} y'_T)^2\ .
\end{equation}

The previous arguments show how the dimension of the operators
leading to $V_H$ depends on the presence of $y'_T$, leading to
logarithmic divergences at 1-loop for $y'_T\neq0$. The presence of the
operators $m_{Y_t}\ \bar Q_L T_R$ and $m'_{Y_t}\ \bar Q_R T_L$
modifies the potential but not its degree of divergence.

One can also understand this result from Feynman diagram
considerations.  For instance, the contribution to the quartic term in
$\Phi$, at leading order in insertions of $m_\psi$ and $\Delta_\psi$
is given by:

\hspace{1.2cm}
\includegraphics[width=0.85\textwidth]{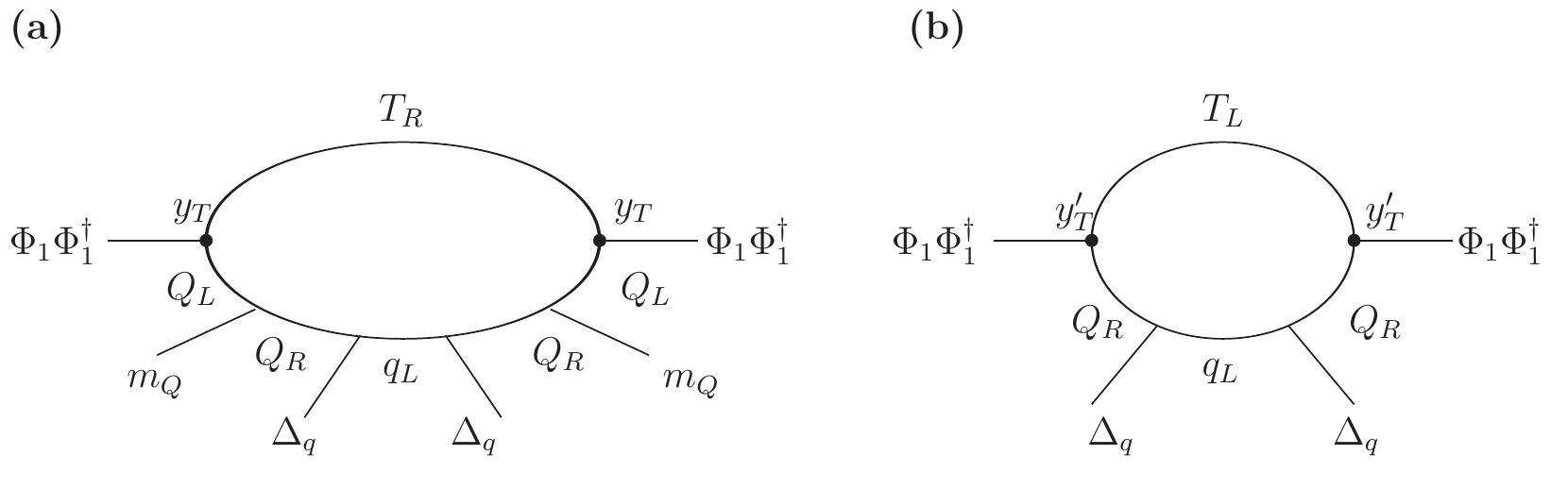}

\noindent and similar diagrams changing $q\leftrightarrow t$ and
$Q_{L,R}\leftrightarrow T_{R,L}$.  These diagrams allow to understand
the superficial degree of divergence of $V_H$ depending on which
operators are present in the theory.

\section{Higgs Phenomenology}
\label{sec:pheno}

We turn now to the Higgs phenomenology of the composite Higgs models
previously described.  We present in this section the results of a
detailed numerical analysis obtained by scanning over a sizeable
region of the parameter space of each model.  The minimization of the
Higgs potential will be fully taken into account.  Note, however, that
we assume that for the light fermion generations both the LH and RH
chiralities have a small degree of compositeness, as opposed to
allowing one of them to have a large mixing angle with the composite
sector, and the other a very suppressed one that accounts for the
small SM fermion
mass~\cite{Redi:2011zi,DaRold:2012sz,Redi:2013eaa,Delaunay:2013pwa}.
This assumption is more natural given the EW precision tests, which
indicate that the light quarks and leptons are mostly elementary,
although one could imagine exploring the second option.  As a result,
the Higgs potential is affected mainly by the top and bottom sectors,
as well as by the gauge sector of the models.  Nevertheless, when
discussing the Higgs decays we will take into account some of the
light fermions, most prominently the $\tau$ lepton, as discussed
below.

\subsection{Numerical Scan}\label{sec-num-scan}

The effective description of a composite Higgs described in the
previous sections depends on a number of parameters.  The gauge sector
is described at the Lagrangian level by the two decay constants
$\{f_\Omega, f_{\Omega_X}\}$ and gauge couplings $\{g_\rho, g_X\}$
associated with the SO(5) and ${\rm U(1)}_X$ (composite) factors,
while in the elementary sector one has the two gauge couplings $g_0$
and $g'_0$ [see Eqs.~(\ref{LOmega}) and (\ref{Lgauge})].  The latter
are related to the SM gauge couplings as given in Eq.~(\ref{SMggp}),
while it is convenient to parametrize the composite gauge couplings in
terms of the elementary/composite mixing angles of the gauge sector:
$t_\theta = g_0 / g_\rho$ and $t_{\theta'_X} = g'_0 / g_X$.  However,
for simplicity, in our scan we will fix $g_X$ by imposing the relation
discussed after Eq.~(\ref{SMggp}), so that there is effectively a
single gauge mixing angle $t_\theta$.  The two decay constants can in
turn be exchanged for the two mass scales $m_\rho$ and $m_X$ defined
in Eq.~(\ref{mrhomX}), but it is more convenient to scan over a subset
of the physical masses after taking into account the
elementary/composite mixing effects (before including EWSB
effects).  Thus, we choose to scan over $m_{\tilde{\rho}} = \sqrt{1 +
t^2_\theta} \, m_\rho = m_\rho / c_\theta$ [see discussion of the last
paragraph of Sec.~\ref{BosonicSector}], and we also choose the
variable $m_{\tilde X} = m_X / c_\theta$.  However, since we focus on
a region of parameter space with $t_\theta \ll 1$, quantitatively
there is not a large difference between $m_{\tilde \rho}$ and $m_\rho$
or $m_{\tilde X}$ and $m_X$.

The fermion sector depends on a set of ``diagonal" masses $m_\Psi$,
one for each composite fermion, and on the ``off-diagonal" masses
$m_{y_u}$ and $m_{y_d}$ of Eq.~(\ref{Lfermions}).  The composite
sector also involves a number of ``Yukawa-like" mass parameters that
we have called $y_u$ and $y_d$ [see Eqs.~(\ref{LY5}), (\ref{LY10}),
(\ref{LY10510}), (\ref{LY5510}), (\ref{LY141410}), (\ref{14y2}) and
(\ref{LY14110}) which define these for each model].  In spite of the
notation, the $y_\psi$ have dimensions of mass, although they
represent interactions with the Higgs field $\Phi$.  Finally, there
are the mixing parameters, $\Delta_q$, $\Delta_u$ and $\Delta_d$,
which also have mass dimension 1.  In practice, the scan will be
restricted to the third generation, so that one should reinterpret the
indices as $u \to t$ and $d \to b$.  We find convenient to exchange
the mixing parameters $\Delta_\psi$ for ``mixing angles" defined by
$t_{\psi} \equiv \tan \theta_\psi = \Delta_\psi / m_\Psi$, where
$\Psi$ is the composite fermion associated with the elementary fermion
$\psi$ [for the MCHM$_5$ we introduce two mixing angles $t_{q^u}$ and
$t_{q^d}$ corresponding to $\Delta_{q^u}$ and $\Delta_{q^d}$; see
comments after Eq.~(\ref{Lfermions})].  Analogously to the gauge
sector above, we also prefer to scan over diagonal fermion masses that
have been rescaled according to $m_{\tilde \Psi} = m_\Psi / c_\psi$,
where $c_\psi = \cos \theta_\psi$ involves the corresponding mixing
angle defined above.  This choice leads to light custodians when the
mixings are large, since their masses are given by $m_{\rm cust} \sim
{\cal O}(m_{\tilde \Psi} c_\psi$)~\cite{Pomarol:2008bh,DaRold:2010as}.
Thus, the parameters for the fermionic sector consist of $\{ m_{\tilde
\Psi}, t_\psi , m_{y_\psi}, y_\psi \}$, where the indices run over the
field content in each model, as described in Sec.~\ref{sec:models} [we
fix $Z_\psi = 1$ in Eq.~(\ref{Lfermions})].

Since one expects that the masses of the various resonances will be of
the same order, for simplicity we have fixed a common mass scale, by
restricting our scan to $m_{\tilde \rho} = m_{\tilde Q} = m_{\tilde U}
= m_{\tilde D}$ (for the MCHM$_5$ we impose the condition on
$m_{\tilde Q^u}$ and $m_{\tilde Q^d}$).  This is not necessary, but we
do not expect that the results will depend on this simplifying
assumption.\footnote{Note that the physical masses are obtained after
taking into account all the mixing effects, as well as EWSB, and will
therefore present a nontrivial spread.  It is also worth noting that
by scanning over $m_{\tilde \rho}$, $m_{\tilde X}$ and $m_{\tilde
\Psi}$, \textit{i.e.}~by factoring out the elementary/composite mixing
angles, we are proceeding in analogy to the extra-dimensional
realizations, where the compactification scale and therefore the
overall Kaluza-Klein (KK) scale is treated as an input parameter.  The
elementary/composite mixing angles of the 4D realization are related
to the 5D localization parameters and boundary conditions for the
various fields.  When obtaining the exact spectrum one can get modes
much lighter than the overall KK scale, typically for large mixing
angles in the third generation fermionic sector.} Thus, the final set
of parameters used in the scan is
\bea
\{ f_h, m_{\tilde \rho}, t_\theta, t_q, t_t, t_b, m_{y_T}, m_{y_B}, y_{T}, y_{B} \}~,
\eea
where we used the notation $y_T$ and $y_B$ instead of $y_t$ and $y_b$
to avoid confusion with the SM top and bottom Yukawa couplings, and we
also included in the list the Higgs decay constant $f_h$ defined by
Eqs.~(\ref{LNGBSimple}) and (\ref{eq:fh}).  We also chose to fix
$m_{\tilde X} = s_{\theta_W} / \sqrt{c_{2\theta_W}} \, m_{\tilde \rho}
\approx 0.65 \, m_{\tilde \rho}$, which amounts to fixing
$f_{\Omega_X} = f_\Omega$ in Eq.~(\ref{LOmega}), given the choice of
$g_X$ described above.  We choose $1/5 \leq t_\theta \leq 1/3$, so
that $g_\rho$ is large but perturbative, and scan over the fermionic
mixing angles according to $s_\psi \in [0.4, 1]$, with a uniform
distribution (but we adjust $s_b$ to reproduce the bottom quark mass
with little effect on the EWSB properties of the parameter point).
For the mass parameters, ($m_{\tilde \rho}, m_{y_T}, m_{y_B}, y_{T}
\textrm{ and } y_{B}$), we scan in units of $f_h$ as follows:
\begin{itemize}

\item $m_{\tilde{\rho}} / f_h \in [2.5, 5]$, which is consistent with
the underlying relation $m_{\rho} \sim g_\rho f_h$ with $g_\rho$ in
the range of interest,

\item $| y_\psi / f_h | < 2\pi$, which encodes the idea of having a
perturbative proto-Yukawa coupling,

\item and $| m_{y_T} / f_h |, | m_{y_B} / f_h | \lesssim 2\pi$~,

\end{itemize}
while $f_h$ is scanned over a wide range, but we choose only points
with $\epsilon < 0.5$, which corresponds to $f_h \gtrsim 500~{\rm
GeV}$.  The final set of points has $f_h$ as large as $\sim 2.5~{\rm
TeV}$ (except for the MCHM$_{5-10-10}$, which has some points with
$f_h$ as large as $\sim 6~{\rm TeV}$).  We also required in the final
set of points that $m_{\tilde \rho} > 2~{\rm TeV}$.  This final set of
numbers already assumes that we have normalized to $m_Z$ (see below).

Having chosen a given point in the parameter space described above, we
minimize the 1-loop Higgs potential to select those points that do
break the EW symmetry.  For each such point, we can rescale all
parameters with dimension of mass so as to reproduce $m_Z$, thereby
normalizing to the EW scale.  We further select those points where the
Higgs mass matches the measured value of $\sim 125~{\rm GeV}$, and
also select those points where the top and bottom quarks match the
experimental observations.  In practice, our final points have $m_h
\in [120-130]~{\rm GeV}$, $m_t \in [140-170]~{\rm GeV}$ and $m_b
\approx 2.7~{\rm GeV}$.\footnote{We note that the relevant masses from
the point of view of the scan should be the running masses at the
scale where the heavy resonances are integrated out.  These would then
be run down to the weak scale with the SM RGE's to make contact with
the experimental measurements.  Since each parameter point has a
different scale for the heavy resonances, we have simply defined
generous windows to capture the spirit of the matching procedure.
Although a more precise analysis is possible, we do not expect that
the conclusions will change.} We can then compute the couplings of the
Higgs to the vector bosons and fermions (both the SM ones as well as
the new resonances), which are then used as input to compute the Higgs
production cross sections and branching fractions.  This is done
numerically without any approximations, as is done for the 1-loop
induced couplings ($hgg$, $h\gamma\gamma$ and $hZ\gamma$) which are
computed using the exact spectrum and couplings to the Higgs.
However, we also compare to the analytical approximation described in
Sec.~\ref{sec:corrections}, which in general gives a qualitative
understanding of the numerical scan.

\begin{figure}[t] 
\centering
\includegraphics[width=0.48\textwidth]{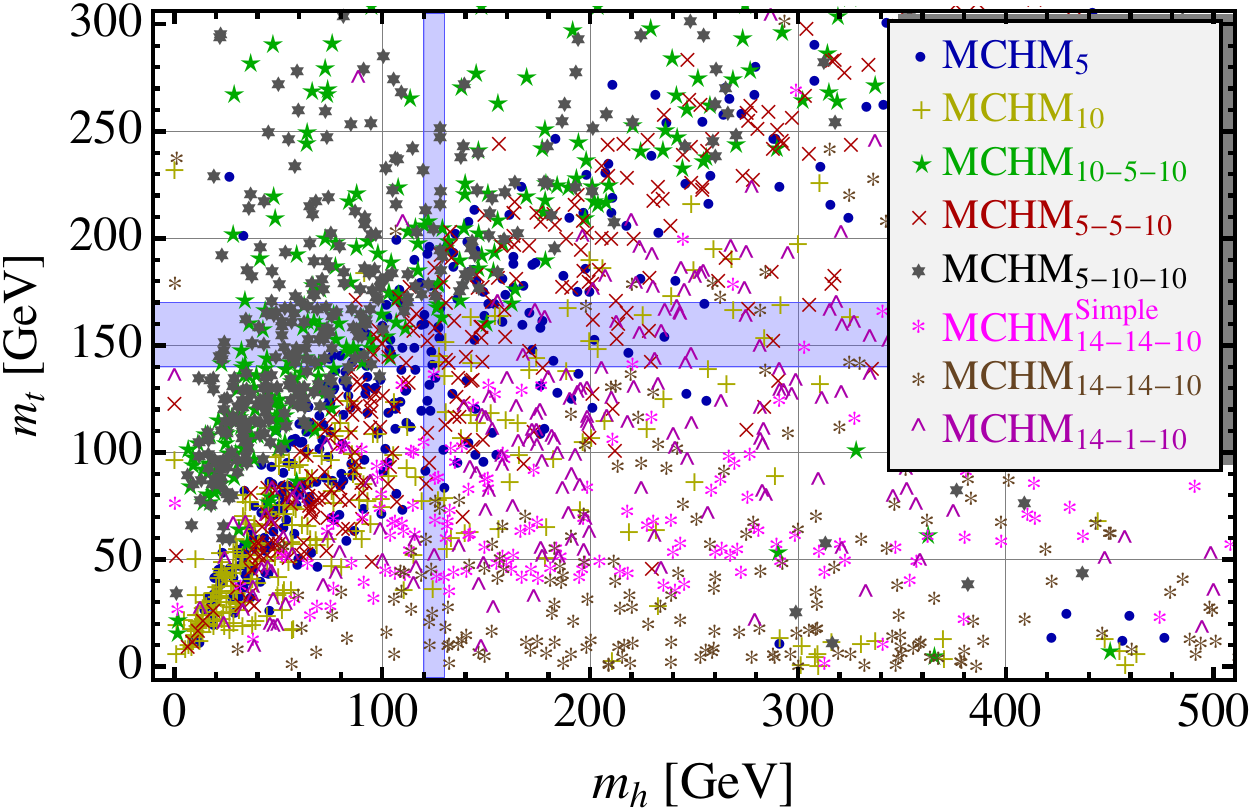}
\hspace{3mm}
\includegraphics[width=0.48\textwidth]{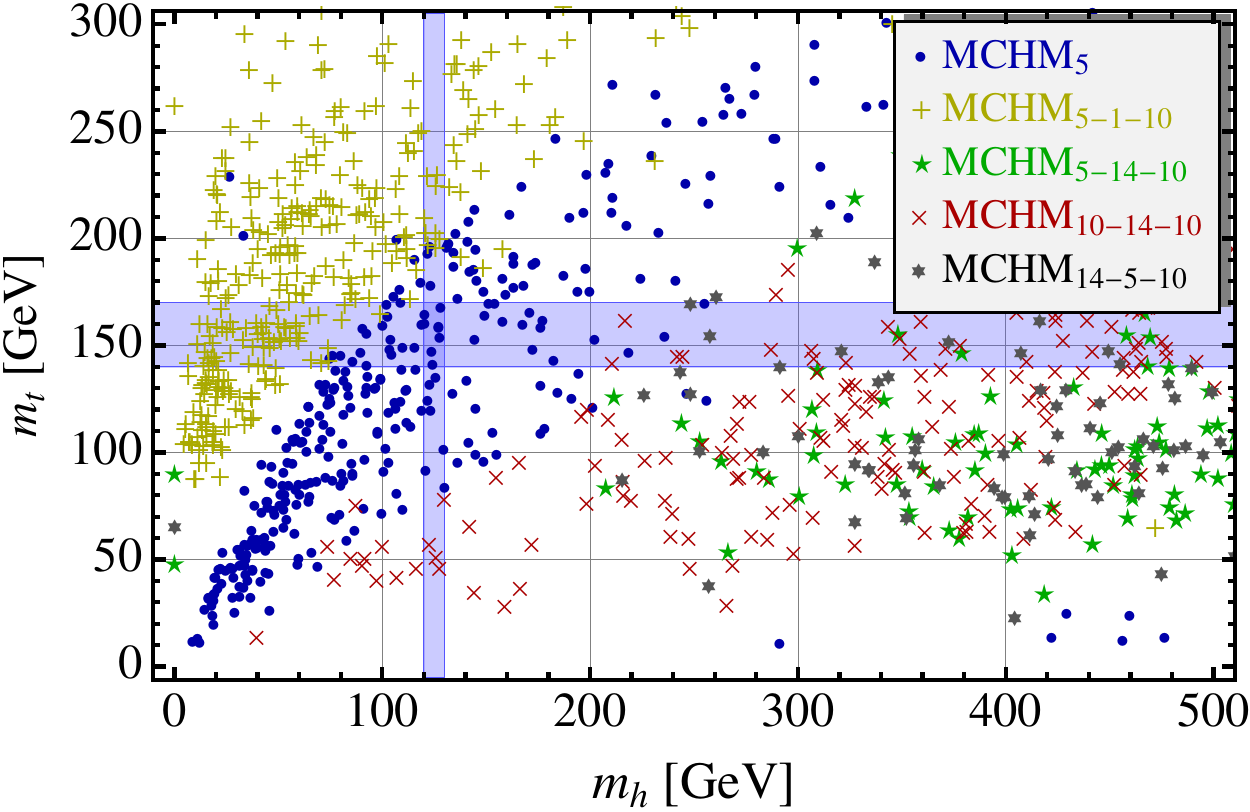}
\caption{A random subset of the points that present electroweak
symmetry breaking, but without requirements on the Higgs, top or
bottom masses (however, we have normalized to $m_Z$).  The vertical
and horizontal bands indicate the windows we have defined for $m_h$
and $m_t$.  In the left panel we show the models we have presented in
detail in Sec.~\ref{sec:models}.  In the right panel we show the
models mentioned in Sec.~\ref{sec:othermodels}, showing again the
MCHM$_5$ for comparison.}
\label{fig:mtvsmh}
\end{figure}

In Fig.~\ref{fig:mtvsmh} we display a random subset of the scanned
points that display EWSB, in the plane of $m_t$ versus $m_h$ (after
normalization to $m_Z$).  We have not imposed here any requirements on
$m_h$, $m_t$ nor $m_b$, only that the desired symmetry breaking
pattern be obtained and that the $b_R$ mixing angle be suppressed (as
is necessary to obtain a light bottom quark in models with just one
operator coupled to $q_L$).  In the left panel we present the (color
coded) models described in detail in Sec.~\ref{sec:models}, showing
that some of the models reproduce more naturally the Higgs and top
masses than others.  In particular, the models involving the 14
representation have a tendency to produce a too large
$m_h$~\cite{Pomarol:2012qf}, although one can find a few points in the
desired range at the price of tuning (the bands correspond to the
windows we have defined in the previous paragraph).

In the right panel, we show the same information for the models
mentioned without details in Sec.~\ref{sec:othermodels}, together with
the MCHM$_5$ for comparison purposes.  We see that these models also
typically do not fall in the phenomenologically desired window: for
the MCHM$_{5-1-10}$ the quartic coupling is usually too small, since
the only source of breaking is the mixing with $q_L$, that leads to a
factor $s_h$ in $\Pi_{u_L}$ and $s_h^2$ in $M_u$, in agreement with
the results found in~\cite{Pomarol:2012qf}.  The MCHM$_{10-14-10}$
leads to a heavy Higgs.  The MCHM$_{14-5-10}$ and MCHM$_{5-14-10}$
allow for two independent proto-Yukawa interactions: ${\cal
L}_y\supset y_u \bar\Psi_5\Psi_{14}\Phi+\tilde y_u
(\bar\Psi_5\Phi)(\Phi^\dagger\Psi_{14}\Phi)$, similar to the
MCHM$_{14-14-10}$.  Both of them generically lead to a heavy Higgs,
while EWSB prefers $\tilde y_u\neq 0$ for the MCHM$_{14-5-10}$ and
$y_u\neq 0$ as well as $\tilde y_u\neq 0$ for the MCHM$_{5-14-10}$.
For the remaining three models we did not find points with the proper
$m_h$ and $m_t$ by performing a random scan.  Finally, the
MCHM$_{14-10-10}$ generically does not lead to EWSB.

In all these models there is a correlation between $m_h$ and
$m_t$~\cite{Contino:2006qr}, that can usually be approximated by:
$m_h^2\sim a \frac{N_c}{\pi^2}\frac{m_t^2}{f_h^2}m_\psi^2$, with
$m_\psi$ the scale of the lightest fermionic resonance cutting off the
1-loop potential and $a$ a factor that is model dependent.  Usually
$a\sim{\cal O}(1)$, however in some cases it can be suppressed
$a\sim{\cal O}(\epsilon^2)$ or enhanced $a\sim{\cal
O}(\epsilon^{-2})$, as shown in~\cite{Pomarol:2012qf}.  The analytical
approximations of~\cite{Pomarol:2012qf} are in qualitative agreement
with the full numerical results of Fig.~\ref{fig:mtvsmh}.

From here on we focus on the models described in detail in
Sec.~\ref{sec:models}, which seem to be phenomenologically preferred
due to the previous observations.  As mentioned earlier, we analyze
the MCHM$_{14-14-10}$ in detail, even though it tends to produce too
heavy a Higgs, as it may serve also to illustrate the situation in
those models we do not elaborate any further.  All the numerical
results of the following sections correspond to points that lie at the
intersection of two bands of Fig.~\ref{fig:mtvsmh}.

\subsection{Corrections to the Gauge and Yukawa Couplings}
\label{sec:couplings}

\begin{figure}[t] 
\centering
\includegraphics[width=0.48\textwidth]{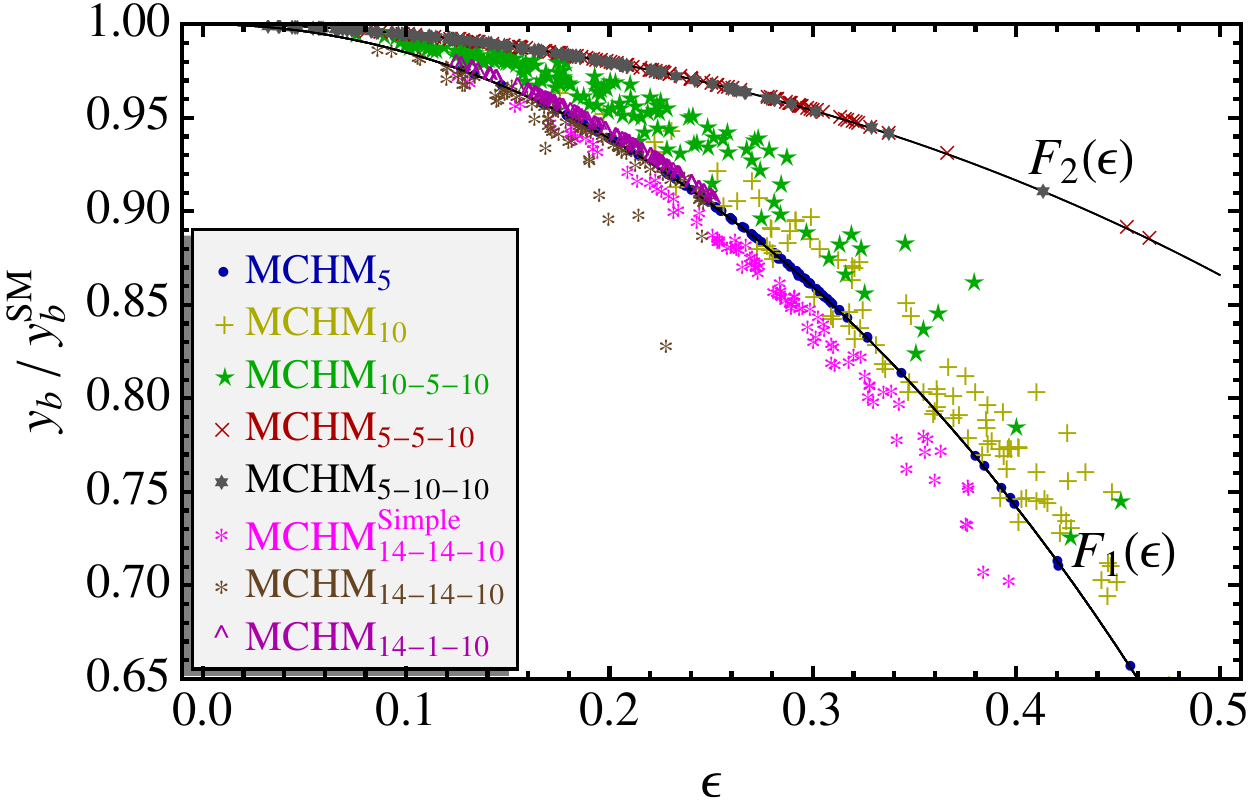}
\hspace{3mm}
\includegraphics[width=0.48\textwidth]{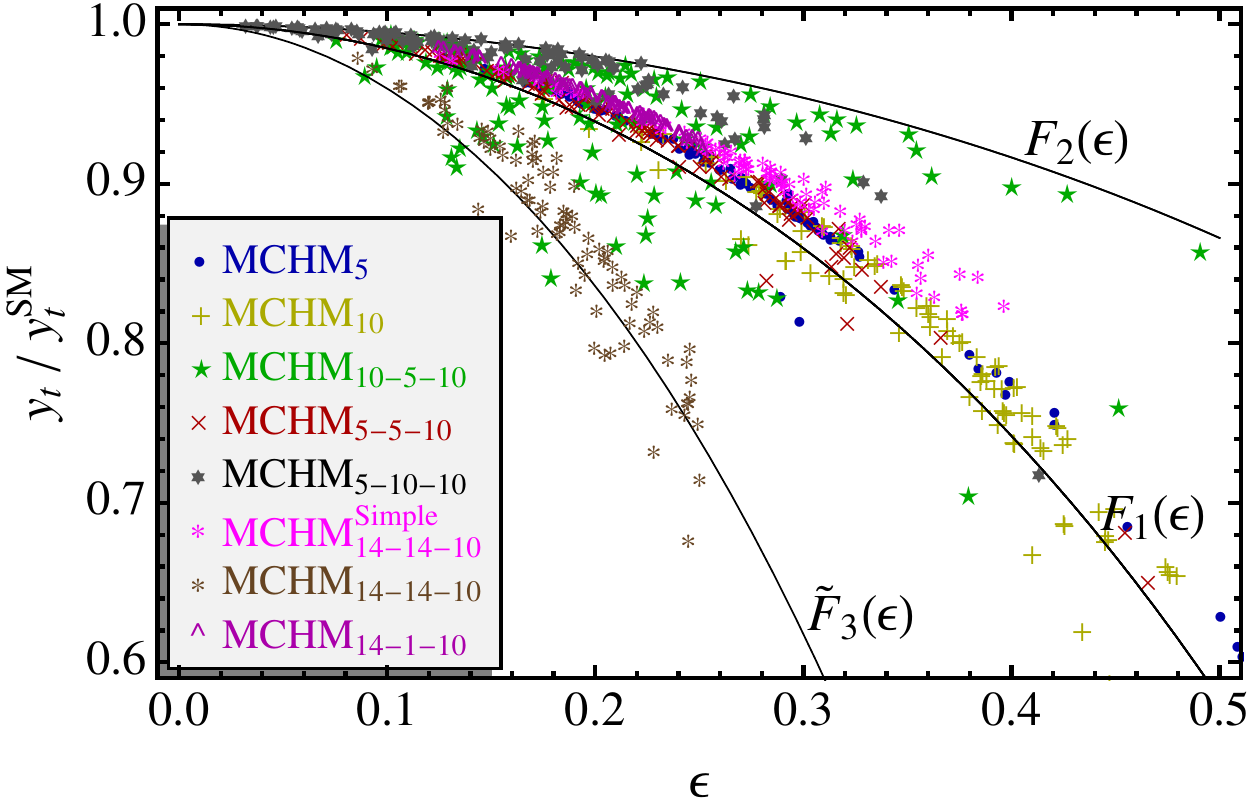}
\caption{Bottom (left panel) and top (right panel) Yukawa couplings in
several models, normalized to the SM (defined as $y_\psi = m_\psi /
v_{\rm SM}$ with $v_{\rm SM} = 246~{\rm GeV}$).  The points correspond
to a random scan in parameter space, while the solid curves correspond
to the analytic approximation discussed in the main text.}
\label{YukApprox}
\end{figure}

We start by comparing the simple analytical approximation described
in Sec.~\ref{sec:corrections} for the deviations in the Higgs
couplings to the SM gauge bosons and fermions w.r.t.~the SM
expectation [see also the discussion after Eq.~(\ref{eq-y})].  As
discussed there, this approximation is expected to work well when the
elementary/composite mixing angles are small, which typically happens
for the light fermions in our scenario.  However, we find that even
for the top quark, the approximation $y_t \approx [F_t(\epsilon) /
(\epsilon f_h)] \, m_t$ is reasonably good, even when the mixing
angles are sizeable, provided there are no ``ultra-light" fermionic
resonances.  This is illustrated in Fig.~\ref{YukApprox}, where we
show the bottom and top Yukawa couplings as a function of $\epsilon$
in several models (normalized to the corresponding SM Yukawa coupling,
$y^{\rm SM}_\psi \equiv m_\psi / v_{\rm SM}$ with $v_{\rm SM} \approx
246~{\rm GeV}$).  The points correspond to a random scan over the
parameter space described in the previous subsection, while the solid
curves correspond to the approximation described in
Sec.~\ref{sec:corrections} (see Table~\ref{table-F}).

We see in the left panel of Fig.~\ref{YukApprox} that the
approximation described in Eq.~(\ref{BasicApproximation}) works very
well for the bottom sector all the way up to relatively large values
of $\epsilon$.  A notable exception occurs for the ${\rm
MCHM}_{10-5-10}$ (green stars), where the analytic expectation, $F_1 =
\cos(2 v /f_h) / \cos(v/f_h)$, systematically overestimates the
suppression in $y_b$ compared to the SM. The sizeable deviation
observed can be understood by considering the next to leading order
term in the expansion of $y_b/y_b^{SM}$ in powers of $\epsilon$, as
shown in Sec.~\ref{sec:models}.  We obtain that, after the selection
of points explained above, the coefficient of the ${\cal
O}(\epsilon^2)$ term for the MCHM$_{10-5-10}$ is of ${\cal O}(0.5)$.
In contrast, the corresponding coefficient for the MCHM$_{10}$,
MCHM$_{14-14-10}$ and MCHM$_{14-14-10}^{\rm simple}$ is of ${\cal
O}(0.1)$,\footnote{MCHM$^{\rm simple}_{14-14-10}$ refers to the model
described in Sec.~\ref{sec14-14-10} with $\tilde{y}_T = 0$ in
Eq.~(\ref{sec14-14-10}) [making $u \to T$].  We refer to the general
model with $y_T$, $y_B$ and $\tilde{y}_T$ turned on as
MCHM$_{14-14-10}$.} for the MCHM$_{5}$ and MCHM$_{14-1-10}$ it is of
${\cal O}(10^{-2})$, and for the MCHM$_{5-5-10}$ and MCHM$_{5-10-10}$
it is ${\cal O}(10^{-4})$, in all the cases increasing with $s_q$ as
expected.  Since $h \to b\bar b$ is the dominant decay mode,
deviations of $y_b$ can have a deep impact in the Higgs phenomenology.

\begin{figure}[t] 
\centering
\includegraphics[width=0.48\textwidth]{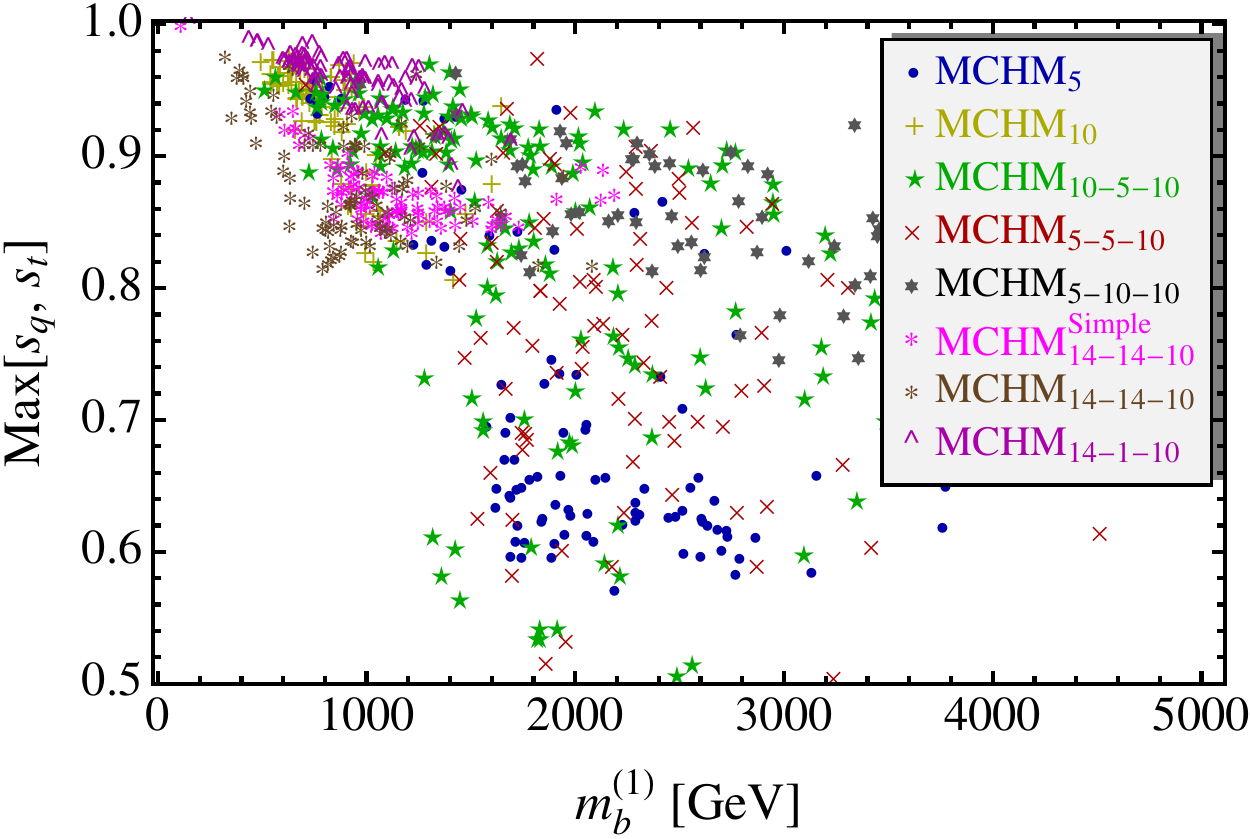}
\hspace{3mm}
\includegraphics[width=0.48\textwidth]{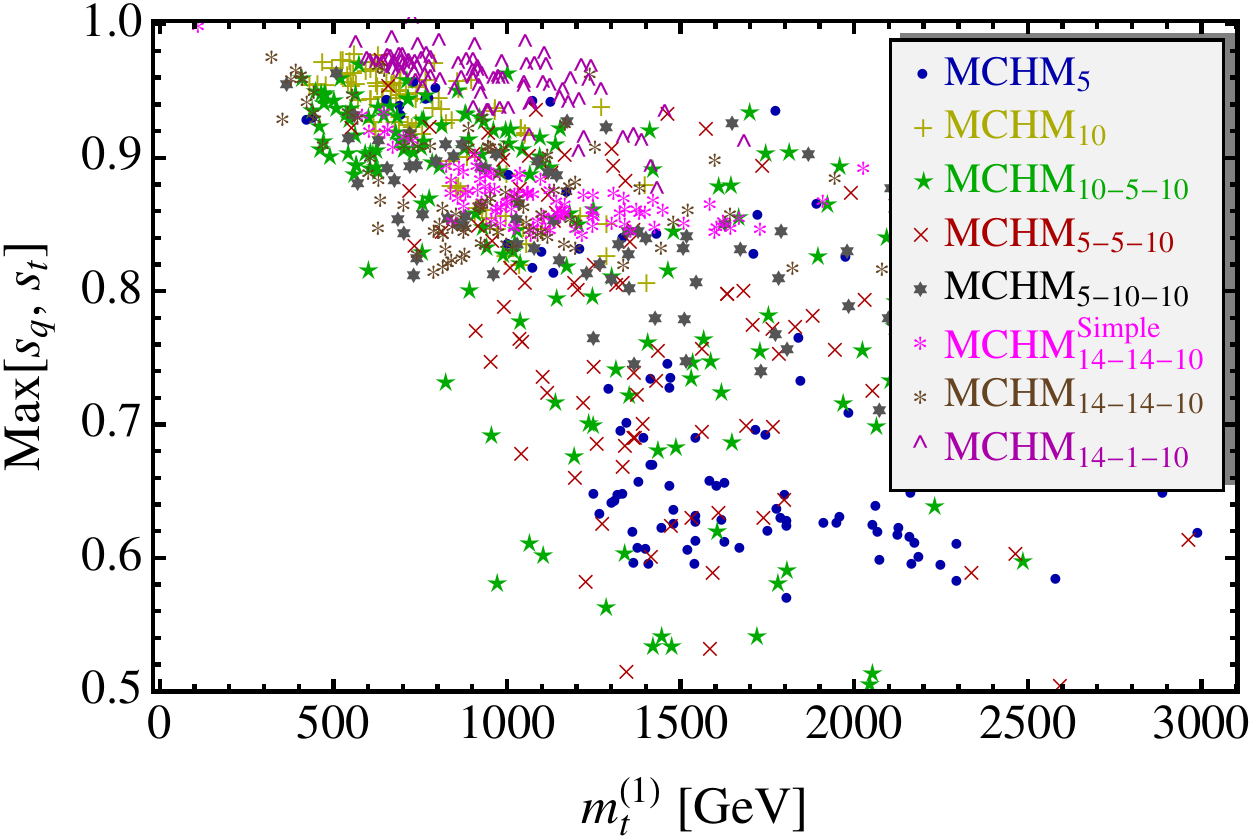}
\caption{The largest of the mixing angles between $s_q$ and $s_t$
versus the lightest $Q = -1/3$ resonance (left panel) and $Q = 2/3$
resonance (right panel) in several models.  For the MCHM$_5$ we plot
the largest between $s_{q_u}$, $s_{q_d}$ and $s_t$.}
\label{sqvsmF}
\end{figure}

It is also interesting to note that the bulk of the points in the ${\rm
MCHM}_{10-5-10}$ display
relatively light ($Q = -1/3$) fermionic resonances, together with
relatively large mixing angles.  We illustrate this in the left panel
of Fig.~\ref{sqvsmF}, where we show the largest of the mixing angles
$(s_q, s_t)$ versus the lightest vectorlike resonance mass in the
bottom sector.  Indeed, most of the green stars (MCHM$_{10-5-10}$)
exhibit resonances below $1~{\rm TeV}$ and $s_q > 0.9$.  Note that the
MCHM$_{10}$ (yellow $+$'s), the MCHM$_{14-14-10}$ (brown $*$'s), and
to a somewhat lesser extent the MCHM$^{\rm simple}_{14-14-10}$
(magenta $*$'s), also contain a subset of points with light states
together with sizeable elementary-composite mixing angles, which is
reflected in the somewhat larger dispersion in Fig.~\ref{YukApprox},
compared to the other models.  However, note that the MCHM$_{14-1-10}$
(dark magenta \^{}'s) has light $Q = -1/3$ resonances together with
large mixing angles, and nevertheless follows the naive approximation
from Eq.~(\ref{BasicApproximation}) for the bottom Yukawa coupling
rather well. 

The right panels of Figs.~\ref{YukApprox} and \ref{sqvsmF} display the
same information for the top sector (using the lightest $Q = 2/3$
fermionic resonance as the relevant variable).  Here, the dispersion
of the points around the continuous curves is larger, but the general
behavior is still well described by the simple analytic formulas given
above, again with the exception of the MCHM$_{10-5-10}$ (green stars),
which all fall below the ``expected curve" given by $F_2 =
\cos(v/f_h)$.  Thus, the analytic approximation underestimates the
suppression in the top Yukawa coupling compared to the SM in this
model.  We also note here that the analytic approximation,
$F_1(\epsilon)$, slightly underestimates the exact result for the
MCHM$_{10}$, MCHM$_{14-14-10}$ and the MCHM$_{14-1-10}$ (with the
effect being more pronounced for the latter two).  Finally, we point
out that after imposing the physical conditions described in the
previous section, the points in the MCHM$_{14-14-10}$ typically have
$y_T \ll \tilde{y}_T$.  This means that the deviations from the SM in
the top sector are reasonably well described by the function
$\tilde{F}_3(\epsilon)$ [see discussion around Eq.~(\ref{F3tilde})],
as can be seen in the right panel of Fig.~\ref{YukApprox}.

Besides the above resonances, one can also find light exotic
resonances with charge $Q=8/3,5/3$ and $-4/3$, depending on the
fermion representations involved.  These resonances are also
custodians, thus their masses are also suppressed if they belong to
SO(5) multiplets with large mixing with the elementary fermions.  They
can have a rich and exciting phenomenology at colliders, although we
will not consider this issue in this work.

The Yukawa couplings of the light fermions should be very well
described by the analytical approximations, at least when both LH and
RH mixing angles are small, as we are assuming.  In particular, all of
them can be expected to deviate from the SM expectation by the same
order as the couplings of the third generation, reflecting the
``universal" character of the leading order deviations found in
composite Higgs scenarios (those parametrized by the $F_i$ functions
of Table~\ref{table-F}).

\subsection{Higgs Production and Decay}

Based on the above observations, we can write simple analytical
expressions for the Higgs branching fractions and production rates
that allow us to understand the qualitative (and often quantitative)
behavior.  However, for the numerical computations in the scan we will
not perform any such approximations, as already mentioned.

For the tree-level Higgs decays, we have
\bea
\Gamma(h \to b\bar{b},\tau\tau) &\approx& \Gamma_{\rm SM}(h \to b\bar{b}, \tau\tau) \times r^2_{b}(\epsilon)~, \\ [0.4em]
\Gamma(h \to c\bar{c}) &\approx& \Gamma_{\rm SM}(h \to c\bar{c}) \times r^2_{c}(\epsilon)~, \\ [0.4em]
\Gamma(h \to WW, ZZ) &\approx& \Gamma_{\rm SM}(h \to WW, ZZ) \times r^2_{V}(\epsilon)~,
\eea
where $\Gamma_{\rm SM}(h \to i)$ is the SM Higgs partial decay width
in the $i$-th channel.  We have assumed here that the leptons (in
particular the $\tau$) are in the same $SO(5)$ representations as the
bottom quark.  Similarly, all up-type quarks (in particular, charm and
top) will be assumed to belong to the same $SO(5)$ representation,
hence $r_{c}(\epsilon) = r_{t}(\epsilon)$, which can be read from
Table.~\ref{table-F} for the different models.\footnote{If different
generations are assigned to different $SO(5)$ representations it is
straightforward to generalize our expressions by simply computing the
corresponding $F_\psi(\epsilon)$ from Eq.~(\ref{BasicRelation}),
although it may happen that this function has additional dependence on
other microscopic parameters.} 

For the loop-level Higgs decays, we write
\bea
\frac{\Gamma(h \to gg)}{\Gamma_{\rm SM}(h \to gg)} &\approx& \frac{| r_t(\epsilon) \, A_{1/2}(m^2_h / 4m^2_t) + r_b(\epsilon) \, A_{1/2}(m^2_h / 4m^2_b) |^2}{| A_{1/2}(m^2_h / 4m^2_t) + A_{1/2}(m^2_h / 4m^2_b) |^2}~, 
\label{Gamma2g} \\ [0.4em]
\frac{\Gamma(h \to \gamma\gamma)}{\Gamma_{\rm SM}(h \to \gamma\gamma)} &\approx& \frac{| r_V(\epsilon) \, A_{1}(\frac{m^2_h}{4m^2_W}) + N_c Q_t^2 \, r_t(\epsilon) \, A_{1/2}(\frac{m^2_h}{4m^2_t}) + N_c Q_b^2 \, r_b(\epsilon) \, A_{1/2}(\frac{m^2_h}{4m^2_b}) |^2}{| A_{1}(m^2_h / 4m^2_W) + N_c Q_t^2 A_{1/2}(m^2_h / 4m^2_t) + N_c Q_b^2 A_{1/2}(m^2_h / 4m^2_b) |^2}~,
\label{Gamma2gamma}
\eea
where $A_{1/2}(\tau)$ and $A_1(\tau)$ are the well-known loop
functions (see App.~\ref{app:loops}), $N_c = 3$ is the number of
colors and $Q_t = 2/3$, $Q_b = -1/3$ are the top and bottom quark
electric charges, respectively.  Note that here we have formally
included only the effects of the zero-modes, since in the limit where
Eq.~(\ref{BasicApproximation}) holds, the contribution of the
associated towers of heavy resonances becomes negligible.  However, to
the extent that $A_{1/2}(\frac{m^2_h}{4m^2_t}) \approx 4/3$ (its
asymptotic value for $4m^2_t \gg m_h^2$), and given the sum rule
Eq.~(\ref{BasicRelation}), the above set of approximations effectively
include the effects of the full top tower.  For the bottom quark
contribution, the situation is different since $|A_{1/2}(m^2_h /
4m^2_b)| \approx 1/16 \ll 1$ for $m_h \approx 125~{\rm GeV}$ and $m_b
\approx 2.7~{\rm GeV}$.  In addition, in some cases (as in the
MCHM$_{10-5-10}$), the contribution of the heavy towers can be as
large as 10\% of the sum in Eq.~(\ref{BasicRelation}).  As a result,
the contribution of the heavy $Q =-1/3$ states to the above
loop-induced processes can be of the same order as the actual
contribution of the bottom quark, since although $y_b / m_b$ still
dominates the sum in Eq.~(\ref{BasicRelation}), it has to be
multiplied by the small $A_{1/2}(m^2_h / 4m^2_b)$ for the physical
processes.  Given that the contribution of the bottom-like resonances
is not included in Eqs.~(\ref{Gamma2g}) and (\ref{Gamma2gamma}), our
approximation could carry an uncertainty of the same order as the
bottom contribution, which can be as large as 10\%.  However, for most
models, the approximation is significantly better.

\begin{figure}[t] 
\centering
\includegraphics[width=0.48\textwidth]{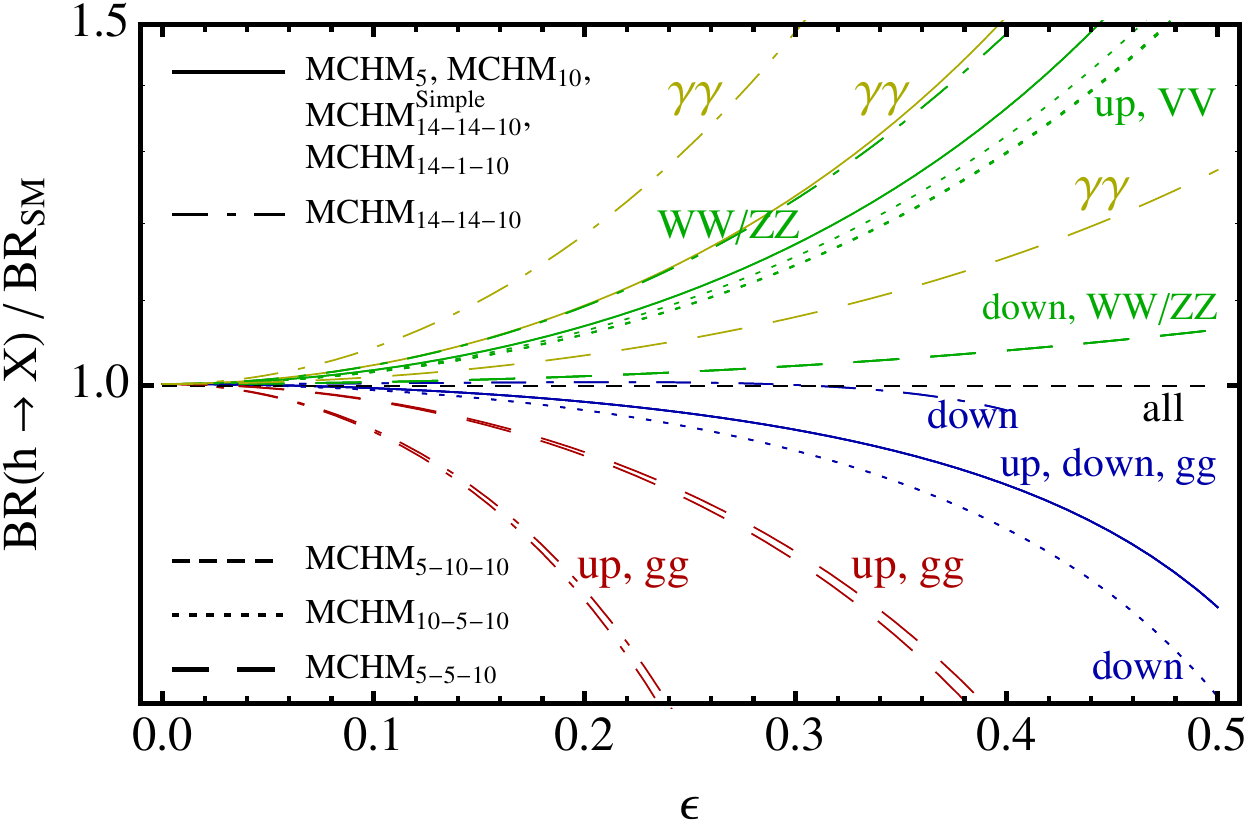}
\hspace{3mm}
\includegraphics[width=0.48\textwidth]{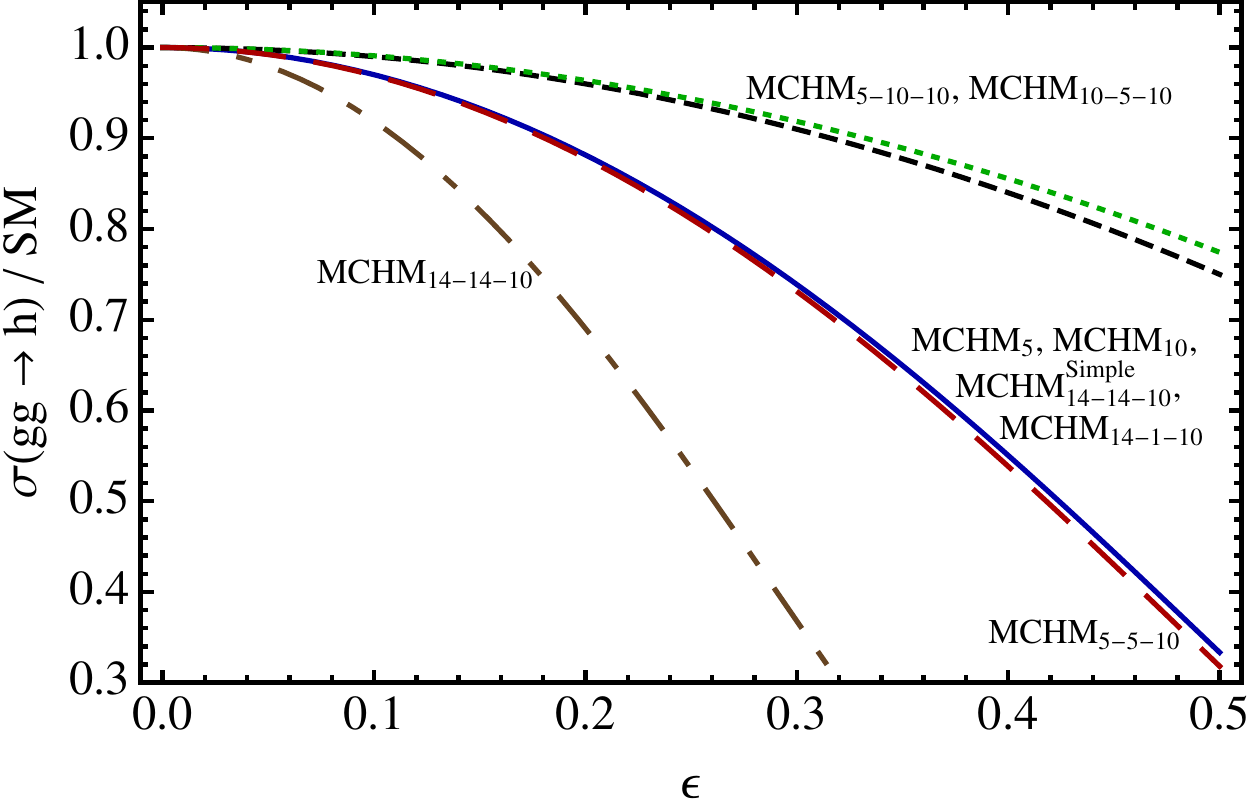}
\caption{Left panel: Branching fractions (normalized to the SM) into
fermions and gauge bosons for several models following from the
approximation in Eq.~(\ref{BasicApproximation}).  Here $VV = WW, ZZ,
\gamma\gamma, gg$.  The color coding of the lines matches the color
coding of the closest legend.  Right panel: gluon fusion production
cross section (normalized to the SM) in those models.  The vector
boson fusion (VBF) cross section coincides with the curve marked as
``MCHM$_{5-10-10}$, MCHM$_{10-5-10}$".}
\label{BRsandProd}
\end{figure}
In the left panel of Fig.~\ref{BRsandProd}, we show the Higgs
branching fractions into fermion and gauge boson pairs in the
MCHM$_{5}$, MCHM$_{10}$, MCHM$^{\rm Simple}_{14-14-10}$,
MCHM$_{14-1-10}$ (solid lines), MCHM$_{14-14-10}$ (dash-dotted lines),
MCHM$_{5-10-10}$ (short dashed lines), MCHM$_{10-5-10}$ (dotted
lines), and MCHM$_{5-5-10}$ (long dashed lines).  We see that in some
cases the BR's are enhanced with respect to the SM while in others
they are suppressed.  One should notice that all partial decay widths
always present a suppression, in particular for the $b\bar{b}$ decay
channel.  As a result the total decay width is suppressed, and the
BR's in some channels can end up being enhanced due to the smaller
denominator.  In contrast, the Higgs production cross sections are
always suppressed with respect to the SM, as shown in the right panel
of Fig.~\ref{BRsandProd} for the gluon fusion Higgs production cross
section, normalized to the SM. We also note that the VBF production
cross section coincides with the upper curve in this plot.

\begin{figure}[t] 
\centering
\includegraphics[width=0.48\textwidth]{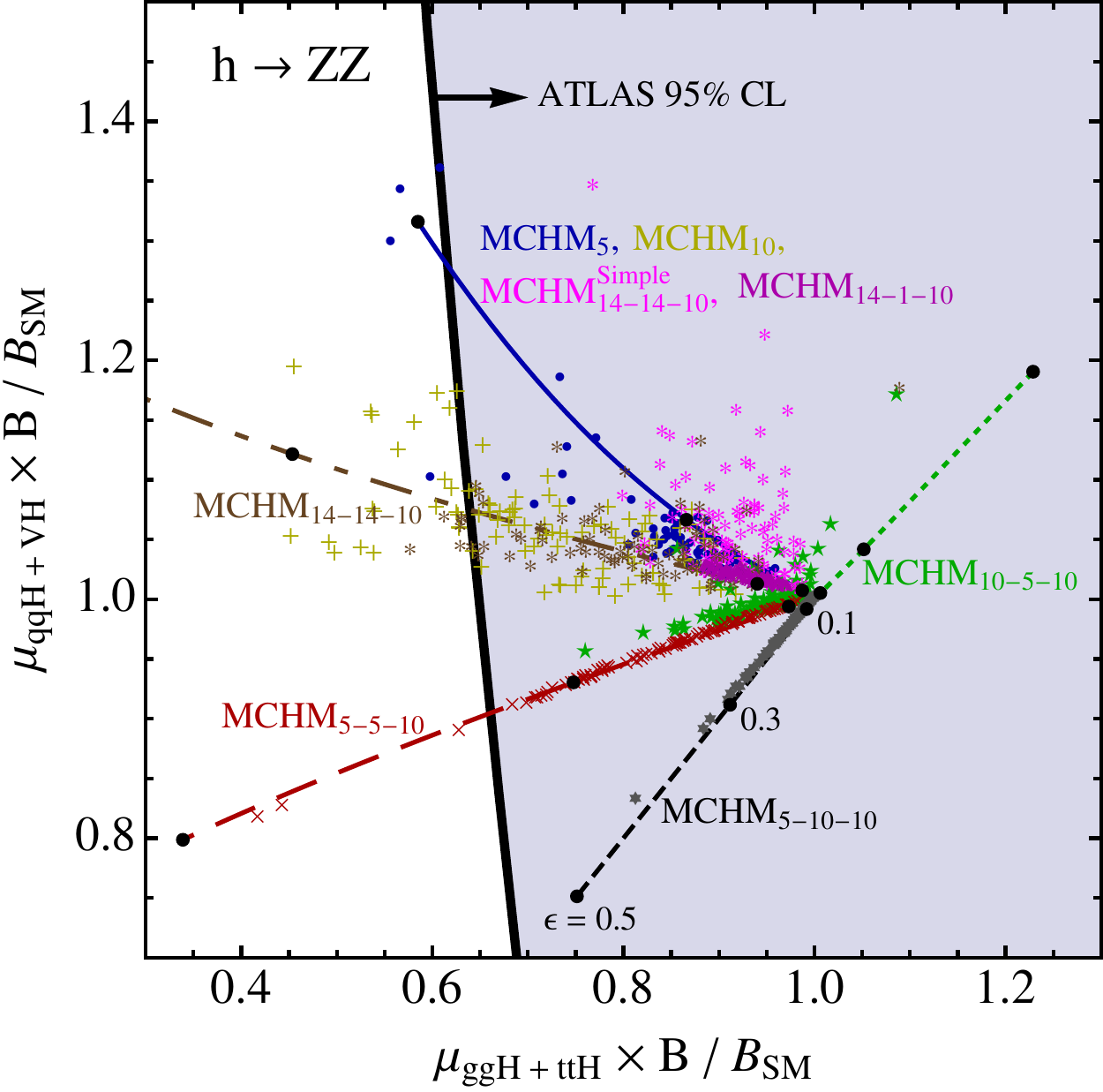}
\hspace{3mm}
\includegraphics[width=0.48\textwidth]{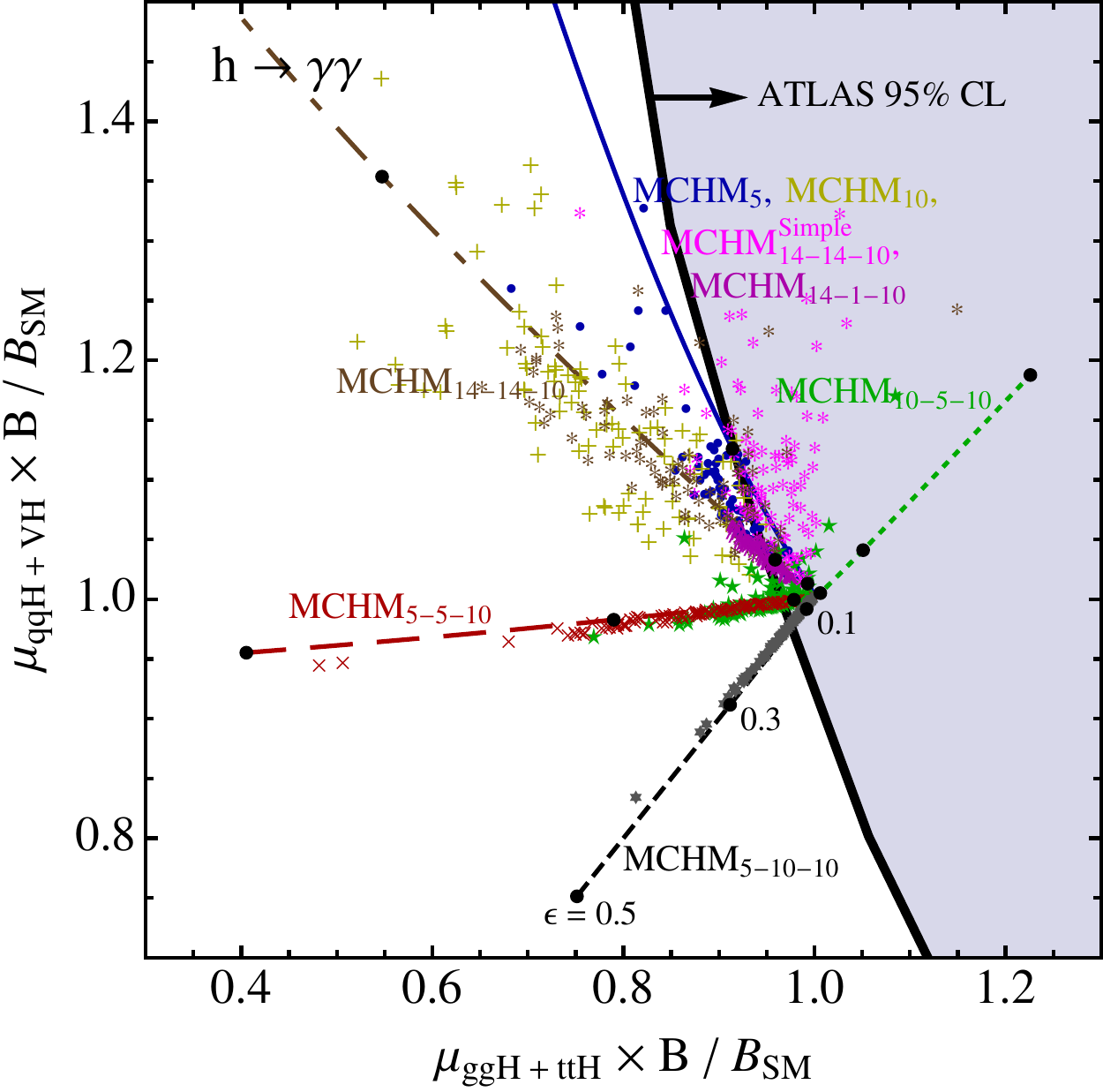}
\caption{Left panel: Rates in the $h \to ZZ$ decay channel separated
according to production mode: gluon fusion ($+ t\bar{t}h$) versus VBF
($+h W/Z$).  The larger black dots indicate the positions of $\epsilon
= 0.1, 0.3, 0.5$.  Right panel: Same for the $h \to \gamma\gamma$
channel.  The solid curves correspond to the analytical approximation
discussed in the main text, while the points correspond to a random
scan that reproduces $m_h \sim 125~{\rm GeV}$, $m_t \sim 160~{\rm
GeV}$ and $m_h \sim 4~{\rm GeV}$.  The shaded region corresponds to
the current 95\% CL curve by ATLAS. The CMS 95\% CL region would cover
the full area of the figure.  The \textit{production} signal strengths
are defined as $\mu_i = \sigma^{\rm Model}(i) / \sigma^{\rm SM}(i)$.
The production cross sections used correspond to the 8~TeV run of the
LHC.}
\label{Exp}
\end{figure}
Consequently, the total cross sections in given channels can be
enhanced or suppressed with respect to the SM, depending on how these
opposing effects play out.  We illustrate this in Fig.~\ref{Exp} for
the $ZZ$ (left panel) and $\gamma\gamma$ (right panel) decay modes,
separating the gluon fusion ($+ t\bar{t} h$) production from VBF ($+
hW/Z$), as done by the ATLAS and CMS collaborations~\cite{gg-VBF}.
The continuous lines correspond to the expectation based on the above
analytical approximation.  We have superimposed the exact predictions
for the scan in the models we consider.  We see that the approximation
tracks well the actual analytical predictions for all models (up to
some dispersion due to the effect of the bottom sector explained
above), except for the MCHM$_{10-5-10}$ on which we comment further
below.  One can understand the behavior of these curves from
Fig.~\ref{BRsandProd}.  For instance, for the MCHM$_{5-10-10}$, since
\textit{all} channels (gauge, down-type and up-type) are suppressed by
exactly the same $r(\epsilon)$, the BR's remain exactly as in the SM,
while the production in all modes is suppressed identically.  Thus,
the curve points at a $45^{\circ}$ angle towards the left-down, as
$\epsilon = \sin(v/f)$ increases and the deviations from the SM
increase.  The MCHM$_{5-5-10}$ shows a very mild enhancement in the
$ZZ$ and $\gamma\gamma$ BR's (see left panel of
Fig.~\ref{BRsandProd}), which is not enough to compensate the
suppression in production.  Since the latter is more significant in
gluon fusion than in VBF, the curve in Fig.~\ref{Exp} points to the
left-down but closer to the horizontal than for the MCHM$_{5-10-10}$.
For the MCHM$_{5}$, MCHM$_{10}$, MCHM$_{14-14-10}$, MCHM$^{\rm
Simple}_{14-14-10}$ and MCHM$_{14-1-10}$, the left panel of
Fig.~\ref{BRsandProd} shows a stronger enhancement in both ${\rm BR}(h
\to ZZ)$ and ${\rm BR}(h \to \gamma\gamma)$, which is sufficient to
compensate the suppression in the VBF production but not enough to
compensate the significant suppression in gluon fusion (see right
panel of Fig.~\ref{BRsandProd}).  As a result, the analytical
prediction curves to the left-up.  Note, however, that the scanned
points for the MCHM$^{\rm Simple}_{14-14-10}$ show a more pronounced
tendency to compensate the suppression in gluon fusion by the
enhancement in the branching fractions than the naive analytical
expectation.  This can be traced to the systematic (albeit small)
deviations exhibited in Fig.~\ref{YukApprox} for the top and bottom
Yukawa couplings.  Finally, we see that the analytical prediction for
the MCHM$_{10-5-10}$ does \textit{not} reproduce the qualitative
behavior of the scan.  While a line at $45^{\circ}$ to the right-up is
expected (from Fig.~\ref{BRsandProd} one can see that the enhancement
in BR's dominates over the suppression in production in all the
modes), most of the points actually present a suppression with respect
to the SM. This can be traced back to our previous comments in regards
to this model: the analytical approximation systematically
overestimates the suppression in the $b\bar{b}$ channel [hence
overestimates the enhancement in ${\rm BR}(h \to ZZ)$ and ${\rm BR}(h
\to \gamma\gamma)$], while it systematically underestimates the
suppression in the top Yukawa coupling, which translates into an
overestimate of the gluon fusion Higgs production rate.  These ${\cal
O}(10\%)$ errors are sufficient within this model to change the
qualitative behavior.  The VBF production is still well described by
the analytic approximation, as is for all the other models, since the
gauge resonances are always heavy.

It is interesting that the different fermionic representations lead to
a different behavior in the plane of Fig.~\ref{Exp}, so that a precise
measurement of these rates could be used to distinguish between
different scenarios (although there could still remain a degeneracy
between the MCHM$_5$, MCHM$_{10}$, MCHM$^{\rm Simple}_{14-14-10}$ and
MCHM$_{14-1-10}$, which in fact could be confused with the more
general MCHM$_{14-14-10}$).  We also show the current 95\%
C.L.~ellipse from the ATLAS analysis~\cite{gg-VBF}, and indicate the
position along the solid line in each model that corresponds to
$\epsilon = 0.1, 0.3, 0.5$.  We see that the experimental
uncertainties still allow for relatively large values of $\epsilon$.
The 95\% C.L.~ellipse from the CMS analysis would fill the region
shown, so we do not indicate it.

The ATLAS and CMS collaborations have measured other properties of the
125~GeV resonance.  For instance, by taking channel by channel ratios
of the $ggH + ttH$ and $qqH + VH$ production modes, and performing a
fit to the data, they can set a bound on $\mu_{qqH + VH} / \mu_{ggH +
ttH}$.  This analysis only assumes that the same boson $H$ is
responsible for all observed Higgs-like signals and that the
separation of gluon-fusion like events and VBF-like events, based on
the event kinematics, is valid.  For instance, the ATLAS collaboration
sets a bound of $\mu_{qqH + VH} / \mu_{ggH + ttH} =
1.2^{+0.7}_{-0.5}$~\cite{Atlas:couplings}.
The models in our scan have $1 \lesssim \mu_{qqH + VH} / \mu_{ggH +
ttH} \lesssim 1.5$, so that they are not yet probed by these analyses.
However, if a ratio below one was established it would disfavor the
pNGB Higgs scenarios based on the lowest dimensional representation of
SO(5).  This is a manifestation of the generally important suppression
in the gluon fusion process w.r.t.~the SM. ATLAS also sets bounds on
the Higgs production by gluon fusion alone, in terms of the rescaling
factor $\kappa_g$.  However, the analysis assumes that all the BR's
are as in the SM and therefore does not apply to the present case.

From the LHC data one can also derive bounds on ratios of branching
ratios, e.g. on $\rho_{\gamma\gamma/ZZ} = [{\rm BR}(\gamma\gamma) /
{\rm BR}(\gamma\gamma)_{\rm SM}] / [{\rm BR}(ZZ) / {\rm BR}(ZZ)_{\rm
SM}]$, etc.  ATLAS finds $\rho_{\gamma\gamma/ZZ} =
1.1^{+0.4}_{-0.3}$~\cite{Atlas:couplings}.  Our scans have $1 \lesssim
\rho_{\gamma\gamma/ZZ} \lesssim 1.1$, so that they are not yet probed
in such measurements.  Similarly, due to the custodial symmetry, we
have $\rho_{WW/ZZ} \approx 1$, and it would be very challenging to
differentiate it from the SM at the LHC; a significant deviation from
the custodial limit would disfavor both the SM and the pNGB scenarios
we have studied.

\begin{figure}[t]
\centering
\includegraphics[width=0.48\textwidth]{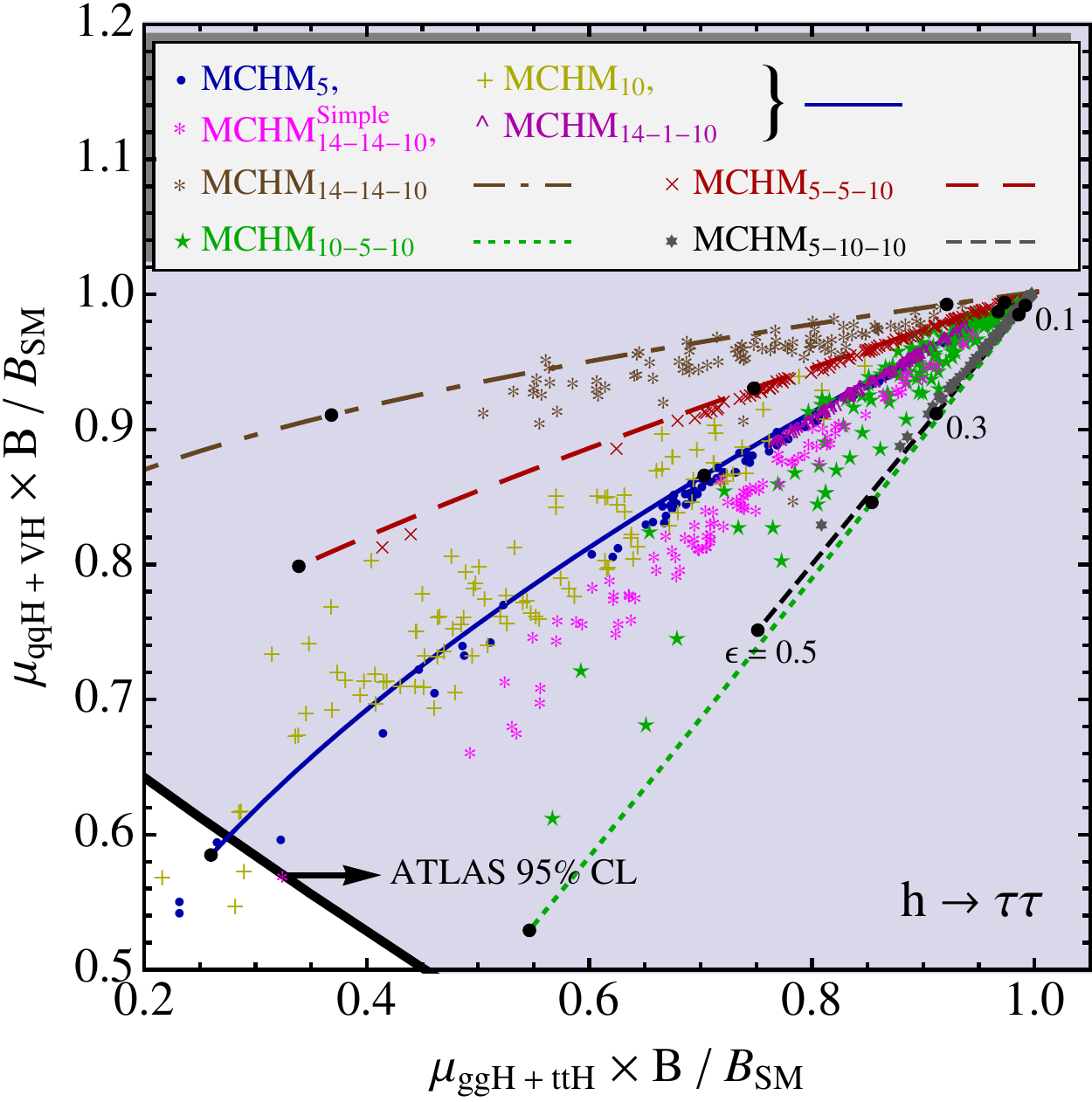}
\hspace{4mm}
\includegraphics[width=0.48\textwidth]{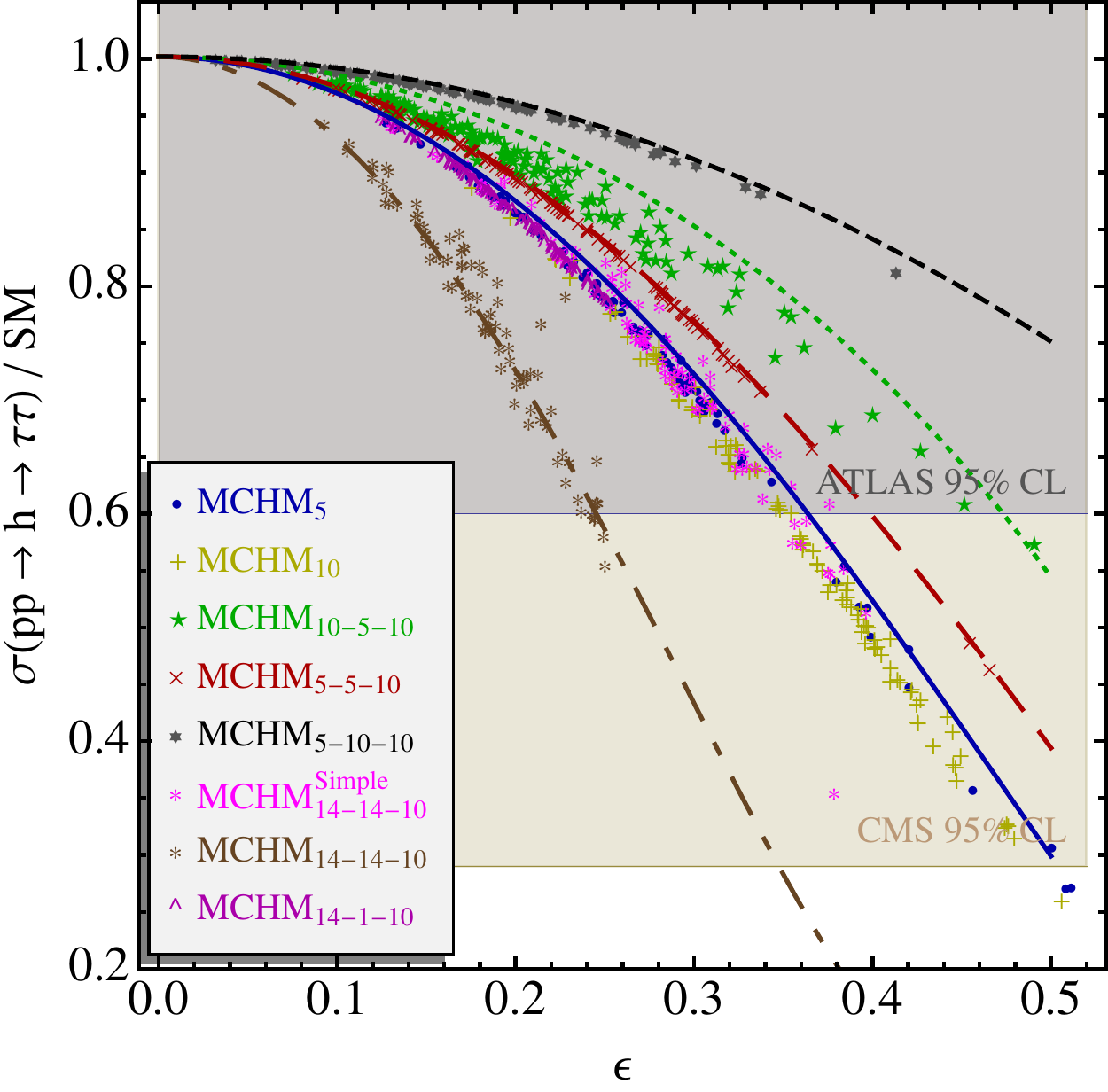}
\caption{Left panel: Similar to Fig.~\ref{Exp}, but for the $h \to
\tau\tau$ channel.  Right panel: we show the total rate
(i.e.~inclusive production) in the $\tau\tau$ channel, normalized to
the SM, versus $\epsilon$.  The horizontal bands correspond to the
95\%~C.L. limit set by the ATLAS~\cite{Atlas:taus} and
CMS~\cite{CMS:taus} collaborations.}
\label{fig:LHC-tautau}
\end{figure}

Apart from the indirect sensitivity to the top quark via the loop
processes above, the fermionic channels, in particular $h \to
\tau\tau$ are starting to be measured with interesting
precision~\cite{Atlas:taus,CMS:taus} for the present work, although
the uncertainties are still sufficiently large to be consistent with
the great majority of our parameter point sample.  In the left panel
of Fig.~\ref{fig:LHC-tautau} we show the expectations for this
channel, discriminating between the Higgs production by gluon fusion
($+ttH$) and VBF ($+VH$), together with the 95\%~C.L. region from
ATLAS. In the right panel we show the signal strength for the
inclusive $h \to \tau\tau$ production as a function of $\epsilon$.
The horizontal bands correspond to the 95\%~C.L. regions from
ATLAS~\cite{Atlas:taus} and CMS~\cite{CMS:taus}.  We note that under
our assumptions, the $\tau\tau$ channel is always suppressed
w.r.t.~the SM. However, one should remember that one may be able to
consider different representations for the $\tau$ sector, without
affecting the properties of the Higgs potential.  Hence, establishing
an enhancement in the $\tau\tau$ channel over the SM would be in
conflict with our assumptions, but we cannot claim that it would rule
out the general framework.

In contrast, in models with a minimal content of composite fermion
multiplets, one expects a robust suppression w.r.t. the SM in the $h
\to b \bar{b}$ decay mode, so that this would be an interesting
channel to probe the scenario.  We find a suppression of $10-20\%$ for
$\epsilon = 0.3$ and $20-40\%$ for $\epsilon = 0.5$, with smaller
dispersion between different models than in the $\tau\tau$ channel.
This is because at the LHC one must consider $pp \to h + X \to b
\bar{b} + X$ in order to be able to discriminate against the large QCD
background, so that only $VBF$ + $VH$ + $ttH$ contribute, but not
$ggH$ which is most sensitive to the new fermionic resonances that
distinguish between different models.  Unfortunately, at the LHC the
precision may not be sufficient to provide a clear test, but its high
luminosity phase or a linear collider could set useful bounds.

\subsection{$h \to Z\gamma$}

We turn now to the last decay channel we consider: $h \to Z\gamma$,
which has not yet been observed, but could be seen in the near future.
The decay of a pNGB Higgs to $Z\gamma$ has received considerable
attention recently.  Ref.~\cite{Azatov:2013ura} has shown that there
can be large corrections to this decay, while being simultaneously
compatible with precision EW measurements, thus providing a very
interesting test.  In order to obtain a large effect in this decay the
composite sector itself must break the $P_{LR}$ symmetry, otherwise
the only source of $P_{LR}$ breaking is the interaction between the
elementary and composite fields, and the effect is
suppressed~\cite{Elias-Miro:2013mua}.  We have not considered breaking
of $P_{LR}$ by the composite sector in our work, so that we expect
small corrections in the $h \to Z\gamma$ channel.  We have computed
the corrections to this rate in the models presented in the previous
sections.  Below we discuss the main features of this decay and show
our results.

In the SM the interaction $hZ\gamma$ is a radiative effect, generated
at 1-loop by virtual $W$'s and fermions.  Similar to
$h\gamma\gamma$, the bosonic and fermionic contributions have opposite
sign.  The first one dominates over the second one by a factor $\sim
10$, and the fermionic loop is dominated by the top contribution.  In
the MCHM one can distinguish the corrections from the new particles in
the loop from those arising from the modified couplings between the
Higgs and the SM gauge and fermion fields, as was the case for the
$h\gamma\gamma$ process.  However, unlike in the $h \to \gamma\gamma$
diagrams, there can be two different particle species running in the
loop, since only one of the external particles is a gauge field of an
unbroken symmetry.  Therefore, in theories with extra $W$'s, besides
the loop with a single heavy field there are 1-loop effects
involving two different virtual states.  We will refer to these
contributions as ``diagonal" and ``non-diagonal", respectively.
Similarly, in theories with new fermions there are 1-loop effects
involving a single new fermion as well as effects involving
propagators of two different fermion species.  We will clarify below
which diagrams give the leading contributions.

As in the SM, in the models we are considering there are no tree-level
contributions to the $h \to Z\gamma$ process, so we focus on the
1-loop effects, starting with those due to bosonic fields.  Each
diagonal contribution is suppressed by a factor $(m_W/m_{W_n})^2 \sim
{\cal O}(10^{-3})$.  Although there are several charged vectors, whose 
contributions add up, we find that the total effect is less
than 1\% of that of the $W$ gauge boson in the SM. Next we consider
the corrections from a loop with a SM-$W$ and a heavy charged vector.
The product of the non-diagonal couplings $ZWW_n$ and $hWW_n$ are
suppressed by a factor $\lesssim{\cal O}(10^{-2})$ compared with the
SM coupling, thus they can be neglected as well.  For the non-diagonal
contributions involving heavy fields the product of the couplings
$ZW_mW_n$ and $hW_mW_n$ can be of the same order as in SM. However, as
in the diagonal contribution, in this case there is also an extra
factor $(m_W/m_{W_n})^2\sim{\cal O}(10^{-3})$.  Therefore, the leading
correction to $hZ\gamma$ mediated by loops of vector bosons is
captured by the correction to the couplings $hW^+W^-$ and $ZW^+W^-$.
The correction to the first one can be approximated by $F_2(\epsilon)
= \sqrt{1-\epsilon^2}$, whereas the correction to $ZW^+W^-$ is very
small.  Thus, one can expect the bosonic 1-loop
correction to $c_{hZ\gamma}$ [see comments after Eq.~(\ref{OpsDim6})]
to be modulated by $F_2(\epsilon)$, leading to a suppression in the
amplitude compared with the SM.

\begin{figure}[t]
\centering
\includegraphics[width=0.48\textwidth]{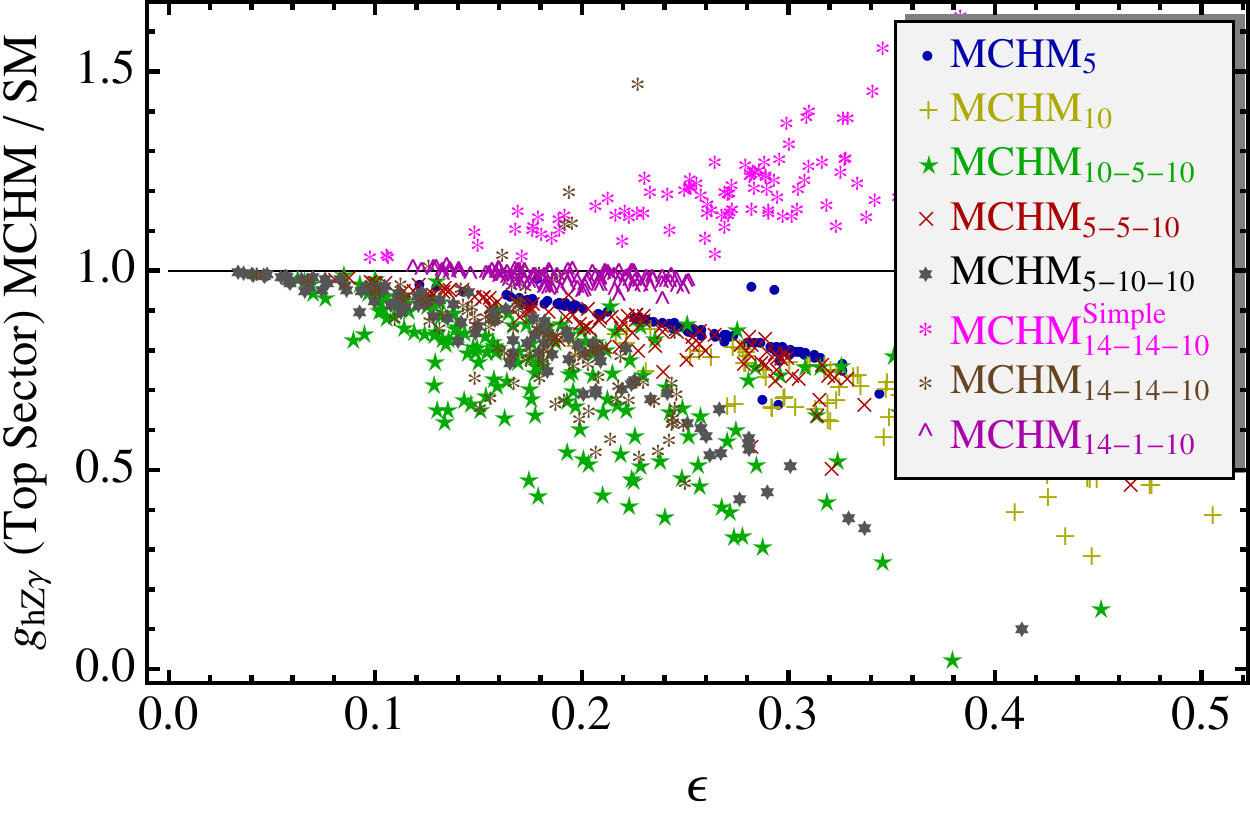}
\hspace{3mm}
\includegraphics[width=0.48\textwidth]{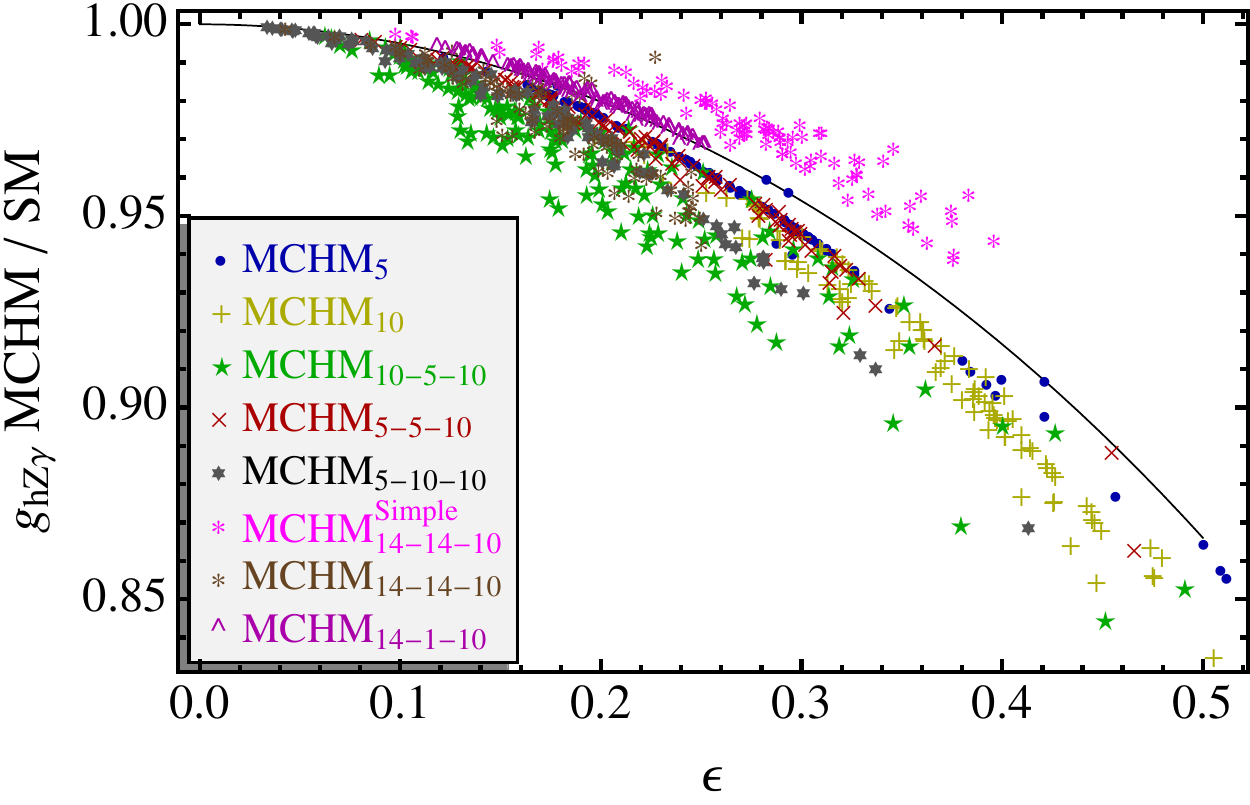}
\caption{Left panel: the amplitude for $h \to Z\gamma$ due to the top
sector in all the models, normalized to the top-mediated amplitude in
the SM. Right panel: the full amplitude (absolute value), arising from
vector bosons and fermions in the models of Sec.~\ref{sec:models},
normalized to the $hZ\gamma$ amplitude in the SM. The continuous line
corresponds to the SM-W loop, with its modified coupling to the Higgs
as encoded in $F_2(\epsilon)$.}
\label{fig:Wt-hZgamma}
\end{figure}

The correction from the fermionic sector is dominated by the top quark
and its partners.  The resonances associated to the light SM fermions
decouple and do not contribute.  This can be understood from the fact
that $hZ\gamma$ requires breaking of $P_{LR}$, and in the present
models that breaking arises only from the mixing between the two
sectors of the theory.  Since we are assuming that the light fermions
have small mixing for both chiralities, the explicit $P_{LR}$ breaking
is suppressed by these small mixings.  The effect from the top
partners can have different signs for different representations.  In
the left panel of Fig.~\ref{fig:Wt-hZgamma}.  we show the corrections
to the amplitude coming from the top sector of all the models,
normalized to the top contribution in the SM. We have included all the
diagonal and non-diagonal contributions.  The corrections to the SM
top result can be of order 50\%, or even larger for $\epsilon\sim 0.5$
and for most of the models there is a suppression.  However one should
remember that the bosonic contribution is one order of magnitude
larger that the fermionic one.  

In the right panel of Fig.~\ref{fig:Wt-hZgamma}, we show the total
amplitude in the MCHM models normalized to the SM, where we have used
the full diagonalization of the mass matrices and couplings to take
into account all the fermionic contributions, the diagonal spin-1
contributions, and the (small) modification of the $ZW^+W^-$ coupling.
However, we do not include the non-diagonal gauge contributions (which
have been argued to be negligible above).  See App.~\ref{app:loops}
for further details of this computation.

\begin{figure}[t]
\centering
\includegraphics[width=0.48\textwidth]{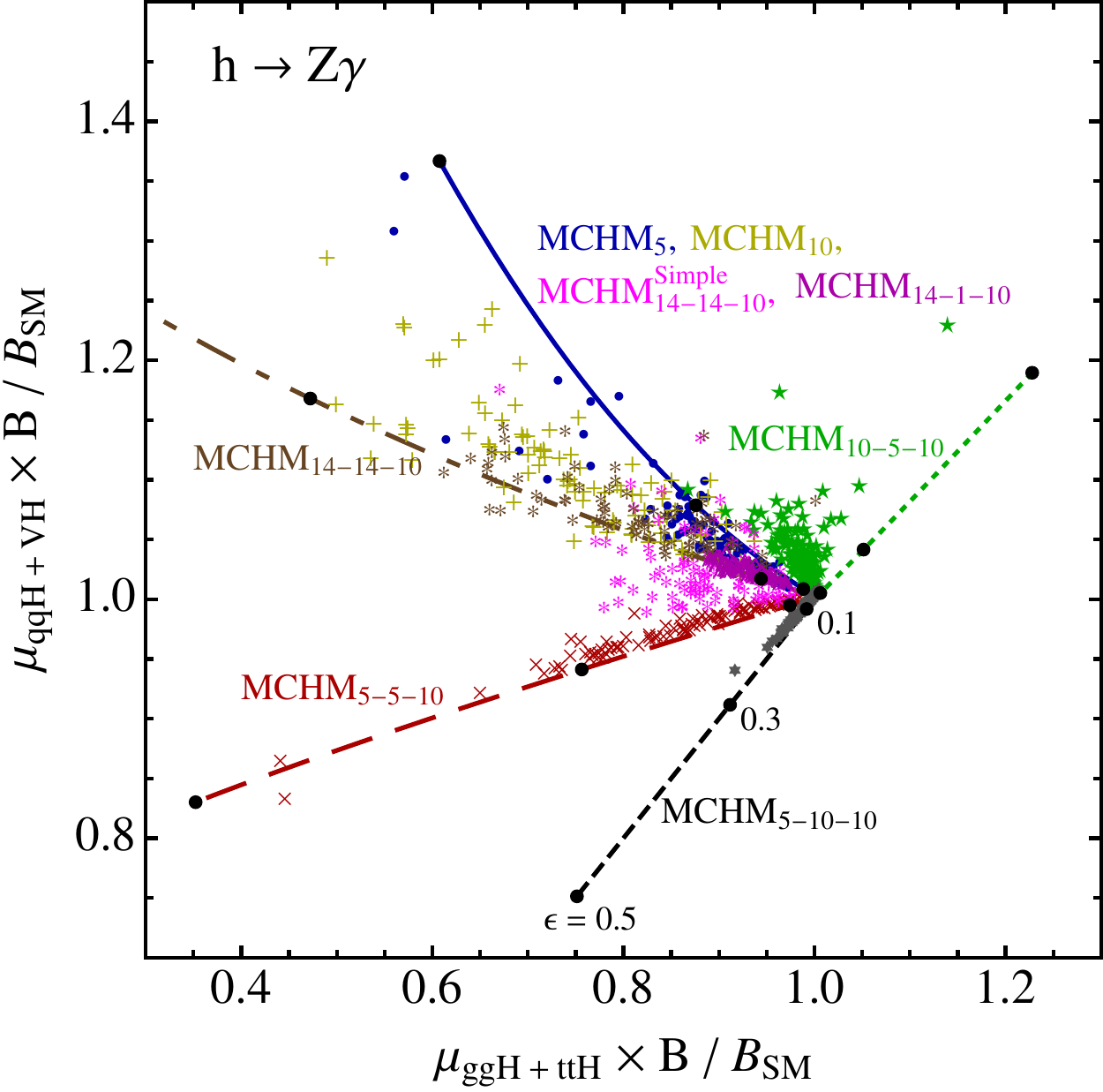}
\hspace{4mm}
\includegraphics[width=0.48\textwidth]{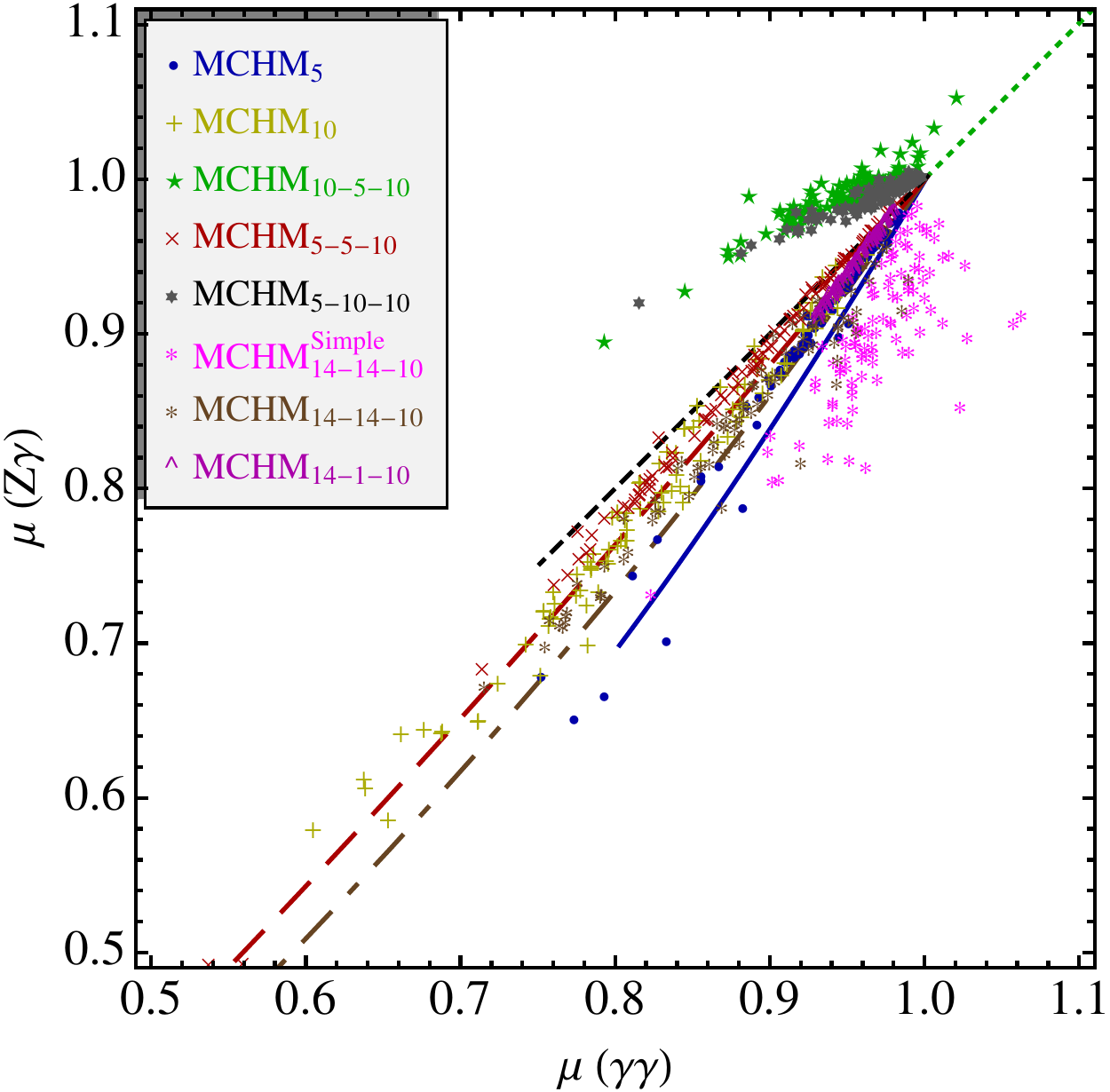}
\caption{Left panel: Similar to Fig.~\ref{Exp}, but for the $h \to
Z\gamma$ channel.  Right panel: we show the total signal strength
$\mu(i) = \sigma(pp \to h \to i) / SM$ (i.e.~inclusive production) in
the $Z\gamma$ channel versus the $\gamma\gamma$ channel, showing a
high degree of correlation.  The larger deviations from $(1,1)$
correspond to larger values of $\epsilon$.}
\label{fig:LHC-hZgamma}
\end{figure}

Since for most of the models and regions of parameter space the
leading order effect is captured by the lightest states running in the
loop, either bosons or fermions, the corrections to $c_{hZ\gamma}$ can
be approximated by the corrections to the Higgs couplings with $W$ and
$t$.  The left panel of Fig.~\ref{fig:LHC-hZgamma} shows that this
approximation works rather well for the models we are considering.
The deviations arise mainly from the diagonal and non-diagonal
contributions of the top partners.  In the right panel we exhibit the
correlation between the rates into $Z\gamma$ and $\gamma\gamma$.  We
see that this correlation is slightly different between the
MCHM$_{10-5-10}$ and MCHM$_{5-10-10}$ on the one hand, and the other
models on the other, which could allow for a distinction if sufficient
precision is achieved, depending on the size of the deviations from
the SM. We note that for the MCHM$_{5-10-10}$ there is a very good
agreement between the analytical approximation and the full numerical
result in $\gamma\gamma$.  However, the top sector gives contributions
of order 5-10\% to $Z\gamma$, as can be seen in the right panel of
Fig.~\ref{fig:Wt-hZgamma}.  Those corrections lead to the small
disagreement between the analytical approximation and the full
numerical result for the MCHM$_{5-10-10}$ seen in the right panel of
Fig.~\ref{fig:LHC-hZgamma} (a similar effect but in the opposite
direction is present for the MCHM$_{14-14-10}^{\rm simple}$).

\section{Tuning in the MCHM}
\label{sec:tuning}

In this section we comment on the degree of fine-tuning associated
with the phenomenologically viable points found above.  Consistency
with the EW precisions tests (EWPT) in these models, mainly the
$S$-parameter and the $Zb_L\bar b_L$ coupling, require
$\epsilon\lesssim 0.3$~\cite{Agashe:2004rs}.\footnote{Although we do
not perform a detailed analysis of the EWPT on all the models we
consider, we recall that the presence of light fermionic resonances
can play an important role in opening up the viable region of
parameter space, as studied in~\cite{Anastasiou:2009rv}.} However the
Higgs potential generically leads to no EWSB, $\epsilon=0$, or to
maximal EWSB, $\epsilon=1$.\footnote{For $v=f_h$, besides the problems
with EWPT, many models lead to massless SM fermions, as can be seen
from the cancellation of the $LR$ correlator $M_\psi$.  This is a
consequence of the restoration of an accidental chiral
symmetry~\cite{Contino:2006qr}.} A careful analysis of the structure
of the Higgs potential shows that the MCHM requires some tuning in the
parameter space of the theory to produce $\epsilon\lesssim 0.5$, and
the amount of tuning depends on the fermion
embedding~\cite{Agashe:2004rs,Panico:2012uw}.  Besides these
conditions, the Higgs potential must also lead to a light Higgs.
Since the top contribution to the 1-loop Higgs potential is cut off by
the fermionic resonances mixing with the top, a light Higgs prefers
light top partners.  Ref.~\cite{Contino:2006qr} has shown the
correlation between $m_h$ and the mass of the lightest resonance for
MCHM$_5$ and MCHM$_{10}$.  Ref.~\cite{Panico:2012uw} has also
discussed the impact of light fermions in the tuning of the MCHM,
arriving to similar results.  In our setup, similar to models in a
slice of AdS$_5$, large compositeness of the SM fermions automatically
lead to light custodians that can alleviate the tuning (see the
discussion in~\ref{sec-num-scan}).  Below we show our results for the
tuning of the models presented in the previous sections.

Following Refs.~\cite{Barbieri:1987fn, Anderson:1994dz, Panico:2012uw}
we use the sensitivity parameter
\begin{equation}
\label{eq-tuning}
\Delta={\rm max}_i \left|\frac{\partial \log m_Z|_{\rm phys}}{\partial
\log x_i}\right|
\end{equation}
as a measure of fine-tuning.  Here $x_i$ are the parameters of the
effective theory and $m_Z$, as given in Eq.~(\ref{eq-vSM}), depends
explicitly on $f_h$ and $\epsilon$, with $\epsilon$ a function of all
the parameters of the theory.  By $m_Z|_{\rm phys}$ we mean that we
have selected a region of parameters of the theory that leads to the
observed Higgs and SM masses.  We have followed the procedure of
Ref.~\cite{Panico:2012uw} which has shown that Eq.~(\ref{eq-tuning})
can be rewritten in terms of the Higgs potential, allowing for a
simple calculation of $\Delta$.  As explained at the beginning of
Sec.~\ref{sec:pheno}, we have considered the dependence of the
potential on the following parameters: the mass scale of the composite
resonances $m_\rho$, the decay constant of the pNGB $f_h$, the
composite proto-Yukawa couplings, the masses mixing composite fermions
$m_y$, the fermion mixing angles $s_\psi$ and the ratio of gauge
couplings $\tan\theta$.

We have computed the tuning of the models presented in the previous
sections, evaluating $\Delta$ in those points of the parameter space
that were selected after the random scan, as explained at the
beginning of Sec.~\ref{sec:pheno}.  We find that the gauge
contribution is subdominant, and the tuning is usually dominated by
the top mixings $s_{q}$, $s_{t}$, the Yukawa $y_T$ and the mixing mass
$m_{y_T}$ when present.  Below we comment on the size of the tuning
for the different models and discuss some details about its parameter
dependence.

We find that the MCHM$_5$ and MCHM$_{5-5-10}$ have generically
$\Delta\sim 5-40$, with the sensitivity parameter dominated by
$m_{y_T}$ and sometimes by $s_t$.  The second model shows some regions
of parameter space with $\Delta\sim 100$ as well as some points where
$s_q$ dominates the tuning.  Notice that the MCHM$_{5-5-10}$ has less
freedom, since there is no $m_{y_B}$ and the $b_L$ mixing is
controlled by the same parameter that controls the $t_L$ mixing,
namely $s_q$, whereas for the MCHM$_5$ there are two mixing
parameters, $s_{q^d}$ and $s_{q^u}$.  The MCHM$_{10}$ has $\Delta\sim
5-80$, although there are points with $\Delta\sim 300$.  The larger
tuning of the MCHM$_{10}$ could be related with the Clebsch-Gordan
coefficient $\sqrt{2}$ suppressing $m_t$ in the latter model, that
requires larger mixing and Yukawa coupling.  In this model $\Delta$ is
usually dominated by $s_q$ and sometimes by $s_t$, $y_T$ or $m_{y_T}$.
As explained in the previous sections, MCHM$_{5-10-10}$ and
MCHM$_{10-5-10}$ require a large degree of compositeness of at least
one of the chiralities of the top, leading to the largest tuning of
the models that we have studied with fermions in representations ${\bf
5}$ and ${\bf 10}$.  We find $\Delta\sim 100-1000$, usually dominated
by $s_q$ and sometimes by $s_t$.  MCHM$_{14-14-10}^{\rm simple}$ and
MCHM$_{14-1-10}$ have $\Delta\sim 80-300$, dominated by $s_t$ for the
first model and by $s_q$ for the second one.  The main reason for the
larger tuning of these models compared with MCHM$_5$ and MCHM$_{10}$
is that they generically predict a larger $m_h$~\cite{Panico:2012uw}.
Thus, requiring $m_h\simeq 125$ GeV selects special regions of the
parameter space with non-natural cancellations in the Higgs potential.
On the other hand, for the MCHM$_{14-14-10}$ that has an extra
proto-Yukawa coupling in the top sector, we find $\Delta\sim 10-150$,
with the tuning dominated by $s_t$ and sometimes by $m_{y_t}$ or
$f_h$.

We find that, after applying our selection criteria over the random
scan, the models with larger tuning also show many points in a region
of the parameter space with large composite scale, $f_h\gtrsim 4$ TeV.
In fact, for these models there are some points where the tuning is
dominated by $f_h$.

\section{Conclusions}
\label{sec:conclusions}

We have used a simple two-site realization of the composite Higgs
scenario~\cite{DeCurtis:2011yx} to systematically investigate the
consequences of several fermion representations of the spontaneously
broken symmetry leading to the Higgs as a pNGB. We have restricted
ourselves to the $SO(5) \to SO(4)$ symmetry breaking pattern, which is
denoted here as the ``Minimal Composite Higgs Model", but we have
explored several combination of the lowest-dimensional representations
of SO(5) in the composite fermion sector.  In particular, we have
fully taken into account the dynamically generated Higgs potential,
which receives crucial contributions from the states associated with
the third family, especially the top quark.  We can therefore
consistently incorporate the measured mass of the resonance discovered
at the LHC in 2012, interpreted as a SM-like Higgs boson, and
investigate the restrictions imposed by the experimental information.
We have also taken into account the effects of the bottom quark
sector, which, although subdominant in determining the dynamics of
EWSB, can have a non-negligible effect on the resulting Higgs
phenomenology.  We have assumed that the light families are mostly
elementary, and therefore have a negligible effect on the Higgs
potential.  However, the couplings of a composite Higgs to all
fermions can receive sizeable corrections leading to important
deviations from the SM expectations.  This can be important in the
near future, as decays such as those into a $\tau$ pair are being
measured with better precision~\cite{Atlas:taus,CMS:taus}.

By including the ``first level" of heavy (spin-1 and spin-1/2)
resonances, we can also compute in detail the effects on loop-induced
processes, such as the Higgs production through gluon fusion and the
Higgs decays into $\gamma\gamma$ and $Z\gamma$.  Such processes
consist of two conceptually different, but related parts.  First,
the couplings of the Higgs to the SM fermions are modified w.r.t.~the
SM, and therefore when they run in the loop the corresponding
contribution is different from the SM one.  Second, the heavy
resonances give an additional non-SM contribution to the loop diagrams.
At zeroth-order and in the simplest models, the sum of the two effects
for the dominant contributions (from the top-related states, as well
as from the W-related ones in the case of $\gamma\gamma$ or $Z\gamma$)
results in a ``universal modification" that depends on the microscopic
parameters only through $\epsilon = \sin v/f_h$.  However, we find
that the corrections to this leading order result, in particular those
of the bottom sector, can have a qualitative impact on the Higgs
properties.

Importantly, we find a generic suppression of the gluon fusion process
in all the models we investigated.  This is also the case for the
MCHM$_{14-14-10}$, which presents a richer structure of invariants and
leads in general to a sum rule that has dependence on microscopic
parameters beyond $\epsilon$.  Although a priori there exists the
potential for finding regions of parameter space with an enhanced
gluon fusion Higgs production cross section~\cite{Azatov:2011qy}, we
find that all the phenomenologically viable points exhibit a rather
significant suppression instead.

Due to the generic suppression of the various decay widths,
in particular $\Gamma(h \to \bar b b)$ which dominates the total Higgs
decay width, one can often find branching fractions that are larger
than those in the SM. The experimental rates then result from
competing effects between production and decay, and can present
enhancements or suppressions in given channels, depending on the model
under consideration.  This offers an interesting handle --were a
robust deviation from the SM to be established-- to get indirect
information about the composite fermion representations, which would
constrain the nature of the underlying strongly interacting theory.

Another interesting decay channel is $h\to Z \gamma$.  We have shown
that the deviations are small and dominated by the corrections from
loops of SM weak bosons, as expected if the $P_{LR}$ symmetry is not
broken by the composite sector~\cite{Elias-Miro:2013mua}.  Moreover,
the contributions from the heavy resonances are small and the
deviations can be approximated at leading order by the corrections to
the $hW^+W^-$ coupling, that are given by a simple function of
$\epsilon$.

We have also investigated the degree of fine-tuning, which is in
general considerable but seems in most cases to compare favorably
against the simplest SUSY scenarios (although this statement should
not be taken as a rigorous one, given the lack of a proper UV
completion for the composite Higgs scenarios).  Interestingly, we find
examples where the sensitivity of the weak scale to the underlying
model parameters is below 10\%.  However, models such as the
MCHM$_{5-10-10}$ and the MCHM$_{10-5-10}$ present a sensitivity at the
few per mille level.  We also note that the models based on the 14
representation, which have been claimed to present little
tuning~\cite{Panico:2012uw} actually are tuned at the per cent or
worse level (although we have not considered a purely composite
$t_R$).  These considerations may be suggestive of which case is more
likely to be realized in nature, although of course experimentally the
approach should be open-minded.

As the LHC and the experimental collaborations prepare for the (close
to) 14 TeV and higher luminosity run, the Higgs sector offers a unique
window into physics beyond the SM. The possibility that the Higgs
boson is a pNGB of some underlying strong dynamics remains as an
attractive framework for understanding the breaking of the EW
symmetry, and the opportunity of learning something about the detailed
properties of such a theory from Higgs measurements can be a realistic
one, as illustrated in this work.  Eventually one should be able to
produce the strong resonances, studying their properties directly, and
start cross-checking against the previous low-energy information.

\subsection*{Acknowledgements}

We thank Carlos Wagner for enlightening discussions in the beginning
of this work and \'Alex Pomarol for useful suggestions.  We also thank
Abdelhak Djouadi, Joseph Lykken, Giuliano Panico, Gilad Perez and
Francesco Riva for discussions.  E.P. and L.D. wish to thank the
Fermilab Theory Group for hospitality during various stages of this
work.  M.C.~would like to thank the Aspen Center for Physics, where
part of this work was completed.  M.C.~would like to thank ICTP-SAIFR
and Centro At\'omico Bariloche for hospitality.  M.C.~and E.P.~would
like to thank the MITP at the Johannes Gutenberg-University Mainz, where part of this work was
completed.  Fermilab is operated by Fermi Research Alliance, LLC under
contract no.  DE-AC02-07CH11359 with the United States Department of
Energy.  L.D. is partly supported by FONCYT-Argentina under the
contract PICT-2010-1737 and CONICET-Argentina under the contract PIP
114220100100319.  This work was supported by the S\~ao Paulo Research
Foundation (FAPESP) under grant \#~2011/11973.

\appendix

\section{Representations of SO(5)}
\label{app:generators}

We consider the following $5\times5$ matrix representation of the
generators $T^B$ of SO(5):
\begin{eqnarray}
T^{1}_L=\left(
\begin{array}{ccccc}
 0 & 0 & 0 & -\frac{i}{2} & 0 \\
 0 & 0 & -\frac{i}{2} & 0 & 0 \\
 0 & \frac{i}{2} & 0 & 0 & 0 \\
 \frac{i}{2} & 0 & 0 & 0 & 0 \\
 0 & 0 & 0 & 0 & 0
\end{array}
\right)\ , \ 
T^{2}_L=\left(
\begin{array}{ccccc}
 0 & 0 & \frac{i}{2} & 0 & 0 \\
 0 & 0 & 0 & -\frac{i}{2} & 0 \\
 -\frac{i}{2} & 0 & 0 & 0 & 0 \\
 0 & \frac{i}{2} & 0 & 0 & 0 \\
 0 & 0 & 0 & 0 & 0
\end{array}
\right)\ , \ 
T^{3}_L=\left(
\begin{array}{ccccc}
 0 & -\frac{i}{2} & 0 & 0 & 0 \\
 \frac{i}{2} & 0 & 0 & 0 & 0 \\
 0 & 0 & 0 & -\frac{i}{2} & 0 \\
 0 & 0 & \frac{i}{2} & 0 & 0 \\
 0 & 0 & 0 & 0 & 0
\end{array}
\right)\ , \nonumber
\end{eqnarray}
\begin{eqnarray}
T^{1}_R=\left(
\begin{array}{ccccc}
 0 & 0 & 0 & \frac{i}{2} & 0 \\
 0 & 0 & -\frac{i}{2} & 0 & 0 \\
 0 & \frac{i}{2} & 0 & 0 & 0 \\
 -\frac{i}{2} & 0 & 0 & 0 & 0 \\
 0 & 0 & 0 & 0 & 0
\end{array}
\right)\ , \ 
T^{2}_R=\left(
\begin{array}{ccccc}
 0 & 0 & \frac{i}{2} & 0 & 0 \\
 0 & 0 & 0 & \frac{i}{2} & 0 \\
 -\frac{i}{2} & 0 & 0 & 0 & 0 \\
 0 & -\frac{i}{2} & 0 & 0 & 0 \\
 0 & 0 & 0 & 0 & 0
\end{array}
\right)\ , \ 
T^{3}_R=\left(
\begin{array}{ccccc}
 0 & -\frac{i}{2} & 0 & 0 & 0 \\
 \frac{i}{2} & 0 & 0 & 0 & 0 \\
 0 & 0 & 0 & \frac{i}{2} & 0 \\
 0 & 0 & -\frac{i}{2} & 0 & 0 \\
 0 & 0 & 0 & 0 & 0
\end{array}
\right)\ , \nonumber
\end{eqnarray}
\begin{eqnarray}
T^{\hat 1}=\left(
\begin{array}{ccccc}
 0 & 0 & 0 & 0 & -\frac{i}{\sqrt{2}} \\
 0 & 0 & 0 & 0 & 0 \\
 0 & 0 & 0 & 0 & 0 \\
 0 & 0 & 0 & 0 & 0 \\
 \frac{i}{\sqrt{2}} & 0 & 0 & 0 & 0
\end{array}
\right)\ , \ 
T^{\hat 2}=\left(
\begin{array}{ccccc}
 0 & 0 & 0 & 0 & 0 \\
 0 & 0 & 0 & 0 & -\frac{i}{\sqrt{2}} \\
 0 & 0 & 0 & 0 & 0 \\
 0 & 0 & 0 & 0 & 0 \\
 0 & \frac{i}{\sqrt{2}} & 0 & 0 & 0
\end{array}
\right)\ ,\nonumber
\end{eqnarray}
\begin{eqnarray}
T^{\hat 3}=\left(
\begin{array}{ccccc}
 0 & 0 & 0 & 0 & 0 \\
 0 & 0 & 0 & 0 & 0 \\
 0 & 0 & 0 & 0 & -\frac{i}{\sqrt{2}} \\
 0 & 0 & 0 & 0 & 0 \\
 0 & 0 & \frac{i}{\sqrt{2}} & 0 & 0
\end{array}
\right)\ , \ 
T^{\hat 4}=\left(
\begin{array}{ccccc}
 0 & 0 & 0 & 0 & 0 \\
 0 & 0 & 0 & 0 & 0 \\
 0 & 0 & 0 & 0 & 0 \\
 0 & 0 & 0 & 0 & -\frac{i}{\sqrt{2}} \\
 0 & 0 & 0 & \frac{i}{\sqrt{2}} & 0
\end{array}
\right)\ .\label{T5}
\end{eqnarray}

The generators $T^B$ act on the fundamental representation {\bf 5} of
SO(5) as: $T^B\psi_{\bf 5}$.  One can label the components of a {\bf
5} by their transformation properties under $T^{3}_{L}$ and
$T^{3}_{R}$.  The following are eigenvectors of those generators:
\begin{align}
&v_{(--)}=\frac{1}{\sqrt{2}}\left(\begin{array}{c}i\\1\\0\\0\\0\end{array}\right)\ , \
v_{(-+)}=\frac{1}{\sqrt{2}}\left(\begin{array}{c}0\\0\\i\\1\\0\end{array}\right)\ , \\ \nonumber \\
&v_{(+-)}=\frac{1}{\sqrt{2}}\left(\begin{array}{c}0\\0\\-i\\1\\0\end{array}\right)\ , \
v_{(++)}=\frac{1}{\sqrt{2}}\left(\begin{array}{c}-i\\1\\0\\0\\0\end{array}\right)\ , \
v_{(00)}=\left(\begin{array}{c}0\\0\\0\\0\\1\end{array}\right)\ ,
\end{align}
with the subindices $(i,j)$ labeling the $T^{3}_{L,R}$ value, $\pm$
for $\pm1/2$.  Thus, a fermion $\psi$ in the fundamental
representation can be written as:
\begin{equation}
\psi_{\bf 5}=\psi_i \, v_{(i)} = \frac{1}{\sqrt{2}}
\left(
\begin{array}{c} 
i(\psi_{--}-\psi_{++}) \\ [0.3em]
\psi_{--}+\psi_{++} \\ [0.3em]
i(\psi_{-+}-\psi_{+-}) \\ [0.3em]
\psi_{-+}+\psi_{+-}) \\ [0.3em]
\sqrt{2}\psi_{00} 
\end{array}\right) ~.
\end{equation}

By using the $5\times5$ matrix representation, the generators $T^B$
act on the adjoint representation {\bf 10} of SO(5) as: $T^B\psi_{\bf
10}=[T^B,\psi_{\bf 10}]$.  The matrices defined in Eq.~(\ref{T5})
provide a basis for this representation.  Other useful basis is one
which can be labeled by the $T^{3}_{L,R}$ eigenvalues
$v_{(t^{3}_{L},t^{3}_{R})}$.  Since a {\bf 10} decomposes under ${\rm
SO(4)} \simeq {\rm SU(2)}_L \times {\rm SU(2)}_R$ as {$\bf 10\sim
(3,1) \oplus (1,3) \oplus (2,\bar 2)$}, we obtain:
\begin{align}
&({\bf 3},{\bf 1}): & v_{(\pm1,0)}=\frac{1}{\sqrt{2}}(T^{1}_{L}\pm iT^{2}_{L}) ~, & \qquad v_{(0,0)}=T^{3}_{L} ~, \nonumber \\ 
&({\bf 1},{\bf 3}): & v_{(0,\pm1)}=\frac{1}{\sqrt{2}}(T^{1}_{R}\pm iT^{2}_{R}) ~, & \qquad v_{(0,0)}=T^{3}_{R} ~, \\
&({\bf 2},{\bf 2}): 
& v_{(-1/2,-1/2)}=\frac{1}{\sqrt{2}}(T^{\hat 1}-i T^{\hat 2})~, \qquad
& v_{(+1/2,+1/2)}=\frac{1}{\sqrt{2}}(T^{\hat 1}+i T^{\hat 2}) ~,  \nonumber \\
& & v_{(-1/2,+1/2)}=\frac{1}{\sqrt{2}}(T^{\hat 3}-i T^{\hat 4})~, \qquad
& v_{(+1/2,-1/2)}=\frac{1}{\sqrt{2}}(T^{\hat 3}+i T^{\hat 4}) ~. \nonumber
\end{align}
A field in the adjoint of SO(5) can be written as:
%
\begin{align}
&\psi_{\bf 10}=\psi_i \, v_{(i)}
\end{align}
%

Similar to the ${\bf 10}$ representation, the ${\bf 14}$
representation of SO(5) can be written in terms of a $5\times5$
symmetric and traceless matrix.  $T^B$ acts on the ${\bf 14}$ as:
$T^B\psi_{\bf 14}=[T^B,\psi_{\bf 14}]$.  A {\bf 14} decomposes under
${\rm SU(2)}_L \times {\rm SU(2)}_R$ as {$\bf 14\sim (3,\bar
3)+(2,\bar 2)+(1,1)$}.  One basis for this representation is
\begin{eqnarray}
&({\bf 3},{\bf 3}): &T^{ab}_{ij}=\frac{1}{\sqrt{2}}(\delta^a_i\delta^b_j+\delta^a_j\delta^b_i) ~, \qquad \qquad a<b ~, \ a,b=1,\dots 4 ~, \nonumber \\
& &T^{aa}_{ij}=\frac{1}{\sqrt{2}}\left(\delta^a_i\delta^a_j-\delta^{a+1}_i\delta^{a+1}_j\right) ~, \qquad a=1,2,3 ~, \nonumber \\
&({\bf 2},{\bf 2}): &T^{\hat a}_{ij}=\frac{1}{\sqrt{2}}\left(\delta^a_i\delta^5_j+\delta^{a}_j\delta^5_i\right) ~, \qquad \qquad a=1,\dots 4 ~,\nonumber \\
&({\bf 1},{\bf 1}): & T^0_{ij}=\frac{1}{2\sqrt{5}}{\rm diag}(1,1,1,1,-4) ~.
\end{eqnarray}
Using this basis one can define a new one labeled by the $T^{3}_{L,R}$
eigenvalues: $\{v_{(t^{3}_{L},t^{3}_{R})}\}$,
\begin{align*}
&({\bf 3},{\bf 3}): & \nonumber\\
&v_{(1,1)}=\frac{1}{2\sqrt{2}}(2iT^{12}+T^{11}-T^{22}) ~, & 
v_{(1,0)}=\frac{1}{2}(-T^{13}-iT^{23}-iT^{14}+T^{24}) ~, \nonumber\\
&v_{(1,-1)}=\frac{1}{2\sqrt{2}}(2iT^{34}+T^{33}) ~, & 
v_{(0,1)}=\frac{1}{2}(-T^{13}-iT^{23}+iT^{14}-T^{24}) ~, \nonumber\\
&v_{(0,0)}=\frac{1}{2\sqrt{2}}(-T^{11}-T^{22}+T^{33}) ~, & 
v_{(0,-1)}=\frac{1}{2}(T^{13}-iT^{23}+iT^{14}+T^{24}) ~,\nonumber\\
&v_{(-1,1)}=\frac{1}{2\sqrt{2}}(-2iT^{34}+T^{33}) ~, & 
v_{(-1,0)}=\frac{1}{2}(T^{13}-iT^{23}-iT^{14}-T^{24}) ~, \nonumber\\
&v_{(-1,-1)}=\frac{1}{2\sqrt{2}}(-2iT^{12}+T^{11}-T^{22}) ~, & \nonumber\\
&({\bf 2},{\bf 2}): & \nonumber\\
&v_{(+1/2,+1/2)}=\frac{1}{\sqrt{2}}(-T^{\hat1}-iT^{\hat2}) ~,
&v_{(+1/2,-1/2)}=\frac{1}{\sqrt{2}}(T^{\hat3}+iT^{\hat4}) ~, \nonumber\\
&v_{(-1/2,+1/2)}=\frac{1}{\sqrt{2}}(T^{\hat3}-iT^{\hat4}) ~,
&v_{(-1/2,-1/2)}=\frac{1}{\sqrt{2}}(T^{\hat1}-iT^{\hat2}) ~, \nonumber\\
&({\bf 1},{\bf 1}): \ v'_{(0,0)}= T^0 ~. &
\end{align*}
A field in the ${\bf 14}$ representation can be written as
\begin{align}
&\psi_{\bf 14}=\psi_i \, v_{(i)}\ .
\end{align}
%

\section{Bosonic mass matrices}
\label{app:masses}

The charged (squared) mass matrix (in the basis $\{ w_L^+, A_L^+,
A_R^+, A^{\hat{+}} \}$ versus $\{ w_L^-, A_L^-, A_R^-, A^{\hat{-}}
\}$, where $w_L^\pm = (w^1_L \mp i \, w^2_L) / \sqrt{2}$, etc.), is
\small
\bea
M^2_C &=& \left(
\begin{array}{cccc}
\frac{1}{2} g_0^2 f_{\Omega }^2 & -\frac{1}{2} g_0 g_{\rho } f_{\Omega }^2 & 0 & 0 \\
-\frac{1}{2} g_0 g_{\rho}f_{\Omega }^2 & \frac{1}{4} g_{\rho }^2 \left( 2 f_{\Omega }^2 + f_1^2
s_h^2 \right) & -\frac{1}{4} g_{\rho}^2 f_1^2 s_h^2 & \frac{g_{\rho }^2 f_1^2 s_h c_h}{2 \sqrt{2}} \\
0 & -\frac{1}{4} g_{\rho }^2 f_1^2 s_h^2 & \frac{1}{4} g_{\rho }^2 \left(2 f_{\Omega }^2 + f_1^2
s_h^2 \right) & -\frac{g_{\rho }^2 f_1^2 s_h c_h}{2 \sqrt{2}} \\
0 & \frac{g_{\rho }^2 f_1^2 s_h c_h}{2 \sqrt{2}} & -\frac{g_{\rho }^2 f_1^2 s_h c_h}{2 \sqrt{2}} & \frac{1}{2} g_{\rho }^2 \left(f_{\Omega }^2 + f_1^2 c_h^2 \right)
\end{array}
\right)~, 
\label{M2C}
\eea
\normalsize
and the neutral (squared) mass matrix (in the basis $\{ w^3_L, b,
A^3_L, A^3_R, A^{\hat{3}}, A^{\hat{4}}, X \}$) is
\small
\begin{equation}
\left(
\begin{array}{ccccccc}
\frac{g_0^2}{2} f_{\Omega }^2 & 0 & -\frac{g_0 g_{\rho }}{2}
f_{\Omega }^2 & 0 & 0 & 0 & 0 \\
0 & \frac{g_0^2 g_x^2 \left(f_{\Omega }^2 + f_{\Omega_X}^2\right)}{2
\left(g_0^2 + g_x^2\right)} & 0 & -\frac{g_0 g_x g_{\rho } f_{\Omega
}^2}{2 \sqrt{g_0^2 + g_x^2}} & 0 & 0 & -\frac{g_0 g_x g_X f_{\Omega_X}^2}{2 \sqrt{g_0^2 + g_x^2}} \\
-\frac{g_0 g_{\rho }}{2} f_{\Omega }^2 & 0 & \frac{g_{\rho }^2}{2}
\left(f_{\Omega }^2 + \frac{f_1^2 s_h^2}{2} \right) & -\frac{g_{\rho
}^2}{4} f_1^2 s_h^2 & \frac{g_{\rho }^2 f_1^2 s_h c_h}{2 \sqrt{2}} & 0
& 0 \\
0 & -\frac{g_0 g_x g_{\rho } f_{\Omega }^2}{2 \sqrt{g_0^2 + g_x^2}} &
-\frac{g_{\rho }^2}{4} f_1^2 s_h^2 & \frac{g_{\rho }^2 }{2}
\left(f_{\Omega }^2 + \frac{f_1^2 s_h^2}{2} \right) & -\frac{g_{\rho }^2
f_1^2 s_h c_h}{2 \sqrt{2}} & 0 & 0 \\
0 & 0 & \frac{g_{\rho }^2 f_1^2 s_h c_h}{2 \sqrt{2}} & -\frac{g_{\rho
}^2 f_1^2 s_h c_h}{2 \sqrt{2}} & \frac{g_{\rho }^2 }{2} \left(f_{\Omega }^2 + f_1^2
c_h^2 \right) & 0 & 0 \\
0 & 0 & 0 & 0 & 0 & \frac{g_{\rho }^2 }{2} \left(f_{\Omega
}^2+f_1^2\right) & 0 \\
0 & -\frac{g_0 g_x g_X f_{\Omega_X}^2}{2 \sqrt{g_0^2 + g_x^2}} & 0 & 0
& 0 & 0 & \frac{g_X^2 }{2} f_{\Omega_X}^2
\end{array}
\right)~.
\nonumber \\
\end{equation}
\normalsize
%

\section{Correlators}
\label{sec:correlators}

In this appendix we express the fermionic correlators of all the
models in the SO(4) symmetric phase in terms of the following general
functions.
\begin{align}
&A_L(m_1,m_2,m_3,m_4,\Delta)= \Delta^2 \left[m_1^2 m_2^2 + m_1^2 m_4^2+m_2^2 m_3^2-p^2 (m_1^2+m_2^2+m_3^2+m_4^2)+p^4\right]\ ; \nonumber \\
&A_R(m_1,m_2,m_3,m_4,\Delta)= \Delta^2 \left[m_1^2 m_2^2+m_2^2 m_3^2-p^2 (m_1^2+m_2^2+m_3^2+m_4^2)+p^4\right]\ ; \nonumber \\
&A_M(m_1,m_2,m_3,m_4,\Delta_1,\Delta_2)= \Delta_1 \Delta_2 \ m_1\ m_2\ m_4 (m_3^2 - p^2)\ ; \nonumber \\
&B(m_1,m_2,m_3,m_4,m_5)= m_1^2 m_2^2 m_3^2 - p^2 (m_1^2 m_2^2+m_1^2 m_3^2+m_2^2 m_3^2+m_2^2 m_5^2+m_3^2 m_4^2) \nonumber \\ 
& \qquad \qquad \qquad \qquad \qquad + p^4 (m_1^2+m_2^2+m_3^2+m_4^2+m_5^2) -p^6 \ .
\end{align}

In the following expressions we use the notation $y_T = y_u$, $y_B =
y_d$, $m_{y_T} = m_{y_u}$, $m_{y_B} = m_{y_d}$, $\Delta_T = \Delta_u$
and $\Delta_B = \Delta_b$ [where the Lagrangian parameters were
defined for each model in Sec.~\ref{sec:models}] to emphasize the role
of the third generation.

\subsection{MCHM$_5$}
\label{app:MCHM5}

\vspace*{-6mm}
\begin{align}
&\hat\Pi_{q^{u(1)}}=\frac{A_L(m_T,0,m_{y_T}+y_T,0,\Delta_{q^u})}{B(m_{Q^u},m_T,0,m_{y_T}+y_T,0)} \ ,  \qquad 
&\hat\Pi_{q^{u(4)}}&=\frac{A_L(m_T,0,m_{y_T},0,\Delta_{q^u})}{B(m_{Q^u},m_T,0,m_{y_T},0)} \ , \nonumber \\
&\hat\Pi_{q^{d(1)}}=\frac{A_L(m_B,0,m_{y_B}+y_B,0,\Delta_{q^d})}{B(m_{Q^d},m_B,0,m_{y_B}+y_B,0)} \ ,  \qquad 
&\hat\Pi_{q^{d(4)}}&=\frac{A_L(m_B,0,m_{y_B},0,\Delta_{q^d})}{B(m_{Q^d},m_B,0,m_{y_B},0)} \ , \nonumber \\
&\hat\Pi_{u^{(1)}}=\frac{A_R(m_{Q^u},0,m_{y_T}+y_T,0,\Delta_t)}{B(m_{Q^u},m_T,0,m_{y_T}+y_T,0)} \ ,  \qquad 
&\hat\Pi_{u^{(4)}}&=\frac{A_R(m_{Q^u},0,m_{y_T},0,\Delta_t)}{B(m_{Q^u},m_T,0,m_{y_T},0)} \ , \nonumber \\
&\hat\Pi_{d^{(1)}}=\frac{A_R(m_{Q^d},0,m_{y_B}+y_B,0,\Delta_b)}{B(m_{Q^d},m_B,0,m_{y_B}+y_B,0)} \ ,  \qquad 
&\hat\Pi_{d^{(4)}}&=\frac{A_R(m_{Q^d},0,m_{y_B},0,\Delta_b)}{B(m_{Q^d},m_B,0,m_{y_B},0)} \ , \nonumber \\
&\hat M_{u^{(1)}}=\frac{A_M(m_{Q^u},m_T,0,m_{y_T}+y_T,\Delta_{q^u},\Delta_t)}{B(m_{Q^u},m_T,0,m_{y_T}+y_T,0)} \ ,  \qquad 
&\hat M_{u^{(4)}}&=\frac{A_M(m_{Q^u},m_T,0,m_{y_T},\Delta_{q^u},\Delta_t)}{B(m_{Q^u},m_T,0,m_{y_T},0)} \ , \nonumber \\
&\hat M_{d^{(1)}}=\frac{A_M(m_{Q^d},m_B,0,m_{y_B}+y_B,\Delta_{q^d},\Delta_b)}{B(m_{Q^d},m_B,0,m_{y_B}+y_B,0)} \ ,  \qquad 
&\hat M_{d^{(4)}}&=\frac{A_M(m_{Q^d},m_B,0,m_{y_B},\Delta_{q^d},\Delta_b)}{B(m_{Q^d},m_B,0,m_{y_B},0)} \ , 
\nonumber
\end{align}
\vspace*{-12mm}

\subsection{MCHM$_{10}$}
\label{app:MCHM10}

\vspace*{-6mm}
\begin{align}
&\hat\Pi_{q^{(4)}}=\frac{A_L(m_T,m_B,m_{y_T}+y_T/2,m_{y_B}+y_B/2,\Delta_{q})}{B(m_{Q},m_T,m_B,m_{y_T}+y_T/2,m_{y_B}+y_B/2)} \ ,  \qquad 
&\hat\Pi_{q^{(6)}}&=\frac{A_L(m_T,m_B,m_{y_T},m_{y_B},\Delta_{q})}{B(m_{Q},m_T,m_B,m_{y_T},m_{y_B})} \ , \nonumber \\ 
&\hat\Pi_{u^{(4)}}=\frac{A_R(m_{Q},m_B,m_{y_T}+y_T/2,m_{y_B}+y_B/2,\Delta_t)}{B(m_{Q},m_T,m_B,m_{y_T}+y_T/2,m_{y_B}+y_B/2)} \ ,  \qquad 
&\hat\Pi_{u^{(6)}}&=\frac{A_R(m_{Q},m_B,m_{y_T},m_{y_B},\Delta_t)}{B(m_{Q},m_T,m_B,m_{y_T},m_{y_B})} \ , \nonumber \\ 
&\hat\Pi_{d^{(4)}}=\frac{A_R(m_{Q},m_T,m_{y_B}+y_B/2,m_{y_T}+y_T/2,\Delta_b)}{B(m_{Q},m_T,m_B,m_{y_T}+y_T/2,m_{y_B}+y_B/2)} \ ,  \qquad 
&\hat\Pi_{d^{(6)}}&=\frac{A_R(m_{Q},m_T,m_{y_B},m_{y_T},\Delta_b)}{B(m_{Q},m_T,m_B,m_{y_T},m_{y_B})} \ , \nonumber \\ 
&\hat M_{u^{(4)}}=\frac{A_M(m_{Q},m_T,m_B,m_{y_T}+y_T/2,\Delta_{q},\Delta_t)}{B(m_{Q},m_T,m_B,m_{y_T}+y_T/2,m_{y_B}+y_B/2)} \ ,  \qquad 
&\hat M_{u^{(6)}}&=\frac{A_M(m_{Q},m_T,m_B,m_{y_T},\Delta_{q},\Delta_t)}{B(m_{Q},m_T,m_B,m_{y_T},m_{y_B})} \ , \nonumber \\ 
&\hat M_{d^{(4)}}=\frac{A_M(m_{Q},m_B,m_T,m_{y_B}+y_B/2,\Delta_{q},\Delta_b)}{B(m_{Q},m_T,m_B,m_{y_T}+y_T/2,m_{y_B}+y_B/2)} \ ,  \qquad 
&\hat M_{d^{(6)}}&=\frac{A_M(m_{Q},m_B,m_T,m_{y_B},\Delta_{q},\Delta_b)}{B(m_{Q},m_T,m_B,m_{y_T},m_{y_B})} \ .
\nonumber
\end{align}
\vspace*{-12mm}

\subsection{MCHM$_{10-5-10}$}
\label{app:MCHM10510}

\vspace*{-6mm}
\begin{align}
&\hat\Pi_{q^{(4)}}=\frac{A_L(m_T,m_B,y_T/\sqrt{2},m_{y_B}+y_B/2,\Delta_{q})}{B(m_{Q},m_T,m_B,y_T/\sqrt{2},m_{y_B}+y_B/2)} \ ,  \qquad 
&\hat\Pi_{q^{(6)}}&=\frac{A_L(0,m_B,0,m_{y_B},\Delta_{q})}{B(m_{Q},0,m_B,0,m_{y_B})} \ , \nonumber \\ 
&\hat\Pi_{u^{(4)}}=\frac{A_R(m_{Q},m_B,y_T/\sqrt{2},m_{y_B}+y_B/2,\Delta_t)}{B(m_{Q},m_T,m_B,y_T/\sqrt{2},m_{y_B}+y_B/2)} \ ,  \qquad 
&\hat\Pi_{u^{(1)}}&=\frac{A_R(0,0,0,0,\Delta_t)}{B(0,m_T,0,0,0)} \ , \nonumber \\ 
&\hat\Pi_{d^{(4)}}=\frac{A_R(m_{Q},m_T,m_{y_B}+y_B/2,y_T/\sqrt{2},\Delta_b)}{B(m_{Q},m_T,m_B,y_T/\sqrt{2},m_{y_B}+y_B/2)} \ ,  \qquad 
&\hat\Pi_{d^{(6)}}&=\frac{A_R(0,m_{Q},0,m_{y_B},\Delta_b)}{B(m_{Q},0,m_B,0,m_{y_B}} \ , \nonumber \\ 
&\hat M_{u^{(4)}}=\frac{A_M(m_{Q},m_T,m_B,y_T/\sqrt{2},m_{y_B},\Delta_{q},\Delta_t)}{B(m_{Q},m_T,m_B,y_T/\sqrt{2},m_{y_B}+y_B/2)} \ ,  \qquad 
& \nonumber \\ 
&\hat M_{d^{(4)}}=\frac{A_M(m_{Q},m_B,m_T,m_{y_B}+y_B/2,\Delta_{q},\Delta_b)}{B(m_{Q},m_T,m_B,y_T/\sqrt{2},m_{y_B}+y_B/2)} \ ,  \qquad 
&\hat M_{d^{(6)}}&=\frac{A_M(m_{Q},m_B,0,m_{y_B},\Delta_{q},\Delta_b)}{B(m_{Q},0,m_B,0,m_{y_B})} \ .
\nonumber
\end{align}
\vspace*{-12mm}

\subsection{MCHM$_{5-5-10}$}
\label{app:MCHM5510}

\vspace*{-6mm}
\begin{align}
&\hat\Pi_{q^{(4)}}=\frac{A_L(m_T,m_B,m_{y_T},y_B/\sqrt{2},\Delta_{q})}{B(m_{Q},m_T,m_B,m_{y_T},y_B/\sqrt{2})} \ ,  \qquad 
&\hat\Pi_{q^{(1)}}&=\frac{A_L(m_T,0,m_{y_T}+y_T,0,\Delta_{q})}{B(m_{Q},m_T,0,m_{y_T}+y_T,0)} \ , \nonumber \\ 
&\hat\Pi_{u^{(4)}}=\frac{A_R(m_{Q},m_B,m_{y_T},y_B/\sqrt{2},\Delta_t)}{B(m_{Q},m_T,m_B,m_{y_T},y_B/\sqrt{2})} \ ,  \qquad 
&\hat\Pi_{u^{(1)}}&=\frac{A_R(m_{Q},0,m_{y_T}+y_T,0,\Delta_t)}{B(m_{Q},m_T,0,m_{y_T}+y_T,0)} \ , \nonumber \\ 
&\hat\Pi_{d^{(4)}}=\frac{A_R(m_{Q},m_T,y_B/\sqrt{2},m_{y_T},\Delta_b)}{B(m_{Q},m_T,m_B,m_{y_T},y_B/\sqrt{2})} \ ,  \qquad 
&\hat\Pi_{d^{(6)}}&=\frac{A_R(0,0,0,0,\Delta_b)}{B(0,0,m_B,0,0} \ , \nonumber \\ 
&\hat M_{u^{(4)}}=\frac{A_M(m_{Q},m_T,m_B,m_{y_T},\Delta_{q},\Delta_t)}{B(m_{Q},m_T,m_B,m_{y_T},y_B/\sqrt{2})} \ ,  \qquad 
&\hat M_{u^{(1)}}&=\frac{A_M(m_{Q},m_T,0,m_{y_T}+y_T,\Delta_{q},\Delta_t)}{B(m_{Q},m_T,0,m_{y_T}+y_T,0)} \ ,  \nonumber \\ 
&\hat M_{d^{(4)}}=\frac{A_M(m_{Q},m_B,m_T,y_B/\sqrt{2},\Delta_{q},\Delta_b)}{B(m_{Q},m_T,m_B,m_{y_T},y_B/\sqrt{2})} \ ,  \qquad 
& \ .
\nonumber
\end{align}
\vspace*{-12mm}

\subsection{MCHM$_{5-10-10}$}
\label{app:MCHM51010}

\vspace*{-6mm}
\begin{align}
&\hat\Pi_{q^{(4)}}=\frac{A_L(m_T,m_B,y_T/\sqrt{2},y_B/\sqrt{2},\Delta_{q})}{B(m_{Q},m_T,m_B,y_T/\sqrt{2},y_B/\sqrt{2})} \ ,  \qquad 
&\hat\Pi_{q^{(1)}}&=\frac{A_L(0,0,0,0,\Delta_{q})}{B(m_{Q},0,0,0,0)} \ , \nonumber \\ 
&\hat\Pi_{u^{(4)}}=\frac{A_R(m_{Q},m_B,y_T/\sqrt{2},y_B/\sqrt{2},\Delta_t)}{B(m_{Q},m_T,m_B,y_T/\sqrt{2},y_B/\sqrt{2})} \ ,  \qquad 
&\hat\Pi_{u^{(6)}}&=\frac{A_R(0,0,0,0,\Delta_t)}{B(0,m_T,0,0,0)} \ , \nonumber \\ 
&\hat\Pi_{d^{(4)}}=\frac{A_R(m_{Q},m_T,y_B/\sqrt{2},y_T/\sqrt{2},\Delta_b)}{B(m_{Q},m_T,m_B,y_T/\sqrt{2},y_B/\sqrt{2})} \ ,  \qquad 
&\hat\Pi_{d^{(6)}}&=\frac{A_R(0,0,0,0,\Delta_b)}{B(0,0,m_B,0,0)} \ , \nonumber \\ 
&\hat M_{u^{(4)}}=\frac{A_M(m_{Q},m_T,m_B,y_T/\sqrt{2},\Delta_{q},\Delta_t)}{B(m_{Q},m_T,m_B,y_T/\sqrt{2},y_B/\sqrt{2})} \ , 
& \nonumber \\ 
&\hat M_{d^{(4)}}=\frac{A_M(m_{Q},m_B,m_T,y_B/\sqrt{2},\Delta_{q},\Delta_b)}{B(m_{Q},m_T,m_B,y_T/\sqrt{2},y_B/\sqrt{2})} \ ,  \qquad 
& \ .
\nonumber
\end{align}
\vspace*{-12mm}

\subsection{MCHM$_{14-1-10}$}
\label{app:MCHM14110}

\vspace*{-6mm}
\begin{align}
&\hat\Pi_{q^{(9)}}=\frac{A_L(0,0,0,0,\Delta_{q})}{B(m_{Q},0,0,0,0)} \ ,  \qquad 
&\hat\Pi_{q^{(4)}}&=\frac{A_L(0,m_B,0,y_B/2,\Delta_{q})}{B(m_{Q},0,m_B,0,y_B/2)} \ , \nonumber \\ 
&\hat\Pi_{q^{(1)}}=\frac{A_L(m_T,0,y_T \sqrt{4/5},0,\Delta_{q})}{B(m_{Q},m_T,0,y_T \sqrt{4/5},0)} \ ,  \nonumber \\
&\hat\Pi_{u^{(1)}}=\frac{A_R(m_{Q},0,y_T \sqrt{4/5},0,\Delta_t)}{B(m_{Q},m_T,0,y_T \sqrt{4/5},0)} \ , \nonumber \\ 
&\hat\Pi_{d^{(4)}}=\frac{A_R(m_{Q},0,y_B/2,0,\Delta_b)}{B(m_{Q},0,m_B,0,y_B/2)} \ ,  \qquad 
&\hat\Pi_{d^{(6)}}&=\frac{A_R(0,0,0,0,\Delta_b)}{B(0,m_B,0,0,0)} \ , \nonumber \\ 
&\hat M_{u^{(4)}}=-\frac{A_M(m_{Q},m_T,0,y_T\sqrt{4/5},\Delta_{q},\Delta_t)}{B(m_{Q},m_T,0,y_T\sqrt{4/5},0)} \ , \nonumber \\ 
&\hat M_{d^{(4)}}=-i\frac{A_M(m_{Q},m_B,0,y_B/2,\Delta_{q},\Delta_b)}{B(m_{Q},0,m_B,0,y_B/2)} \ ,  
\nonumber
\end{align}
\vspace*{-12mm}

\subsection{MCHM$_{14-14-10}$}
\label{app:MCHM141410}

\vspace*{-2mm}
Here we use the notation $\bar{y}_T = y_T + \tilde{y}_T$ that includes the two Yukawa structures displayed in Eqs.~(\ref{14y1}) and (\ref{14y2}), which enters in the singlet terms below:
\begin{align}
&\hat\Pi_{q^{(9)}}=\frac{A_L(m_T,0,m_{y_T},0,\Delta_{q})}{B(m_{Q},m_T,0,m_{y_T},0)} \ ,  \qquad 
&\hat\Pi_{q^{(4)}}&=\frac{A_L(m_T,m_B,m_{y_T}+y_T/2,y_B/2,\Delta_{q})}{B(m_{Q},m_T,m_B,m_{y_T}+y_T/2,y_B/2)} \ , \nonumber \\ 
&\hat\Pi_{q^{(1)}}=\frac{A_L(m_T,0,m_{y_T}+\bar{y}_T 4/5,0,\Delta_{q})}{B(m_{Q},m_T,0,m_{y_T}+\bar{y}_T 4/5,0)} \ ,  \nonumber \\
&\hat\Pi_{u^{(9)}}=\frac{A_R(m_{Q},0,m_{y_T},0,\Delta_t)}{B(m_{Q},m_T,0,m_{y_T},0)} \ ,  \qquad 
&\hat\Pi_{u^{(4)}}&=\frac{A_R(m_{Q},m_B,m_{y_T}+y_T/2,y_B/2,\Delta_t)}{B(m_{Q},m_T,m_B,m_{y_T}+y_T/2,y_B/2)} \ , \nonumber \\ 
&\hat\Pi_{u^{(1)}}=\frac{A_R(m_{Q},0,m_{y_T}+\bar{y}_T 4/5,0,\Delta_t)}{B(m_{Q},m_T,0,m_{y_T}+\bar{y}_T 4/5,0)} \ , \nonumber \\ 
&\hat\Pi_{d^{(4)}}=\frac{A_R(m_{Q},m_T,y_B/2,m_{y_T}+y_T/2,\Delta_b)}{B(m_{Q},m_T,m_B,m_{y_T}+y_T/2,y_B/2)} \ ,  \qquad 
&\hat\Pi_{d^{(6)}}&=\frac{A_R(m_{Q},0,0,0,\Delta_b)}{B(m_{Q},m_B,0,0,0)} \ , \nonumber \\ 
&\hat M_{u^{(9)}}=\frac{A_M(m_{Q},m_T,0,m_{y_T},\Delta_{q},\Delta_t)}{B(m_{Q},m_T,0,m_{y_T},0)} \ ,  \qquad 
&\hat M_{u^{(4)}}&=\frac{A_M(m_{Q},m_T,m_B,m_{y_T}+y_T/2,\Delta_{q},\Delta_t)}{B(m_{Q},m_T,m_B,m_{y_T}+y_T/2,y_B/2)} \ , \nonumber \\ 
&\hat M_{u^{(1)}}=\frac{A_M(m_{Q},m_T,0,m_{y_T}+\bar{y}_T 4/5,\Delta_{q},\Delta_t)}{B(m_{Q},m_T,0,m_{y_T}+\bar{y}_T 4/5,0)} \ , \nonumber \\ 
&\hat M_{d^{(4)}}=-i\frac{A_M(m_{Q},m_B,m_T,y_B/2,\Delta_{q},\Delta_b)}{B(m_{Q},m_T,m_B,m_{y_T}+y_T/2,y_B/2)} \ ,
\nonumber  
\end{align}
%
  
\section{Loop-level Processes}
\label{app:loops}

We collect here the expressions for the processes $h \to gg$, $h \to
\gamma\gamma$ and $h \to Z\gamma$.  We focus on the amplitudes only,
since the decay rates are obtained by rescaling the SM rates.  This
allows one to include the state of the art QCD corrections, under the
assumption that the K-factors for the SM and new physics diagrams are
common.  The full details for the SM expressions can be found, for
instance, in~\cite{Djouadi:2005gi}.

For the $h \to gg$ and $h \to \gamma\gamma$ amplitudes the relevant
loop functions are
\bea
A_{1/2}(\tau) &=& 2 [\tau + (\tau -1) f(\tau)] \tau^{-2}~, \\ [0.3em]
A_{1}(\tau) &=& - [2 \tau^2 + 3 \tau + 3(2\tau - 1) f(\tau)] \tau^{-2}~,
\eea
where
\bea
f(\tau) &=& 
\left\{
\begin{array}{ll}
{\rm arcsin}^2 \sqrt{\tau}  &  \tau \leq 1   \\ [0.4em]
\normalsize
- \frac{1}{4} \left[ \log \frac{1 + \sqrt{1 - \tau^{-1}}}{1 - \sqrt{1 - \tau^{-1}}} - i \pi \right]^2 &  \tau > 1
\end{array}
\right.
~.
\label{floops}
\eea
We obtain our amplitude for the gluon fusion process from
\bea
{\cal A}(h \to gg) &\propto& v_{\rm SM} \, \sum_{\psi = t,b} \left\{ \frac{4}{3}\left[\tr(Y_\psi M_\psi^{-1})-\frac{y^{(0)}_\psi}{m^{(0)}_\psi}\right] + \frac{y^{(0)}_\psi}{m^{(0)}_\psi} \, A_{1/2}\left(\frac{m_h^2}{4 m^{(0) \, 2}_\psi} \right) \right\}~,
\label{Ahgg}
\eea
where $y^{(0)}_\psi$ and $m^{(0)}_\psi$ are the Yukawa coupling and
mass of the lightest state (identified with the SM fermion) in the
corresponding tower, obtained by numerical diagonalization of the full
Yukawa and mass matrices, $Y_\psi$ and $M_\psi$, respectively.  The
traces can be read for each model from Table.~\ref{table-F}.  The SM
amplitude, in the same normalization as Eq.~(\ref{Ahgg}), reads ${\cal
A}(h \to gg)^{\rm SM} \propto A_{1/2}(m_h^2 / 4 m^2_t) + A_{1/2}(m_h^2
/ 4 m^2_b)$.

For the diphoton channel, we use
\bea
\frac{{\cal A}(h \to \gamma\gamma)}{v_{\rm SM}} &\propto& 
-7 \left[ \frac{\cot(v/f_h)}{f_h} - \frac{g^2_W v_{\rm SM}}{4 m^2_W} \right] + \frac{g^2_W v_{\rm SM}}{4 m^2_W} \, A_{1}(m_h^2 / 4 m^2_W)
\nonumber \\ [0.3em]
& & \mbox{} + \sum_{\psi = t,b} N_c Q_\psi^2 \left\{ \frac{4}{3}\left[\tr(Y_\psi M_\psi^{-1})-\frac{y^{(0)}_\psi}{m^{(0)}_\psi}\right] + \frac{y^{(0)}_\psi}{m^{(0)}_\psi} \, A_{1/2}\left(\frac{m_h^2}{4 m^{(0) \, 2}_\psi} \right) \right\}~,
\label{Ah2gamma}
\eea
where for the $W$-tower we used that the analogue of the fermion trace
is $\frac{1}{2} \, d \log ({\rm det} M^2_C) / d v = \cot(v/f_h) / f_h$
with $M^2_C$ the squared mass matrix in the charged sector [see
Eq.~(\ref{M2C}) of App.~\ref{app:masses}].  The $W$-mass squared,
$m^2_W$, corresponds to the lightest eigenvalue of $M^2_C$, and the
coupling $g^2_W$ is defined as the diagonal entry corresponding to
this lightest state in the matrix $G^2_C = (2 / v_{\rm SM}) \, d M^2_C
/ dv$, after rotating to the mass eigenbasis.  Both $m_W$ and $g^2_W$
are obtained numerically.  The SM amplitude in the normalization of
Eq.~(\ref{Ah2gamma}) is ${\cal A}(h \to \gamma\gamma)^{\rm SM} \propto
A_1(m^2_h / 4 m^2_W) + N_c Q_t^2 A_{1/2}(m_h^2 / 4 m^2_t) + N_c Q_b^2
A_{1/2}(m_h^2 / 4 m^2_b)$.

The new feature in the $h \to Z\gamma$ process compared to the
previous ones~\cite{Djouadi:1996yq} is that there can be two different
particle species running in the loop (since the $Z$ vertex corresponds
to a broken gauge symmetry).  For the fermionic contributions we use
the general formulas presented in App.~F of
Ref.~\cite{Azatov:2013ura}, which allow to include such
``non-diagonal" contributions.  These expressions are written in terms
of the Passarino-Veltman 1-loop functions and we use the package
LoopTools~\cite{Hahn:1998yk} to evaluate them numerically.

For the charged $W$ and heavy partner loops in $h \to Z\gamma$ there
is no analogue general formula for the non-diagonal contributions.
Since we expect such effects to be negligible due to the large masses
involved, we are satisfied with including only the diagonal gauge
effects, which are completely dominated by the SM-$W$ loop itself (but
the result is different from the SM one due to the modified
couplings).  The diagonal terms can be written in terms of
\bea
A_{1}(\tau,\lambda) &=& 4 \left( 3 - \frac{s^2_W}{c^2_W}  \right) I_2(\tau,\lambda) 
+ \left[ \left( 1 + \frac{2}{\tau} \right) \frac{s^2_W}{c^2_W} - \left( 5 + \frac{2}{\tau} \right) \right] I_1(\tau,\lambda)~,
\eea
where
\bea
I_1(\tau,\lambda) &=& \frac{\tau \lambda}{2(\tau - \lambda)} + \frac{\tau^2 \lambda^2}{2(\tau - \lambda)^2} \left[ f(\tau^{-1}) - f(\lambda^{-1}) \right] + \frac{\tau^2 \lambda}{(\tau - \lambda)^2} \left[ g(\tau^{-1}) - g(\lambda^{-1}) \right]~,
\\ [0.4 em]
I_2(\tau,\lambda) &=& - \frac{\tau \lambda}{2(\tau - \lambda)} \left[ f(\tau^{-1}) - f(\lambda^{-1}) \right]~,
\eea
with $f(\tau)$ as defined in Eq.~(\ref{floops}) and 
\bea
g(\tau) &=& 
\left\{
\begin{array}{ll}
\sqrt{\tau^{-1} - 1} \, {\rm arcsin}^2 \sqrt{\tau}  &  \tau \leq 1   \\ [0.4em]
\normalsize
\frac{\sqrt{1 - \tau^{-1}}}{2} \left[ \log \frac{1 + \sqrt{1 - \tau^{-1}}}{1 - \sqrt{1 - \tau^{-1}}} - i \pi \right] &  \tau > 1
\end{array}
\right.
~.
\label{gloops}
\eea
The expression for the $h \to Z\gamma$ amplitude, keeping only the
fermionic and the $W$ contributions, is:
\bea
\frac{{\cal A}(h \to Z\gamma)}{v_{\rm SM}} &\propto& 
\left( \frac{g^2_W v_{\rm SM}}{4 m^2_W} \right) g_{W^+ W^- Z} \times A_{1}(m_h^2 / 4 m^2_W, m^2_Z / 4 m^2_W)
\nonumber \\ [0.3em]
& & \mbox{} + \sum_{\psi = t_L, t_R, b_L, b_R} \sum_{ij} 4N_c Q_\psi \lambda^h_{\psi, ij} \lambda^Z_{\psi, ji} F(m_i, m_j, m_h, m_Z)~,
\label{AhZgamma}
\eea
where $g_W^2$ was defined above and $g_{W^+ W^- Z}$ is the coupling of
the $Z$ to a $W^+ W^-$ pair in the given model (in the SM one has
$g_{W^+ W^- Z} = g \, c_W$).  It is obtained by projecting the
appropriate mass eigenstates after diagonalization of the full system,
and therefore includes the effects of mixing with the heavy spin-1
resonances.  For the fermionic contribution: $\lambda^h_{\psi, ij}$
and $\lambda^Z_{\psi, ji}$ are the couplings to the Higgs and $Z$ of
the fermion mass eigenstates $i$ and $j$ (see conventions in Eq.~(F.1)
of Ref.~\cite{Azatov:2013ura}), with masses $m_i$ and $m_j$.  These
are obtained numerically by writing the corresponding coupling
matrices in the mass eigenbasis.  The function $F(m_i, m_j, m_h, m_Z)$
is given in Eq.~(F.3) of Ref.~\cite{Azatov:2013ura}, and in the limit
$m_1 = m_2 \equiv m$, reduces to $(1/2m) A_{1/2}(m^2_h / 4m^2, m^2_Z /
4m^2)$, where $A_{1/2}(\tau,\lambda)$ is the standard fermionic loop
function for this process (see e.g.~\cite{Djouadi:2005gi}):
\bea
A_{1/2}(\tau,\lambda) &=& I_1(\tau,\lambda) - I_2(\tau,\lambda)~.
\eea
Using the same normalization as above, we have ${\cal A}(h \to
Z\gamma)^{\rm SM} \propto g c_W A_1(m^2_h / 4 m^2_W, m^2_Z / 4 m^2_W)
+ \sum_{i = t,b} 2 N_c Q_i (g/c_W) (\frac{1}{2} - 2 Q_i s^2_W)
A_{1/2}(m_h^2 / 4 m^2_i, m^2_Z / 4 m^2_i)$.  


\end{document}